# Title: The Impact of US Medical Product Regulatory Complexity on Innovation: Preliminary Evidence of Interdependence, Early Acceleration, and Subsequent Inversion

Author: Iraj Daizadeh, PhD, Global Regulatory Affairs, Takeda Pharmaceuticals, 40 Landsdowne St. Cambridge, MA, 02139, iraj.daizadeh@takeda.com, ORCID: 0000-0003-3648-023X

Abstract: Is the complexity of medical product (medicines and medical devices) regulation impacting innovation in the US? If so, how? Here, this question is investigated as follows: Various novel proxy metrics of regulation (FDA-issued guidelines) and innovation (corresponding FDA-registrations) from 1976-2020 are used to determine interdependence, a concept relying on strong correlation and reciprocal causality (estimated via variable lag transfer entropy and wavelet coherence). Based on this interdependence, a mapping of regulation onto innovation is conducted and finds that regulation seems to accelerate then supports innovation until on or around 2015; at which time, an inverted U-curve emerged. If empirically evidentiary, an important innovation-regulation nexus in the US has been reached; and, as such, stakeholders should (re)consider the complexity of the regulatory landscape to enhance US medical product innovation. Study limitations, extensions, and further thoughts complete this investigation.

# 1. Introduction

## 1.1 An Overview of Regulation and Innovation: Theoretical and Empirical Underpinnings

A common and prevailing discourse in regulated markets – such as those of medical products, energy and electric power, motor vehicles, air transportation, fishing, and telecommunications – relates to the impact of regulation[1] onto innovation[2] (Stewart, 1981). Typically, the dialogue evolves to one of competition, as it is generally thought that (de) regulation may (increase) decrease innovation, most visibly when competition among firms is fierce (viz., 'neck-to-neck') (Aghion et al, 2014). Two quasi-opposing canonical theories describe this association (Aghion et al., 2014; Franco, et al., 2016): The Schumpeterian Effect (Schumpeter, 1942) considers innovation profits gained via those capable of investing in such activities (large actors with market hegemony), while dissimilarly the Escape Competition Effect (Arrow, 1962) regards the economic costs incurred by the entry of disruptive technologies by these same actors (that is, an inertia to adapt to new market entrants due to a cost/revenue stabilizing force of the 'status quo'). Both arguments find some level of reconciliation when the relationship between competition and innovation present as an inverted U-shaped; that is, a threshold exists where the rates of competition (regulation) begin to impede innovation (or vice versa). Supportive empirical evidence has been observed across a select set of geographies and sectors (Aghion, et al., 2005; and most recently, Polemis and Tselekounis, 2021[3] and Pan et al., 2022). See Quignon (2022) and Bentzen et al. (2021) (and references therein) for brief reviews and ancillary extrapolations to further characterize the interdependence.

Corollary inquiries beyond the regulation-competition-innovation axes originate from axiomatic assumptions of the role of regulation and innovation (see Stewart, 1981 footnote 71, page 1278), specifically through placing technical constraints on the firm (what can and cannot be done), (en)forcing compliance investments (in people, processes, and systems), evolving complexity (and thus temporal uncertainty), and perceived delays (e.g., due to internal and external regulatory reviews) (ibid, see Section A, page 1279 and footnote 72). Arguably, these works generally view the problem from both legal and economic lenses unidirectionally. For example, following Bentzen et al. (2021), regulatory flexibility (interpretability of the regulatory landscape) exists as a competitive advantage when the firm may quickly (compared with competitors) make pertinent alterations in its people, processes, and systems to embrace a recent regulatory change (ibid, page 9). However, the flexibility may be challenging if the regulation is complex (and thus difficult to decipher). Market innovations would reflect the complexity of the regulations embedded within it (e.g., new product labeling). Complexity drives competitive advantage in terms of how the firm responds to the complexity of the imposing regulation (ibid, page 10). Ultimately, this perspective feeds into the institutional innovation of the firm (Almalki and Durugbo, 2022).

[1] Herein, defined broadly as a collection of laws, regulations, and guidelines.
[2] Herein, defined broadly as the discovery, development and delivery of novel products and services.
[3] Ribeiro et al. (2022) found a "innovation stimuli regulation" achieving positive response in a similar dataset.

Alternatively, regulation inspiring innovation has also been investigated, particularly when there is interest in nurturing a novel technology. Notably ‘green energies’ and more generally electricity are clear examples of such initiatives (see, e.g., Schreiber, 2022; Clausen, 2022; Ribeiro et al., 2022). In this framework, and using the environmental sector as reference, regulation may be used to “support and steer innovation, if constructed properly (Clausen, 2022, section 1.1, page 2).” Under the thesis of the ‘Porter hypothesis’ strict environmental regulation may actually trigger innovation, albeit there is some general disagreement in the literature in the general case (see, e.g., Prokop et al., 2022). This work has led to the Innovation Triple Helix, a theory placing regulation as the center of innovation – embracing industry, regulatory authorities, and academia/researchers (ibid).

Now, as there are 3.12M publications on scholar.google.com,[4] however pertinent the above presents only a (very small) portion of the literature space of regulation and innovation. To the author's knowledge, there is no one theoretical model that may completely explicate the interplay between regulation and innovation, albeit progress has been made in both the regulatory and innovation landscapes to understands its drivers (see, e.g., the 10 lessons from Prokop et al., 2022). In what follows, the above is leveraged (among other literature) to construct a view of regulation-innovation within the framework of US medical products.

[4] Dated 12-Nov-2022 at 11:34am under 'Any time' 'sort by relevance', not including citations.

### 1.2 An Overview of US Medical Product Development: Medicines and Medical Devices

Medical product (medicine (drug) and device[5] – see Table 1 for definitions) development[6] is a complicated pursuit (unfortunately, no single reference is sufficient to provide a cradle-to-grave view of the ever-evolving global medical product development process; readers may start, however, here: Blass, 2021; Ogrodnik, 2019). While there are extensive discussions on both drug and device development in the literature, at a high level, the author likes to consider the processes in terms of 4 steps: viz., Discovery, Development, Delivery, and Life-Cycle Management (LCM). Following Figure 1, 'Discovery' generally refers to identifying the intended patient population, reflecting on the unmet patient need and any standard of care, documenting the requirements of a successful medical product (aspirational mechanism of action, posology, safety, efficacy, and quality (e.g., storage considerations) profiles), constructing (formulating) the medical product(s), and performing the requisite chemical, biological, and nonclinical (non-human animal) testing (ibid). 'Development' effectively translates into clinical research – starting from first-in-human (either subjects with disease or healthy volunteers) and continuing to a point when the accumulated clinical data comprising the marketing authorization application has been completed. Manufacturing scale-up and completion of various other ancillary (e.g., chronic toxicology) studies occurs during this time (Fregni and Illigens, 2018). 'Delivery' comprises the Food and Drug Administration (FDA, Agency, US Health Authority (HA)) review and registration process and subsequent physical delivery of the FDA-approved medical product to the point-of-care (Mantus and Pisano, 2014). 'Life Cycle Management' seeks to maintain FDA-registration of the medical product, optimize commercial viability, and considers the impact of its demise (Hansen and Ellergy, 2012). This paradigm is applicable to medicines (novel, generic, biosimilars) as well as complex, high risk medical devices (under the Premarket Approval (PMA) scheme). Less complex medical devices may adopt the same model yet omitting certain concepts such as animal / human testing as well as FDA approval, adhering to an FDA clearance process for registration based on extrapolation from prior art (viz., under FDA Premarket Notification (PMN; 510k) scheme). The entirety of the process is 'owned' by one or more sponsors, and may take up to decades to complete, costing significant investment in the hopes of yielding greater returns than investment to (in part) plow-back as reinvestment into continued product development (see, e.g., Kaitin and Gets, 2022).

---

[5] Combination products, either device/device, drug/device, or drug/drug, add an additional layer of technical complexity. In this work, if there is at least one FDA registration package, the concept is included in the dataset.

[6] Note: The term 'development' is used in 2 different ways as is common in the industry: generally, as in 'drug development,' specifically, as in (e.g.,) 'clinical development' (viz., human testing, respectively).

With minor exception, each regulatory stage ('discovery,' 'development,' 'delivery,' and 'life cycle management') is under generally increasing levels of regulatory control, with 'discovery' effectively relatively unburdened until the initiation of animal studies, while that of delivery and/or life cycle having relatively significant regulatory controls. The contemporary drug and/or device development firm may conduct one or more innovation-related activities in serial or in parallel, in one or more jurisdictions, and/or either by itself or through some sort of partnership (see, e.g., Melnychuk et al., 2021, de Villemeur et al., 2022).

As a direct consequence of the complexity, medical product development has an intricate supply and demand dynamic. Supply side influences include explicit (e.g., buildings, instruments, materials, patents; access to infrastructure) and tacit (e.g., know-how and trade-secrets; ability to rapidly learn/adapt) knowledge (see, e.g., Kalcheva et al., 2018; Reinhardt, 2001, Allarakhia and Walsh, 2011). In part, using bio-center / clustering access and global recruitment have been used to reduce the friction (see, e.g., Williams and Pouder, 2020; Kim and Kim, 2018; Pereira et al., 2020; Rosiello and Orsenigo, 2008). Demand pressures include value propositions for healthcare providers to prescribe the medical product as well as ability of the individual patient to access a medical product within the healthcare system (see, e.g., Ellis and McGuire, 1993). As implied in the introduction above, the industrial medical product infrastructure also is coupled to a macro-, meso-, and micro-economic landscape, which has been heavily influenced by various forces including that of an evolving regulatory framework.

### 1.3 An Overview of US Medical Product Regulation: Medicines and Medical Devices

As mentioned, in the United States (not unlike other jurisdictions), the processes to discover, development, deliver, and maintain a medical product are regulated under the auspices of a regional health authority, the FDA. This Agency (among others (e.g., US United States Department of Agriculture (USDA), Education (Education), and Labor (DOL)), in addition to its supervisory department, Health and Human Services (HHS)) follow a process to implement laws and subsequently (working with affected industry) educate its stakeholders of the agency's current thinking on its implementation (GAO, 2015). In the US, as defined within the Administrative Procedure Act (APA) and described in Figure 1 of GAO (2015), congress through its processes enacts laws, which may empower certain agencies; these affected agencies would then construct regulations based on the statutes codified in associated laws: worth nothing that both laws and regulations are legally binding. Importantly, certain agencies, such as the FDA, further explicate their interpretations of these regulations through "Guidance," which are not legally binding. The FDA guidance documentation and promulgation processes are further refined, under 21 U.S.C. § 371(h), that "directed the agency to evaluate the effectiveness of its practices and develop and issue regulations specifying its procedures, (ibid, page 10), clarified its Good Guidance Practices (*ibid*, page 10-11), and introduced a 2-level scheme, whereby Level 1 documents "set forth initial interpretations of statutory or regulatory requirements; (ii) Set forth changes in interpretation or policy that are of more than a minor nature; (iii) Include complex scientific issues; or (iv) Cover highly controversial issues,"; and Level 2 documents the remaining (not Level 1) (see 21 C.F.R. 10.115[7]). Importantly, the "Agency may not use documents or other means of communication that are excluded from the definition of guidance document to informally communicate new or different regulatory expectations to a broad public audience for the first time. These GGP's must be followed whenever regulatory expectations that are not readily apparent from the statute or regulations are first communicated to a broad public audience (21 CFR 10.115(e)).” The process integrates public feedback at certain times during the process (see Section I in Paradise and Bavlsik (2020) for more details; Gusmano, 2013 for general introduction to public feedback on FDA decision making; Executive Order: 13891[8] may have directly affected guidance construction).[9]

[7] Accessible at: https://www.accessdata.fda.gov/scripts/cdrh/cfdocs/cfcfr/cfrsearch.cfm?fr=10.115 accessed on 14-Nov-2022.

As an aside, the above complex process may be simplified through the lens of knowledge economizing theory considering the dimensions of (1) codification: the degree to which data can be written; (2) abstraction: the degree to which the data can be understood; and (3) diffusion: the degree to which the data can be effectively communicated to an audience (Daizadeh, 2006). Laws are generally conceptual, typically requiring an Agency to further define (regulation). Oftentimes, these regulations may require further granularity using case studies and/or other illustrative means to assist the reader in deciphering the law/regulation into actual working practice (guidance). In other words, the process allows concretization (less abstraction) of the concept of the intended law into day-to-day practice via greater codification and more specificity of the audience. Working backwards, the presence of a guideline (or guidelines) is a need to de-complexify a law and its accompanied regulations, directly inferring that the guideline may be construed as a metric of complexity. In some sense, the more guidelines for a given law (or laws), the more complex the regulatory framework enacted.

In the US, the legislative backdrop of medical product regulation rests on the Food, Drug, and Cosmetic Act, which was enacted in the mid-1930s. New or amended medical product laws were enacted intermittently until the 1970s, and then significantly accelerated into the 1990s and the beginning of the 21$^{st}$-century (see [10], Table 1 in Daizadeh (2022) and Daizadeh (2020)). Tangentially, but of significant import to this work, in the 1970s, it may be hypothesized that the perceived complexity of the laws and ensuing regulations drove the FDA to release its first medical product guidelines (medicines in 1975 and medical devices in 1976). Subsequently, the number and variety of MP guidelines evolved dramatically over the years via a cobra-like pattern for medicines and relatively constant manner (until the acceleration in mid-2010s) for medical devices (Daizadeh, 2020, 2021, 2022) (see Figure 3 and Figure 4 (Bottom)). The time lag in this process (from conception to promulgation) of a guidance has not been studied to the author's knowledge but may be of the order of months to years.

[8] Accessible at: https://www.federalregister.gov/documents/2019/10/15/2019-22623/promoting-the-rule-of-law-through-improved-agency-guidance-documents accessed on 14-Nov-2022.

[9] The term ‘guideline’ is used here interchangeably with guidance as it is more succinct, easier to understand to a reader, and used across jurisdictions (e.g., in Europe). The author notes that the FDA collectively defines such documents as: “Guidance Document,” “Compliance Policy Guides,” “Special Controls Documents,” “Small Entity Compliance Guides,” “Information Sheets,” “Good Review Practices,” “Industry Letters,” “Manuals,” “Concept Papers,” and “Memoranda,” on the FDA website: https://www.fda.gov/regulatory-information/search-fda-guidance-documents, and exercises its authority of such documents under 21 CFR 10.115(b1-b3).

As mentioned, the degree of FDA oversight of the process increases over time, mostly bourgeoning at animal testing (when applicable) and peaking at time of application (prior to registration due (in part) to the protection of the population (as the medical product would be 'on the market'). Generally, the registration process is initiated when an application[11], which describes the safety, efficacy, and quality of the medical product, is submitted by the sponsor(s) and received by the FDA. In principle, then, the application itself represents – in the sponsor(s)' view – the cumulative sum of all prior investments (inputs) with the hope of FDA-registration so that the firm may begin to collect economic rents by delivering an FDA-registered medical product to the applicable population within the jurisdiction. Thus, the total number of medical product applications would be the ideal vehicle to track the evolution of medical product development in the US. Unfortunately, however, the FDA does not release specific information related to medical products (either medicines or medical devices) applications unless the application led to successful registration (Huneycutt, et al., 2020[12]). A view on the monthly number of FDA-registrations is presented in Figure 2 and Figure 4 (top).

[10] https://www.fda.gov/about-fda/fda-history-exhibits/brief-history-center-drug-evaluation-and-research, https://www.fda.gov/media/109482/download, https://www.fda.gov/about-fda/fda-history-research-tools/researching-fda-published-secondary-sources#General%20Sources, and https://www.fda.gov/medical-devices/overview-device-regulation/history-medical-device-regulation-oversight-united-states accessed on 11-Nov-2022

[11] PMA: 21 CFR 814 accessible at: https://www.accessdata.fda.gov/scripts/cdrh/cfdocs/cfcfr/cfrsearch.cfm?cfrpart=814 accessed on 14-Nov-2022; PMN: 21 CFR 807 accessible at: https://www.accessdata.fda.gov/scripts/cdrh/cfdocs/cfcfr/cfrsearch.cfm?fr=807.92 accessed on 14-Nov-2022; New Drug Application (NDA): 21 CFR 314 accessible at: https://www.accessdata.fda.gov/scripts/cdrh/cfdocs/cfcfr/CFRSearch.cfm?CFRPart=314 accessed on 14-Nov-2022; Biologics License Application (BLA): 21 CFR 601 accessible at: https://www.accessdata.fda.gov/scripts/cdrh/cfdocs/cfcfr/CFRSearch.cfm?CFRPart=601 accessed on 14-Nov-2022; Abbreviated New Drug Application (ANDA): 21 CFR 314.94 accessible at: https://www.accessdata.fda.gov/scripts/cdrh/cfdocs/cfcfr/cfrsearch.cfm?fr=314.94 accessed on 14-Nov-2022;

**1.4 Hypothesis: Is US medical product development impacted by regulation? If so, how?**

This work considers the following assumptions to build a testable infrastructure to test this hypothesis:

(1) FDA-issued guidelines and FDA-registrations may be used as indices to investigate the impact (if any) of regulation onto medical product development (innovation).

Onur and Söderberg (2020) used “all product market approvals in the high-risk medical device market over the 1978-2007 period,” equating to PMAs and PMA supplements to investigate the “effect of variation in regulatory review time on firms’ choices between radical and incremental innovations.” The authors state that this was the first time to their knowledge that such data was used in the (economics) literature. Daizadeh (2020, 2021a, 2021b) investigated structural temporal breaks and business cycles latent in FDA-registrations (medicines and medical devices) and showed cointegration between FDA-registered medicines and FDA-issues guidances. To the author’s knowledge, there are no other works that use FDA-registration data to pursue an investigation into the innovation dynamics of medical product development.

As presented in Table 1, there are several types of FDA approval pathways – each pathway specifying a unique input and output, which may have economic value. Onur and Söderberg (2020)’s reasonably used PMA data to investigate radical innovations, as the legal nature of the PMA reflects the innovative structure of novel (non-substantially equivalent and thereby 'radical') medical devices. Similarly, in this work, the index 'innovation' is a natural reflection of the innovative (e.g., incremental versus radical) spirit behind the legal mechanisms yielding each type of FDA registrational process. The monthly value of the innovation index is the monthly sum of FDA-registrations of medical devices and medicines. The former is defined as the sum of the monthly number of PMAs and PMNs; that is, the monthly number of medical devices developed under either the PMA (radical) and PMN (incremental) pathway. The latter is defined as the sum of the monthly number of new (radical) and supplemental (incremental) biologic (radical), drug (radical), generic (incremental), and biosimilar (incremental) applications[13].

[12] No specific reference was found to medical device FDA transparency measures specifically on public disclosure of negative registrational outcomes.

Importantly, these innovation-related data reflect successful (defined as the ability to transform (tangible and intangible) input into a marketed asset (output)) sponsors, irrespective of size, geographical dispersion, market hegemony, number, and kind of partnerships, and so on. These interesting variables may be investigated based on these data, given the large degree of metadata associated with each FDA-registration data record.

Medical product guidelines have received considerable attention by various stakeholders in the literature. Oftentimes, these discussions focus on the impact of one or more guideline(s) on a specific aspect of the process (see, e.g., Samaei, et al., 2022; Tarver and Neuland, 2021) or of the medical product (or class) itself (see, e.g., Renukuntla et al, 2022). Much less often, there is a request for more guidance (see, e.g., Fanse et al., 2022). Daizadeh (2021a, 2022) provided the first commentary to the author's knowledge not only suggesting that an assessment of guidelines should be performed as to better understand the effectiveness of such guidances to the medical product industry but also to resolve any perceived gaps (e.g., when a new guidance would be appreciated or needed). A completely missed opportunity, however, is the use of medical product guidelines as a metric in itself – e.g., as a proxy for regulatory complexity (regulation) – a high (low) number of guidelines over a given time-period may indicate periods of a need for high (low) policy interpretation intensive activities. This is the intention behind the use of this data to support this investigation.

[13] One could argument that the introduction of generics and biosimilars (specifically) were radical in spirit due to the degree of data requirements to obtain FDA registration (particularly, those of biosimilars).

As discussed in the conclusionary section, the author believes that in totality this data set is worthy of significant investment by the research community investigating the effectiveness of market product development in the United States, given its informativeness, ease of access, and degree of metadata.

(2) The time evolution of these metrics (FDA-issued guidelines and FDA-registrations) provide insight into medical product development.

FDA-registration and FDA-issued guidelines data are date-stamped; the application, various administrative actions, and the resultant registration are officially managed by the FDA administrative processes; thereby, creating a reliable flow of data over time – effectively, a time series. Accumulating the data into a per monthly time series allows for – importantly visualizing the flow of registrations as well as significantly easing computation including performing comparability studies.

Recall that it may take decade or longer from ideation to registration of a medical product. Waves of registration therefore may indicate waves of medical product innovativeness; ebbs in registration may indicate innovative abysses. Similarly, for guidances: waves of new guidances may imply waves of new and complex medical product legislation; ebbs in guidelines, concordantly, may imply periods of lack of legal innovation. The requirement for new medical product legislation may be due to accommodating new technologies (e.g., cell and genetic therapies[14]), further research in a specific target population

[14] FDA Guidance for Human Somatic Cell Therapy and Gene Therapy. Guidance for Industry. March 1998. Docket Number: FDA-2009-D-0132-0016. Accessible from https://www.fda.gov/regulatory-information/search-fda-guidance-documents/guidance-human-somatic-cell-therapy-and-gene-therapy. Accessed on 11-Nov-2022

(e.g., Pediatric Research Equity Act of 2003), or enabling novel processes (e.g., real-world evidence[15]). As an aside, it is interesting that while patents have enjoyed much interest as an indicator in the innovation, econometrics, and scientometrics literatures (see, e.g. Nagaoka, 2010), there is (as mentioned) hardly any mention of registration of a medical product, which involves much more innovative capabilities (which unlike a patent (or group of patents) only requires brief innovative attributes (viz., (1) appropriate subject matter, (2) utility, (3) novelty, (4) non-obviousness, and (5) enablement (under 35 United State Code)). The medical product indices (along with other metadata located with the associated records) presented here are truly informative of innovativeness given their role as a recipient of the various discovery, development, delivery, and life-cycle investments.

(3) The interplay of these metrics reflects, as proxies (surrogates), the impact of one metric on the other; that is, their cointegration is causally informative.

Daizadeh (2020) found that medicinal FDA-registrations and FDA-issued guidances were cointegrated implying a potential causal relationship. Given the complexities of the time series (non-stationarity, non-linearity, multiple structural change points), however, causality assessments were not performed at the time. Cognizant of the difference in time lags (a registration may take a decade or so, while a guidance may take months to years), it would make sense for such integration between these (and the other) time series data at a certain level. The reason for this lies in the purpose of the guidance document itself. If guidances are meant (in certain cases) to facilitate medical product development in a certain area (let's say orphan medicines), then it may be predicted that there would be an imbalance in the number of registrations in that specific topic area shortly thereafter. Or, if guidances are meant (in certain cases) to facilitate or expedite compliance of some type (e.g., COVID-19 diagnostic kits), such guidances may be promulgated shortly after such a technology were introduced. Further discussion on this circulate (co)integration is to be presented in the discussion section.

[15] Framework for FDA's Real World Evidence, dated: Dec 2018. Accessible from https://www.fda.gov/media/120060/download. Accessed on 11-Nov-2022

Beyond correlation, cointegration may imply potential causation (Granger, 1988) and thus provides the theoretical foundations for inter-variable interdependency. In our case, interdependency would allow for a direct view (through regression) of regulation and innovation using the proxy (surrogate) metrics (viz., FDA-issued guidances and FDA-registrations), the goal of this work.

This paper is outlined as follows: First, the conceptual model and data collection is outlined. Second, we present the overall approach to analysis. Third, the results of the analysis are presented and interpreted. Lastly, the paper concludes with a summary statement, and a discussion on the limitations of approach and potential directions for future investigations.

**2. Conceptual Model and Empirical Data**

**2.1 Conceptual Model**

Guided by Figure 1, and contextualized by the introductory statements above, a general conceptual framework may be considered as follows:

$$Eq.1.\quad \{MPI(t)\} = \{Extrinsic(t)\} + \{Intrinsic(t)\} + \{Instinsic\ (t)\ |Extrinsic\ (t)\} + \{Stochastic(t)\}$$

Here, the left-hand side of Equation 1 reflects represents the time-dependent total innovative capacity of the medical product (MPI) sector comprised of all sponsors which has successfully prosecuted at least one FDA-registered medical product (medical device or medicine). The right-hand side of Equation 1 is comprised of a combination of states of extrinsic, intrinsic, and stochastic influences. We note that there may exist a state in which there is a co-existing mixture of intrinsic conditional to an extrinsic force. For our specific context, a good example may be a new or updated regulation (extrinsic) that may result in hiring of a certain skill set (intrinsic) to perform a specific task under the compliant process.

The key assumption of this model is that *MPI* would be proportional to a medical product regulatory (MPR) influence of an unknown degree (denoted by $\alpha$), which would be time dependent as both the regulations and the subsequent interpretation and implementation by the firm would be serially correlated (with some time unknown time lag); i.e.,

$$Eq.2 \quad \{MPI(t)\} \propto \{MPR(t)\}^{\alpha}$$

However, Eq. 2 conceptually recognizes that the regulation-innovation relationship is relatively weak relative to other factors influencing the medical product development lifecycle, such as risks to technical success (van der Graaf, 2022) or access to financing (Lo and Thakor, 2022), and so on.

A metric for MPI is assumed to be the sum of FDA-registered medical products, defined as the sum of medicinal (BLA, NDA, ANDA and corresponding supplements) and medical device (PMA, PMN) registrations. It is understood that each FDA-registration equates to an innovative success (as market access is formally not allowed if registration cannot be attained) for given medical product (asset, *a*) in at least one approved indication (*i*) from a named sponsor(s) (firm, *f*) at a given timestamp *t* (which is aggregated over a given month, *t'*), as follows:

$$\begin{aligned} Eq.3 \quad \{MPI\}_{t'} \\ &= \sum_{f,ai}(Medicines + Medical\ Device\ Registrations)_t \\ &= \sum_{f,ai}(BLA + sBLA + NDA + sNDA + ANDA + sANDA)_t \end{aligned}$$

The regulatory influence would be assumed to be defined as the monthly tally of all FDA-issued guidelines, which takes on the form, for a given medical product or process overseeing medical product development:

$$Eq.4 \quad \{MPR\}_{t'} = \sum_{p,ai} (Medicines + Medical\ Devices\ Guidelines)_t$$

The question posed is to empirically test the existence of an interdependency between MPI and MPR, and, if existence is confirmed, the order of the proportionality, α.

**2.1 Data Sources**

The method for data-collection, data-cull, and data-formatting have been previously described in the literature (Daizadeh, 2020; 2021a; 2021b; 2021c; 2022a; 2022b). While the method is summarized below, the data itself is available in the Appendix for reproducibility and extension by the reader.

- MPR: The Medical Product Regulatory metric is constructed through summing the monthly number of FDA-registered medical devices and medicines, obtained via the following algorithm:
    - Medical Devices: PMNs and PMAs data obtained from the FDA websites: https://www.fda.gov/medical-devices/510k-clearances/downloadable-510k-files and https://www.fda.gov/medical-devices/device-approvalsdenials-and-clearances/pma-approvals. The data were read into EXCEL, the monthly counts derived.
    - For medicines, the following field 'All Approvals and Tentative Approvals by Month' was used on the collecting report from the site: https://www.accessdata.fda.gov/scripts/cder/daf/. The report included any original or supplemental biologicals (BLA/sBLA), new (NDA/sNDA) or abbreviated (including tentative ANDA/sANDA) approved by the Center for Drug Evaluation and Research (CDER). Unfortunately, the reports do not include applications or supplements approved by the Center for Biologics Evaluation and Research (CBER), as these are currently unavailable via the FDA websites. Upon entry into the data-repository via the website, the number of monthly approvals from May 1976 to December 2020 was then

determined. The values were then important into Excel, monthly counts derived, and summed with that of the medical device data described above.

- MPI: The Medical Product Innovation metric is constructed from data obtained from FDA repository accessible at: https://www.fda.gov/regulatory-information/search-fda-guidance-documents. For medicines, the Product filter was used to collect those from Biologics and Drugs, or FDA Organization for CDER; for MDs, the FDA Organization for CDRH was used. Each search produced data that was exported to Excel, in which the data was integrated, manually culled, duplications removed, and monthly values tallied.

The totality of the data was then exported from Excel, via a comma-separated file, and imported into the R Program for data analyses (R Core Team, 2022: R version 4.2.1 (2022-06-23 ucrt)).

## 3. Analyses and Results

On a high level, the general algorithmic (heuristic) approach for this analysis is as follows: calculate descriptive and dynamic statistics to inform on methodology for assessing interdependence, which is (weakly) defined as non-spurious correlation with bi-directionally (symmetrically) causality (Back and Tarver, 1959; Pasinetti, 2019). If evidence of interdependence is elucidated, then proceed to regression analysis to estimate α.

### 3.1 Determine Descriptive and Dynamic Statistics of the Variables in the Study

There are 6 time series of interest: The four (bases of the investigational space); i.e., the monthly number of FDA-registered medical devices (MDs) and medicines, and the corresponding FDA-Issued MD and Medicines guidelines. The two derived time series: the monthly number of FDA-registered medical products (MPI) and corresponding FDA-issued medical product guidelines (MPR) (each respectively derived from a sum of the underlying time series). Given that it is anticipated that MPI is weakly associated with MPR, the cumulative sum was used to increase single-to-noise ratio. Structurally, this approach is sensical as ongoing maintenance to retain the registrational status; similarly, guidances continue to be used of reference even though they may be tactical value. It is possible that the registrational status or guidance document is rescinded (withdrawn) by FDA (see, e.g., in the case of accelerated approvals for cancer drugs[16]). Assuming such cases are rare, these approach or results are not considered to be affected, however, additional testing is necessary to check accordingly (as future work). A time-dependent representation of the evolution of the variables presented in the study are presented in Figures 2-5.

The baseline statistics of the time series including descriptive and dynamic characteristics of the variables used in this study are presented in Tables 2-4 (Revelle, 2022; Gross and Ligges, 2015; Trapletti and Hornik, 2022; Qiu, 2015; Garcia, 2022; Hyndman, et al., 2022; Hyndman and Khandakar, 2008; Leschinski, 2019; Ollech, 2021; Wei and Simko, 2021; Harrell, Jr., F, 2022). In summary, it is found that the time series characteristics vary in terms of skew, kurtosis, normality, linearity, and seasonality. All time series were nonstationary with long-memory. There is a significant time lag (on the order of 1-3 years) for both registrations as well as guidelines (Table 3). These time lags suggest latent waves of registrations and guidelines promulgations that may reflect the productivity of industry (see, e.g., Daizadeh, 2020) and/or the FDA (in terms of guidelines).

The complexity of the time series should not be understated. For example, fortunately, while it is challenging to resolve structural breaks for nonstationary and non-linear time series, wavelet transform analysis may be used to gain insight into structural breaks and/or other complexities (Schulte, 2016; Sang, 2012; Torrence and Compo, 1998). The wavelet power spectra (Roesch and Schmidbauer, 2018; Figures 6 and 7) recapitulate and refine the flows latent in the time domain (Figures 2-5), demonstrating when short and long periodicities appear as a function of time. For example, notice the 8-year and 16-

[16] https://www.fda.gov/drugs/resources-information-approved-drugs/withdrawn-cancer-accelerated-approvals

year periodicities appearing from 1990-2020 in the FDA-registered medical products (Figure 7 Top Left), while the FDA-issued medical product guidelines had a concentration of short-term (≤ 2 year) periodicities appearing over blocks of a 5-10 years (1995-2000, 2005-2010, 2015-2020) (Figure 7 Top Right). Of particular interest is the consistently long-periodicities (8 and 16 year) spanning the domain from 1980 to 2020 observed in the charts presenting the cumulative data (Figure 5 and Figure 7 bottom). In the same figures, rapid and strong appearance of additional periodicities are observed in the cumulative FDA-issued medical products guidelines (Figure 7 bottom right) but not in the corresponding cumulative FDA-registered medical products (Figure 7 bottom left). We note this discrepancy as an important finding for later in the discussion.

**3.2 Interdependency: Estimate Correlation and Causality of the Variables in the Study**

Given the complexity (primarily due to non-stationarity) of the time series, a Spearman calculation was used to estimate correlation between the time series (Table 5) (R Core Team, 2022). Interestingly, correlation strengths were observed across all pair-wise variables analyzed. From relatively weak (0.18-0.25) to moderately strong (0.51-0.73). These values are consistent with the various spectra in the time and frequency domains. For example, the correlation coefficient for the cumulative variables (FDA-registered medical products and associated FDA-issued guidelines) indicate that they are strongly coupled (0.999). The weakest of the correlation coefficients were between FDA-registered medicines with that of FDA-issued medical devices guidelines (0.18), which is not unexpected. The small but extant correlation is generally rationalized, however, as a subgroup of medical device guidelines also inform medicinal development (e.g., in the case of combination (medicine-medical device) product development).

Causality estimation was performed in part via Variable-Lag Transfer Entropy (VLTE) (Amornbunchornvej, 2021; R Package: VLTimeCausality). As mentioned, the multivariate time series intrinsic statistics are complex: normalizing or transforming the data (or even attempting to fit within the common Granger causality assessment scheme (1998)) was unrealistic given degree of subtleties such as non-linearity, non-normality, and, importantly, the multiple structural breaks (as descriptively shown in the wavelet spectra). Thus, a transfer entropy-type calculation was preferred (Xuegeng and Pengian, 2017; Dhifaoui et al., 2022), particularly one that could consider the presumed non-fixed (variable) time lag for both the FDA-registered medical products and FDA-issued guidelines (as discussed above and the elucidated long-memory (yet variable) time-lag presented in Figure 3).

Reviewing Table 6 finds several causal relationships, including FDA-registered medical VLTE-causes FDA-registered medicines; FDA-issued medical device guidelines VLTE-causes FDA-registered medical devices and FDA-issued medicines guidelines; FDA-issued medicines guidelines VLTE-causes FDA-registered medical devices and FDA-issued MD Guidelines. Of interest, cumulative FDA-issued medical product guidelines and corresponding FDA-registrations are bi-directional.

Given the strong correlation and bidirectionality (as well as the interesting univariate periodicities as shown in the wavelet spectra) of the derived cumulative FDA-registered medical products and associated FDA-issued guidelines, a wavelet coherence analysis (Roesch and Schmidbauer, 2018) is performed primarily to better understand the nature of the interaction. As shown in Figure 9 (Top and Bottom), long-memory periodicities are well-illustrated. Additional periodicities were gained from the mid-2015s to present. The arrows toward 16 to 32 years tend to be lead/lag indicators. While it is difficult to ascertain, the arrows seem circular suggesting a bidirectional relationship.

In summary, interdependency between MPI and MPR is demonstrated as strong correlation and causal assessments are present in the three complementary analyses. The question now becomes can this be quantified.

### 3.3. Determining the Relationship between Regulation and Innovation: The Exponent α

To calculate the characteristic of the relationship, α, a simple regression is needed. Regression analysis was attempted. First, a curvilinear relationship is found when the cumulative FDA-issued medical product guidelines (MPR) is plotted against the cumulative FDA-registered medical products. Three regression analyses were attempted: linear, quadratic, and cubic (Martins dos Santos, 2022). Regression correlation coefficients is used to determine the best fit. The following was found: Linear $R^2$ = 0.874, quadratic $R^2$ = 0.964, and cubic $R^2$ = 0.978. The quadratic was determined to be the best fit as cubic only gave incremental difference. Thus, α is estimated to be a roughly inverted quadratic scale over the entire time-period analyzed. Visually, α seems to evolve early in a positive manner, saturating and then inverting, a key finding of this work.

.

Table 1: Definitions for Medical Devices and Medicines.

| Medical Product | Statuary Definitions [a] | Types: |
|---|---|---|
| Medical Devices | Section 201(h) of the FD&C Act (21 USC 321(h)):<br><br>The term "device" means: an instrument, apparatus, implement, machine, contrivance, implant, in vitro reagent, or other similar or related article, including any component, part, or accessory, which is (1) recognized in the official National Formulary, or the United States Pharmacopeia, or any supplement to them, (2) intended for use in the diagnosis of disease or other conditions, or in the cure, mitigation, treatment, or prevention of disease, in man or other animals, or (3) intended to affect the structure or any function of the body of man or other animals, and which does not achieve its primary intended purposes through chemical action within or on the body of man or other animals and which is not dependent upon being metabolized for the achievement of its primary intended purposes. | Premarket Notification (PMN; 510(k): A device that is substantially equivalent to a legally marketed device (under section 513(i)(1)(A) FD&C Act) that is not subject to premarket approval.<br><br>Premarket Approval (PMA): A Class III medical devices that supports or sustains human life, are of substantial importance in preventing impairment of human health, or which present a potential, unreasonable risk of illness or injury. SEC. 515. [21 USC §360e] |
| Medicines (Drugs) | Section 201(g) of the FD&C Act (21 USC 321(g)):<br><br>The term "drug" means: (A) articles recognized in the official United States Pharmacopoeia, official Homoeopathic Pharmacopoeia of the United States, or official National Formulary, or any supplement to any of them; and (B) articles intended for use in the diagnosis, cure, mitigation, treatment, or prevention of disease in man or other animals; and (C) articles (other than food) intended to affect the structure or any function of the body of man or other animals; and (D) articles intended for use as a component of any articles specified in clause (A), (B), or (C). . . . | New entities and supplements: Federal Food, Drug, and Cosmetic Act (FD&C Act)<br>Generic: Drug Price Competition and Patent Term Restoration Act (DPC-PTR Act)<br>Biosimilar: Biologics Price Competition and Innovation Act (BPCI Act) |
| [a] There is overlap of the definitions. Generally acknowledged by FDA and (e.g.,) clarified under Final (September 2017) Guidance for Industry and FDA Staff: "Classification of Products as Drugs and Devices and Additional Product Classification Issues." Document Number: FDA-2011-D-0429, accessible from the FDA Website: https://www.fda.gov/regulatory-information/search-fda-guidance-documents/classification-products-drugs-and-devices-and-additional-product-classification-issues | | |

Table 2: Descriptive Statistics of the Variables Used in this Study.

| Variable* | Mean | Standard Deviation | Median | Minimum | Maximum | Range | Skew | Kurtosis | Standard Error |
|---|---|---|---|---|---|---|---|---|---|
| **FDA-Registered MD** | **377.71** | **98.84** | **376.5** | **2** | **700** | **698** | **-0.41** | **0.74** | **4.27** |
| **FDA-Registered Medicines** | **339.09** | **136.15** | **321.5** | **68** | **858** | **790** | **0.78** | **0.63** | **5.88** |
| **FDA-Issued MD Guidelines** | **1.23** | **2.04** | **1.0** | **0** | **17** | **17** | **3.29** | **14.81** | **0.09** |
| **FDA-Issued Medicines Guidelines** | **2.89** | **4.65** | **1.0** | **0** | **25** | **25** | **2.36** | **5.58** | **0.20** |
| **FDA-Registered MPI** | **716.81** | **197.80** | **721.0** | **187** | **1355** | **1168** | **-0.06** | **-0.09** | **8.45** |
| **FDA-Issued MPR** | **4.12** | **6.09** | **2.0** | **0** | **39** | **39** | **2.55** | **7.42** | **0.26** |
| **Cumulative FDA-Registered MPI** | **177557.10** | **113659.38** | **179909.0** | **255** | **384208** | **383953** | **0.05** | **-1.23** | **4909.34** |
| **Cumulative FDA-Issued MPI Guidelines** | **526.80** | **536.86** | **364.0** | **0** | **2209** | **2209** | **1.07** | **0.35** | **23.19** |
| *** N=536: monthly values from May 1976 to December 2020** | | | | | | | | | |

Table 3: Dynamic Qualities of the Variables Used in this Study. All Results Including P Values are Presented in the Appendix.

| Variable[a] | Normality[b] | Non-stationarity[c] | Long-Memory (Lag; Months)[d] | Seasonality[e] |
|---|---|---|---|---|
| **FDA-Registered MD** | **Reject** | **1** | **30** | **TRUE** |
| **FDA-Registered Medicines** | **Reject** | **1** | **16** | **FALSE** |
| **FDA-Issued MD Guidelines** | **Reject** | **1** | **2** | **FALSE** |
| **FDA-Issued Medicines Guidelines** | **Reject** | **1** | **28** | **FALSE** |
| **FDA-Registered MPI** | **Accept** | **1** | **33** | **TRUE** |
| **FDA-Issued MPI Guidelines** | **Reject** | **1** | **25** | **TRUE** |
| **Cumulative FDA-Registered MPI** | **Reject** | **2** | **118** | **TRUE** |
| **Cumulative FDA-Issued MPI Guidelines** | **Reject** | **2** | **108** | **TRUE** |

**[a] n=536: monthly values from May 1976 to December 2020**
**[b] Normality tests: Jarque Bera Test, Anderson-Darling, Cramer-von Mises; null hypothesis of normality; all *p* values << 0.01, with exception as noted**
**[c] Non-stationarity tests: graphical (autocorrelation function (ACF)) and statistical (e.g., Kwiatkowski–Phillips–Schmidt–Shin (KPSS), Augmented Dickey–Fuller (ADF), Ljung-Box (LB)) suggest non-stationarity, with the number of differences required to achieve stationarity of 1.**
**[d] Long memory tests: graphical (autocorrelation) and statistical (Qu test and Multivariate local Whittle Score type test) tests suggest long-memory (with exception of FDA-Registered Medicines); lag time derived from ACF is presented in the table.**
**[e] Data**

Table 4: Linearity of the Variables Used in this Study. All Results Including *p* Values are Presented in the Appendix.

| Variable | Nonlinearity Test[a] | | | | | |
|---|---|---|---|---|---|---|
| | Teraesvirta's Neural Network | White Neural Network | Keenan's One-Degree | McLeod-Li | Tsay's | Likelihood ratio |
| **FDA-Registered Medical Devices** | **Linear** | **Linear** | **Linear** | **Nonlinear** | **Linear** | **Nonlinear** |
| **FDA-Registered Medicines** | **Nonlinear** | **Nonlinear** | **Nonlinear** | **Nonlinear** | **Linear** | **Linear** |
| **FDA-Issued Medical Device Guidelines** | **Nonlinear** | **Nonlinear** | **Linear** | **Nonlinear** | **Nonlinear** | **Linear** |
| **FDA-Issued Medicines Guidelines** | **Nonlinear** | **Nonlinear** | **Linear** | **Nonlinear** | **Nonlinear** | **Linear** |
| **FDA-Registered Medical Products** | **Nonlinear** | **Nonlinear** | **Nonlinear** | **Nonlinear** | **Nonlinear** | **Nonlinear** |
| **FDA-Issued Medical Product Guidelines** | **Nonlinear** | **Nonlinear** | **Linear** | **Nonlinear** | **Nonlinear** | **Nonlinear** |
| **Cumulative FDA-Registered MPI** | **Nonlinear** | **Nonlinear** | **Linear** | **Nonlinear** | **Linear** | **Linear** |
| **Cumulative FDA-Issued MPI Guidelines** | **Nonlinear** | **Linear** | **Linear** | **Nonlinear** | **Nonlinear** | **Linear** |

**[a] Note: All variables were transformed to stationary representation via differencing.**
**[b] Linearity tests: the null hypothesis of some aspect of linearity is accepted if *p* value > 0.008 (i.e., 0.05/6) to account of multiplicity.**

Table 5: **(Spearman) Correlation Values of Variables Used in this Study.**

| Variable | FDA-Registered MD | FDA-Registered Medicines | FDA-Issued MD Guidelines | FDA-Issued Medicines Guidelines | FDA-Registered MPI | FDA-Registered MPI Guidelines | Cumulative FDA-Registered MPI | Cumulative FDA-Issued MPI Guidelines |
|---|---|---|---|---|---|---|---|---|
| **FDA-Registered MD** | **1.0** | **0.44** | **0.25** | **0.34** | **0.75** | **0.35** | **0.51** | **0.51** |
| **FDA-Registered Medicines** | | **1.0** | **0.18** | **0.21** | **0.91** | **0.22** | **0.31** | **0.31** |
| **FDA-Issued MD Guidelines** | | | **1.0** | **0.50** | **0.25** | **0.75** | **0.51** | **0.51** |
| **FDA-Issued Medicines Guidelines** | | | | **1.0** | **0.31** | **0.92** | **0.73** | **0.73** |
| **FDA-Registered MPI** | | | | | **1.0** | **0.33** | **0.45** | **0.45** |
| **FDA-Registered MPI Guidelines** | | | | | | **1.0** | **0.73** | **0.74** |
| **Cumulative FDA-Registered MPI** | | | | | | | **1.0** | **0.999** |
| **Cumulative FDA-Issued MPI Guidelines** | | | | | | | | **1.0** |

Table 6: Causality Assessment (CA) via Variable-Lag Transfer-Entropy (VLTE) of the Variables Used in this Study. [a] A CA Result of TRUE Suggests Row 'VLTE-Causes' Column.

| Variable | FDA-Registered MD | FDA-Registered Medicines | FDA-Issued MD Guidelines | FDA-Issued Medicines Guidelines | FDA-Registered MPI | FDA-Registered MPI Guidelines | Cumulative FDA-Registered MPI | Cumulative FDA-Issued MPI Guidelines |
|---|---|---|---|---|---|---|---|---|
| **FDA-Registered MD** | | **TRUE<br>TER: 1.297<br>p-value: 0.004** | **FALSE<br>TER: 0.7526<br>p-value: 0.002** | **FALSE<br>TER: 0.6022<br>p-value: 0.032** | **TRUE<br>TER: 1.978<br>p-value:0** | **FALSE<br>0.741<br>p-value: 0.002** | **FALSE<br>TER: 0.04626<br>p-value: 0.992** | **FALSE<br>TER: 0.1632<br>p-value: 1** |
| **FDA-Registered Medicines** | **FALSE<br>TER:0.5626<br>p-value: 0.002** | | | **FALSE<br>TER: 0.935<br>p-value:0.018** | **TRUE<br>TER: 3.809<br>p-value:0** | **TRUE:<br>TER: 2.166<br>p-value:0** | **FALSE<br>TER: 0.3122<br>p-value: 0.18** | **FALSE<br>TER: 0.259<br>p-value: 0.284** |
| **FDA-Issued MD Guidelines** | **TRUE<br>TER: 2.336<br>p-value:0** | **FALSE:<br>TER: 2.669<br>p-vlaue:0.054** | | **TRUE<br>TER: 1.417<br>p-value:0** | **TRUE<br>TER: 2.866<br>p-value: 0.012** | **TRUE<br>TER: 4.517<br>p-value:0** | **FALSE<br>TER: 0.05025<br>p-value: 0.998** | **FALSE<br>TER: 0.04843<br>p-value: 1** |
| **FDA-Issued Medicines Guidelines** | **TRUE<br>TER: 1.436<br>p-value:0** | | **TRUE<br>TER: 1.936<br>p-value:0** | | **TRUE<br>TER: 1.252<br>p-value: 0** | **TRUE<br>TER: 4.362<br>p-value:0** | **FALSE<br>TER: 0.1191<br>p-value: 0.53** | **FALSE<br>TER: 0.2228<br>p-value: 0.148** |
| **FDA-Registered MPI** | **TRUE<br>TER: 1.327<br>p-value:0** | **TRUE<br>TER: 4.987<br>p-value:0** | | **FALSE<br>TER: 0.6051<br>p-value: 0.026** | | | **FALSE<br>TER: 0.03936<br>p-value: 0.986** | **FALSE<br>TER: 0.03403<br>p-value: 0.992** |
| **FDA-Registered MPI Guidelines** | **TRUE<br>TER: 1.313<br>p-value: 0** | | **TRUE<br>TER: 7.454<br>p-value: 0** | **TRUE<br>TER: 4.308<br>p-value: 0** | **TRUE<br>TER: 1.127<br>p-value: 0.004** | | **FALSE<br>TER: 0.1173<br>p-value: 0.428** | **FALSE<br>TER: 0.1452<br>p-value: 0.298** |
| **Cumulative FDA-Registered MPI** | **TRUE<br>TER: 38.22<br>p-value: 0** | | **TRUE<br>TER: 8.582<br>p-value: 0** | **TRUE<br>TER: 35.64<br>p-value: 0** | **TRUE<br>TER: 16.81<br>p-value: 0** | **TRUE<br>TER: 35.54<br>p-value: 0** | | **TRUE<br>TER: 215<br>p-value:0** |
| **Cumulative FDA-Issued MPI Guidelines** | **TRUE<br>TER: 18.67<br>p-value: 0** | | **TRUE<br>TER: 8.76<br>p-value: 0** | **TRUE<br>TER: 36.58<br>p-value: 0** | **TRUE<br>TER: 20.84<br>p-value: 0** | **TRUE<br>TER: 36.84<br>p-value: 0** | **TRUE<br>TER: 1.649<br>p-value: 0.004** | |

**[a] Causality Assessment is True If X VLTE-causes Y When TE Ratio (TER) > 1, and *p value* </= α (i.e., 0.05, with 500 bootstraps). TER and *p value* are also presented in the table below. Diagonals or cells with no discernable TER were ignored; thus, their matrix elements are empty. A p-value of 0 implies a p-value below a numerical threshold.**

**Figure 1: An Overview of the Medical Product Discovery, Development, and Delivery process.**

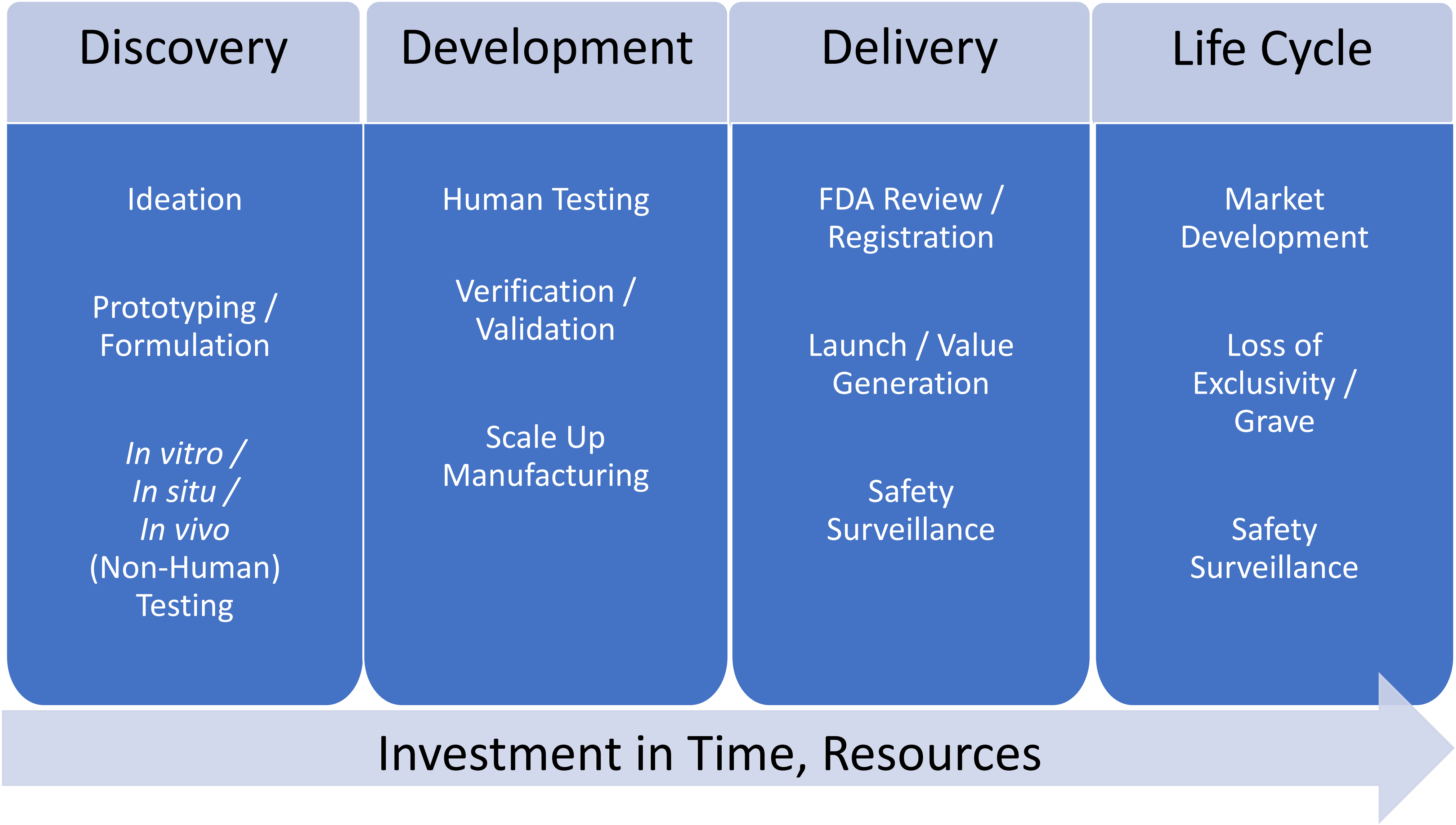

**Figure 2: The Monthly Number of FDA-Registered MDs (TOP) and Medicines (BOTTOM).**

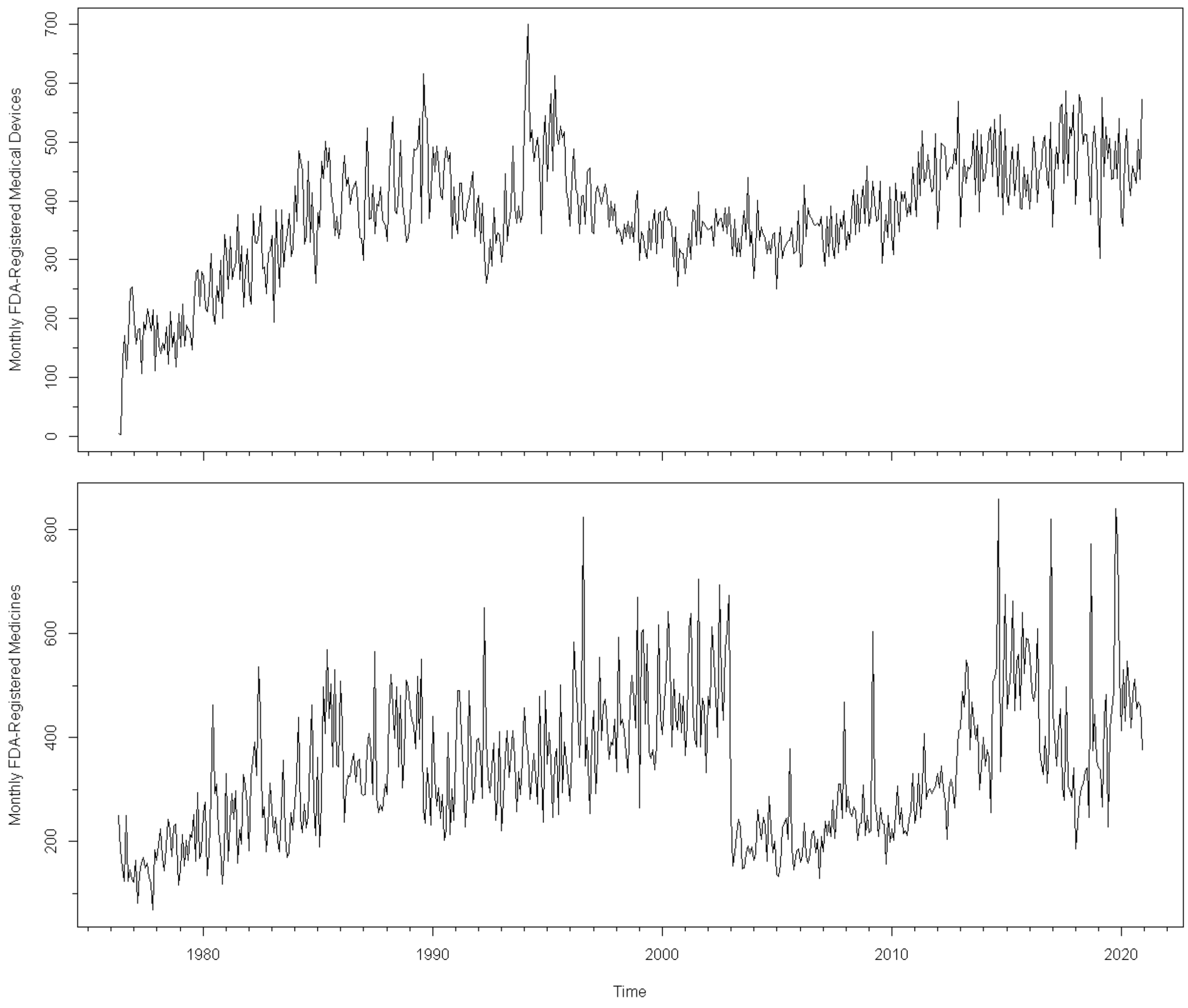

**Figure 3: The Monthly Number of FDA-Issued MDs (TOP) and Medicines (BOTTOM) Guidelines.**

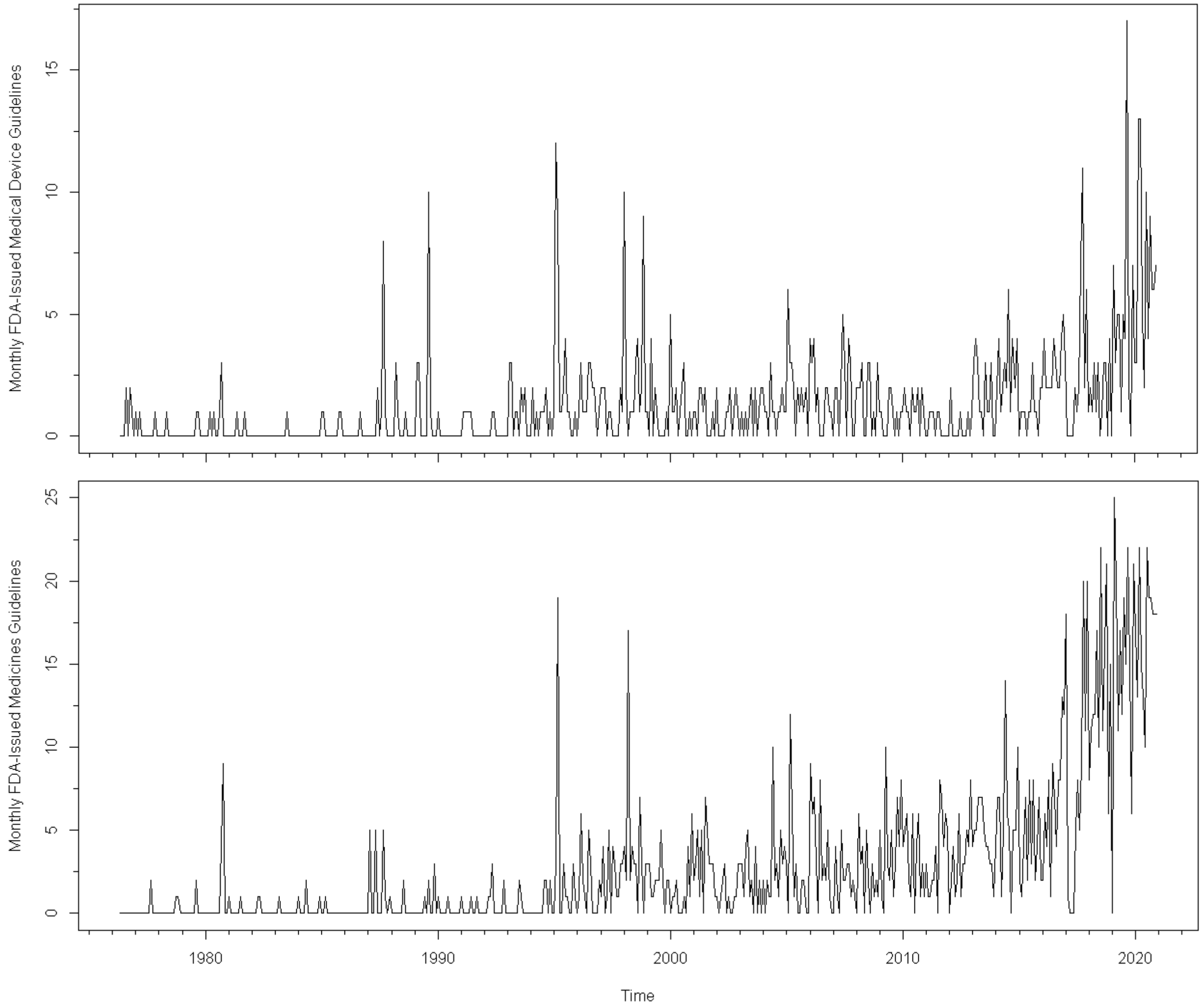

**Figure 4: The Monthly Number of FDA-Registered Medical Products Innovated (TOP) and Respective FDA-Issued Guidelines (BOTTOM).**

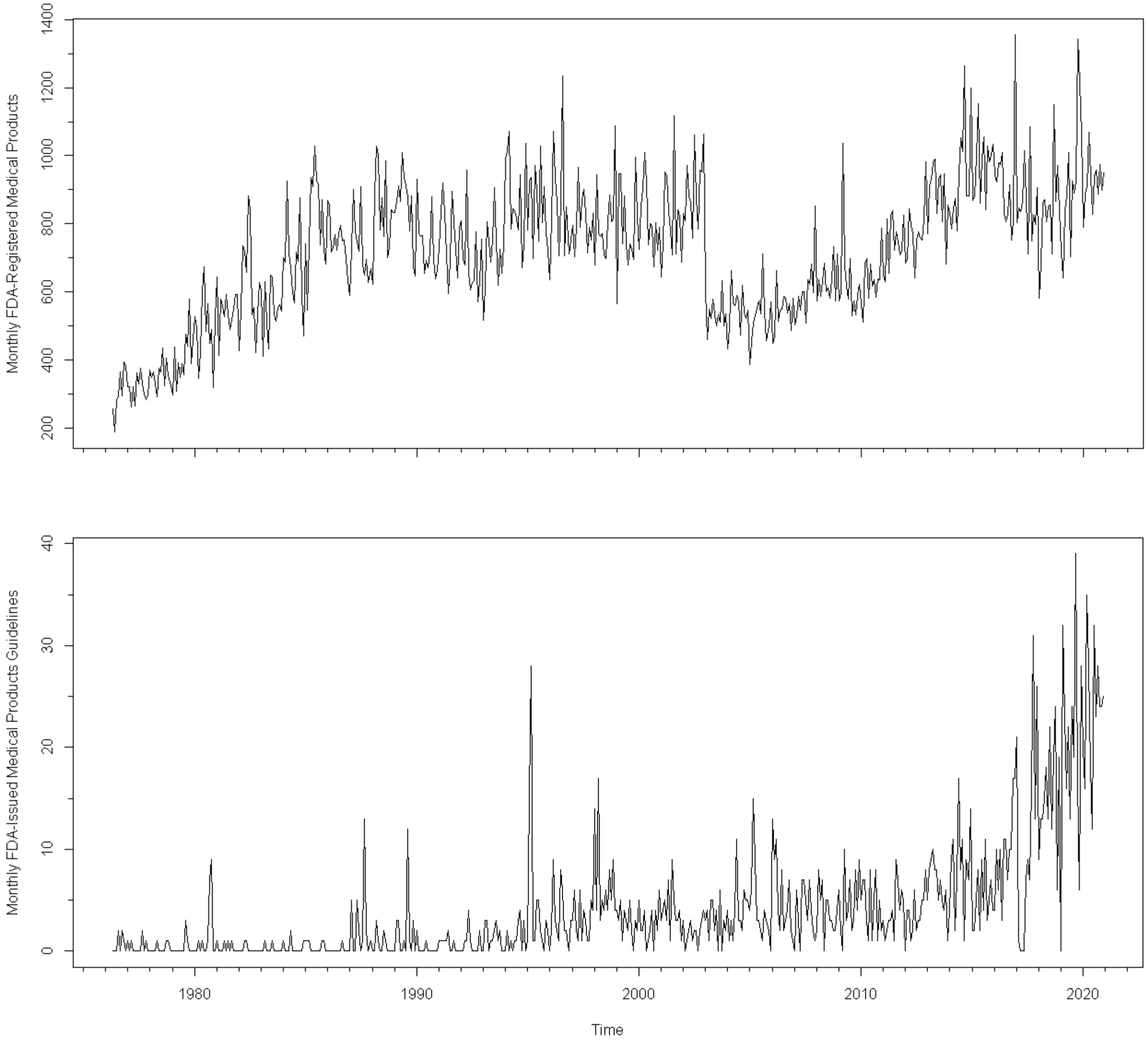

**Figure 5: Cumulative Monthly Number of FDA-Registered Medical Products and Respective FDA-Issued Guidelines.**

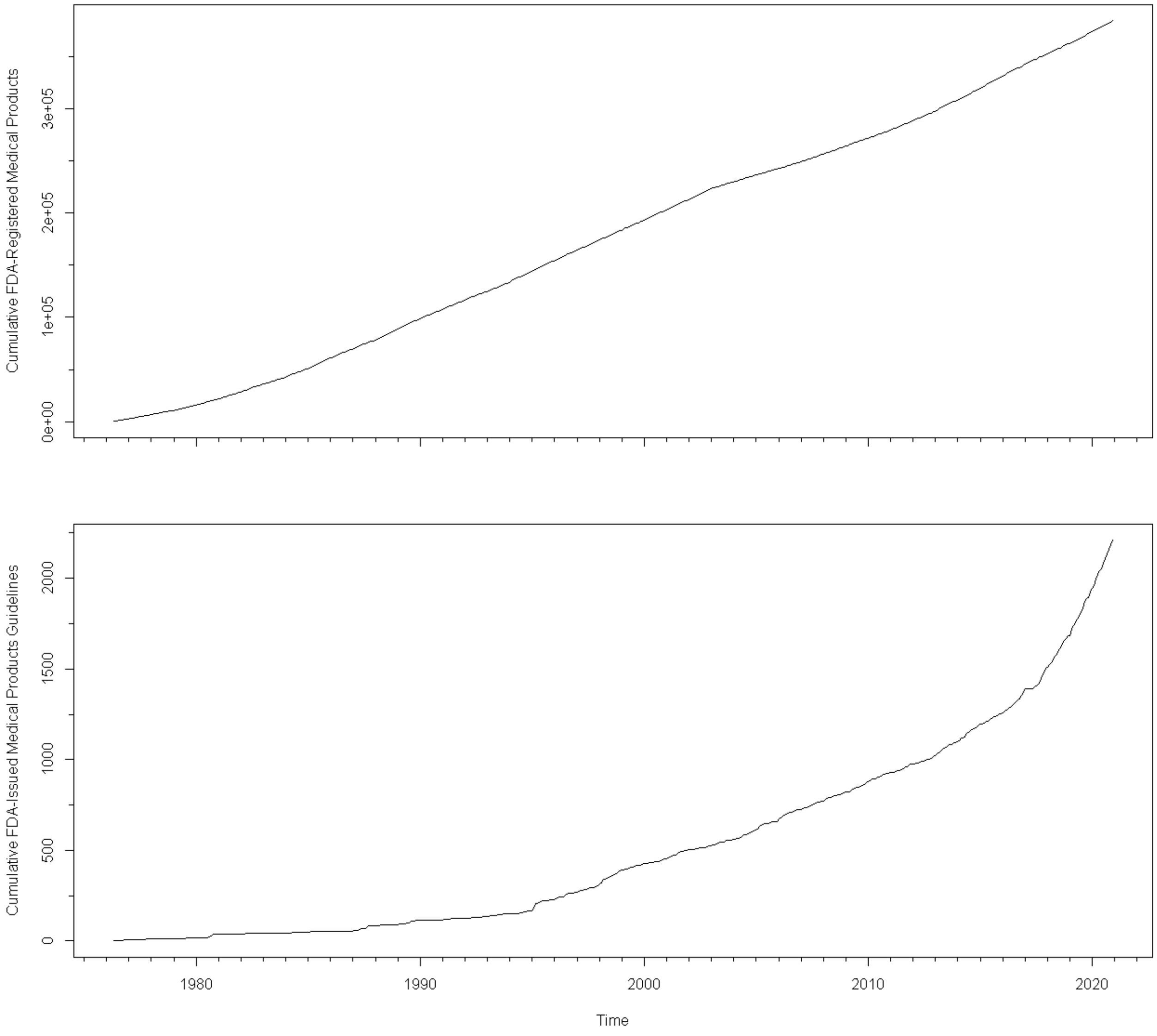

**Figure 6: Wavelet Power of the Number of the Variables Used in this Study. Legend: Wavelet Power is in Units of Quartiles. See Text for Details.**

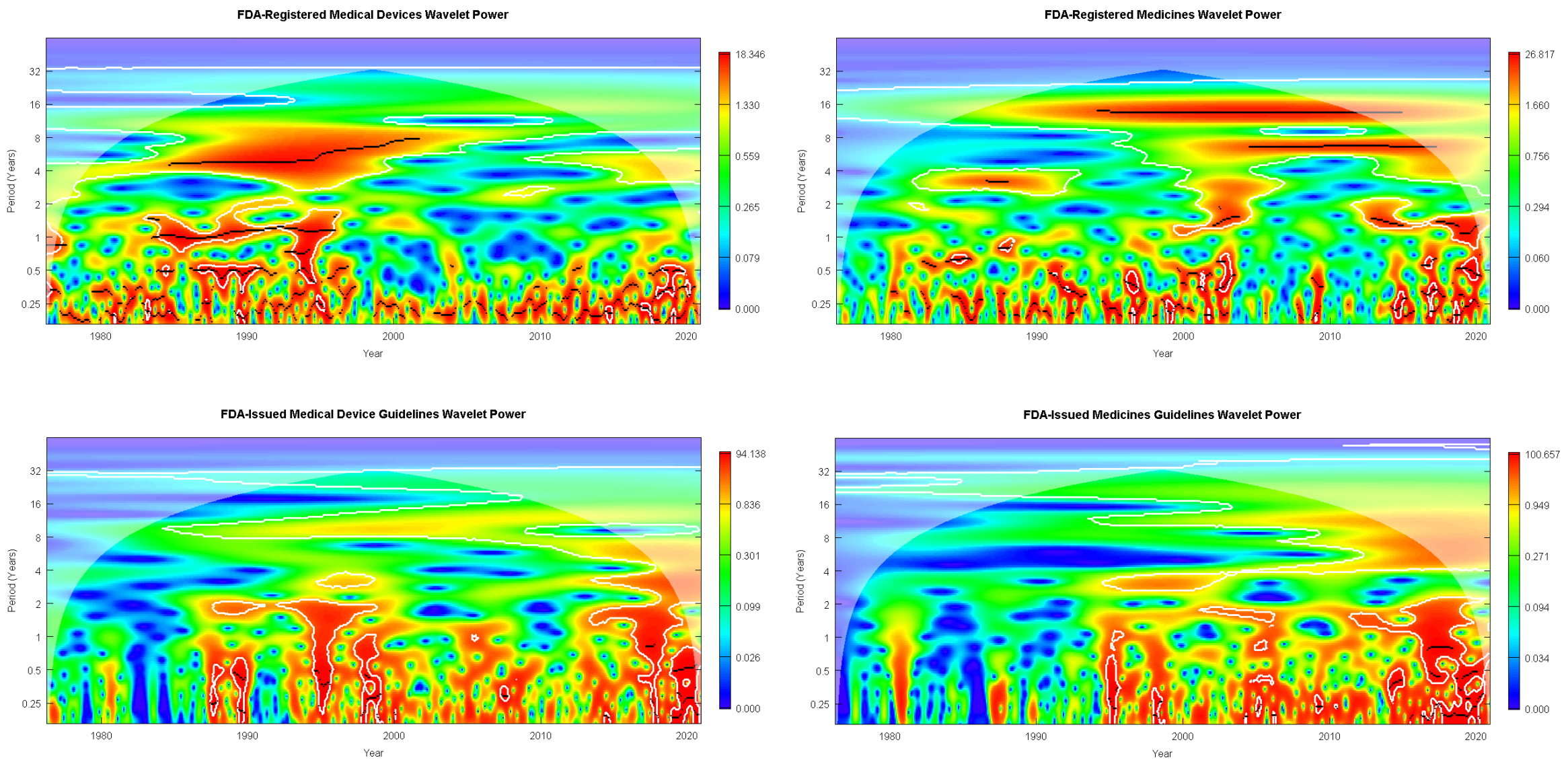

**Figure 7: Wavelet Power of the Number of the Variables Used in this Study. Legend: Wavelet Power is in Units of Quartiles. See Text for Details.**

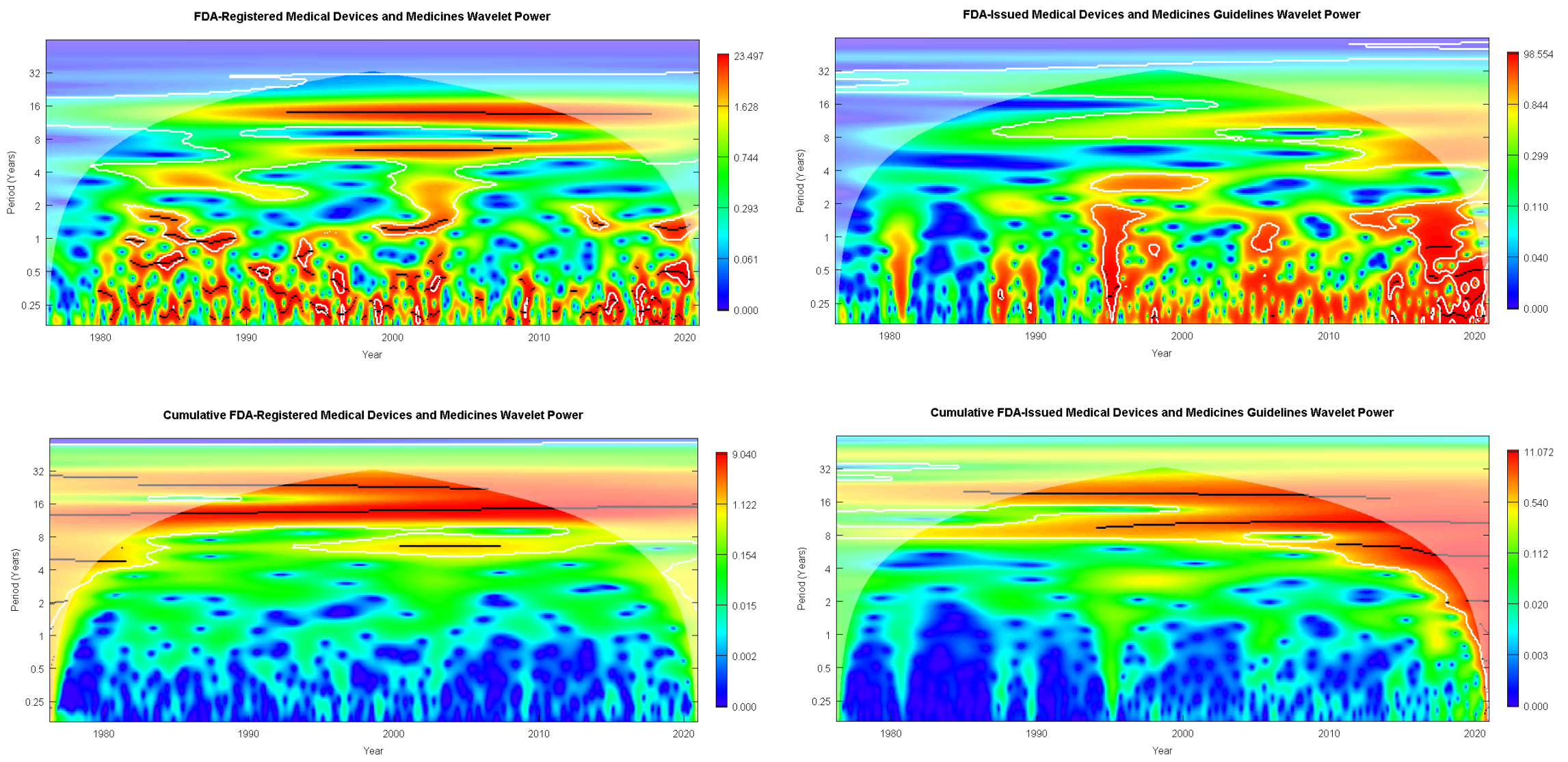

**Figure 8: Quadratic Regression Fit of the Cumulative FDA-Issued Medical Product Guidelines (a Proxy Metric for Regulatory Complexity) and FDA-Cumulative FDA-Registered Medical Products (a Proxy Metric for Innovation). For reference: $R^2 = 0.96$; Regression Equation: $41900.9044+372.599437x-0.107272x^2$.**

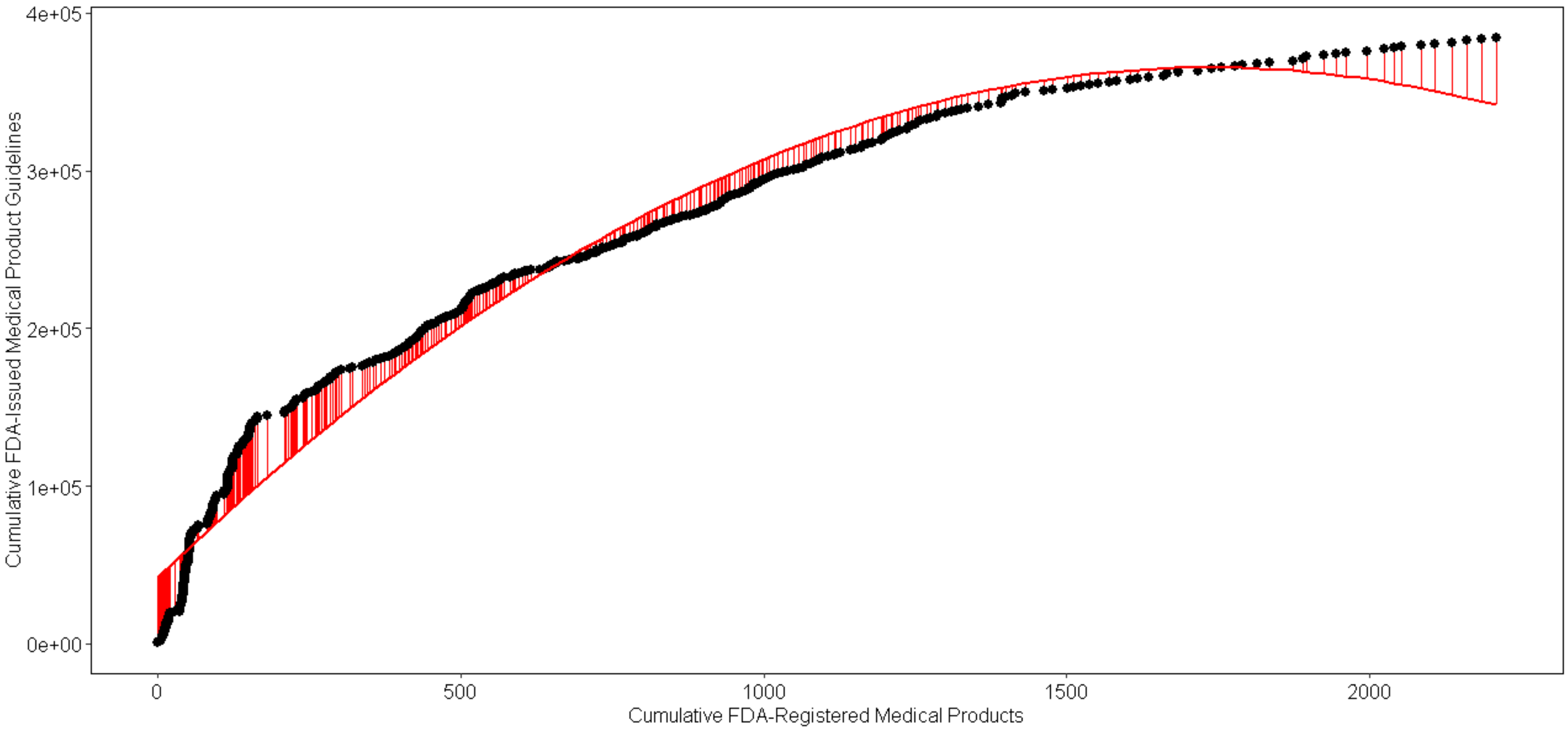

**Figure 9: Wavelet Coherence and Corresponding Cross-Wavelet Power Presenting the Interplay between Cumulative FDA-Issued Medical Product Guidelines (a Proxy Metric for Regulatory Complexity) and FDA-Cumulative FDA-Registered Medical Products (a Proxy Metric for Innovation). Note that the Average Cross-Wavelet Power occurs sharply at or around ~16 years. Legend: Wavelet Power is in Units of Quartiles. See Text for Details.**

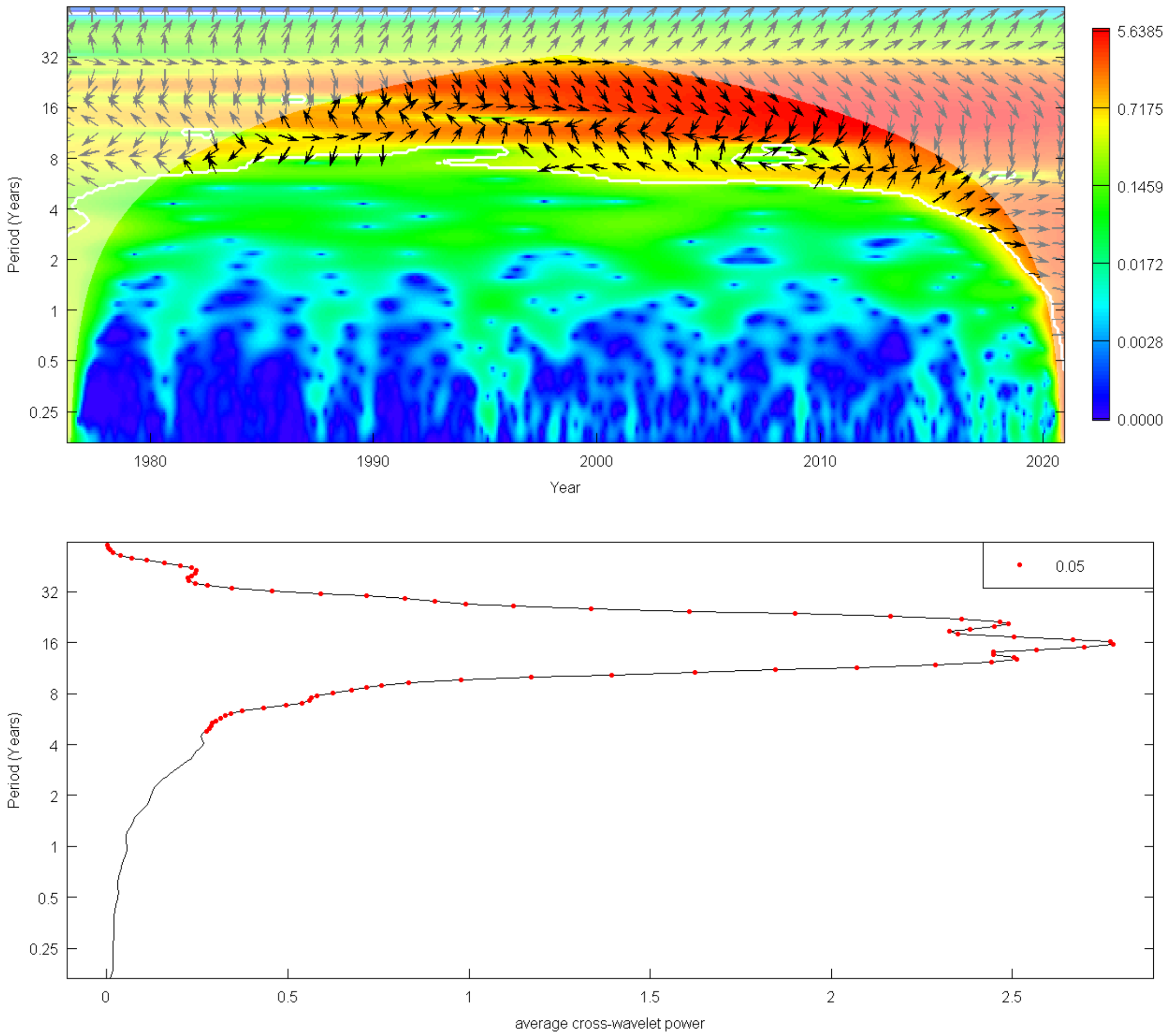

## 4. Discussion and Conclusion

### 4.1 US Medical Product Regulation and Innovation Nexus

An understanding of development – including those factors that may accelerate or hinder their innovation – is crucial given the importance of discovering, developing, delivering, and maintaining quality, safe, and effective medical products. Here, in this report, an attempt is made to describe the impact of regulation and innovation within the United States medical product industry.

To the author's knowledge, while the scholarly literature on this question in the sector is vast[17], it seems to be concentrated on (1) one subsector – either medical devices (see, e.g., Horst and McDonald, 2020) or medicines (see, e.g., Bouchard et al., 2009); (2) proffers either the faults of an implemented regulation (or guidance) or requests additional regulatory influence (see, e.g., Gutierrez, et al., 2020); or (3) shares systematic regulatory knowledge to a specific audience (see, e.g., Amur et al., 2008). Importantly, to the best of the author's knowledge, this is a first research inquiry into the impact of regulation onto innovation that considers (1) the industry holistically (i.e., both medical devices and medicines (drugs)); (2) two direct metrics of innovation and regulation directly applicable to medical products: the former a key surrogate for success for medical product development (viz., an FDA-registration) and the latter a key surrogate for the complexity of regulation (viz., an FDA-issued guideline); (3) the evolution of the FDA as well as the medical device industry as a whole, as it integrates temporal data (from 1976 to 2020); (4) a multi-pronged analysis approach seeking to minimize uncertainty (viz., confirmation of interdependency and is subsequent degree (viz., the exponent α)); and, lastly, (5) the framework of the end-to-end (discovery, development, delivery, and maintenance) paradigm for medical product development, a key factor to accurately estimating the regulation and innovation variables.

[17] An extensive prospective search with a review is probably desirable. On 19-Nov-2022, the author found, for example, the following keyword search on scholar.google.com (1) approximately 2000 records using the search string: '("regulation" and "innovation") and ("medical product" or "medical device" or "medicine" or "drug")'; (2) 27,000 records with '("regulation" and "innovation") and ("medical device")'; (3) 2,070,000 records with '("regulation" and "innovation") and ("medicine").' Some recent examples are placed as text references should the reader be interested.

With the above in mind, this work finds that regulation and innovation (as defined by the surrogates: FDA-issued guidances and FDA-registered medical products) are: (1) Statistically complex (like other econometrics or scientometrics variables). The metrics flow in time in a non-stationary, non-linear, non-normal, and with a long-memory manner reacting (structural changes or change points) with (likely) extrinsic forces. (2) Interdependent; that is, there are strongly correlative and bidirectionally (symmetrically) co-influencing and co-moving variables. (3) Over the history of the data presented in this work (from the mid-1970s to the end of 2020), the relationship has yielded an inverted structure. If confirmed, then this would suggest that regulation is outpacing innovation; that is, the number of monthly guidelines is increasing at a faster rate than the monthly number of registrations.

**4.2 Study Limitations: Data and Analysis**

These results and subsequent interpretations presented here depend on several key assumptions, as follow:

The data was collected from the publicly FDA website, which assumes that the 'system of record' is robust. The number of and accompanied metadata are large; manually culling is required to isolate the variables of interest. Thus, there is some small uncertainty in the integrity of the final datasets.

Importantly, the data collecting method did not take into number of withdrawn regulatory matter for either registrations or guidelines. As stated, the FDA or the sponsor may have withdrawn a registration; or the FDA may have retired or replaced a guidance document. It may be possible but challenging to estimate FDA withdrawn records, since a method may need to be reliant on either FDA or sponsor media communications. However, the author is currently unaware of any vehicle to ascertain those strictly from a sponsor's perspective – notably when it decides to no longer maintain a medical product late in its lifecycle.

Thus, there is residual uncertainty latent in the data. This uncertainty should not be understated as it is critical for the reader to understand that the author has constructed a cumulative medical product data record for both registrations and guidelines, for which they key results rest (viz., interdependency and temporal flow including inversion). The cumulative medical product data assumes that the FDA and sponsor continue their prosecution of the medical product and guideline from its inception onward; that is, the individual metrics accumulate over time. Should the number of guidelines (or registrations) materially decrease, then the ratio of registrations to guidelines would change and such the curve may or may not invert. Confirmation of the database is outstanding and may also comprise corollary investigations.

Lastly, while investigators should feel confident with the descriptive, dynamic, and regression analysis performed from a strictly algorithmic perspective, the causality assessment using variable-lag transfer energy methodology is relatively new and as such additional testing using different econo/sciento-metrics may be appreciated to better understand the algorithm's limitations. Potentially other algorithms may also be used to cross-check the analysis; however, to the author's knowledge, none thus far take into regards both the variable lag as well as structural complexities of the time series to-date.

**4.2 Future Directions**

The use of FDA metrics data presents a unique opportunity to explore the evolution of the US medical product industry from an economic perspective. Like intellectual property, stock market, and/or other metrics, the waves of change in the time series metrics may be directly associated with environment changes. However, unlike such approaches, they may not be as inter-correlated with other metrics (see, e.g., Daizadeh, 2007, 2009) or within themselves (see, e.g., Daizadeh, 2021c, 2021d), and thus react relatively early to changes to various micro-, meso-, macro-economic changes, given their relatively long lag-times (on the order of years to decades). Further, the metrics proposed are generally restricted to FDA-regulated industries. Thus, further avenues to consider, beyond reconfirming the results presented:

- While medicinal lag is well known (Wardell, 1973), determining (a median or average) time lags from promulgation of new legislation to guideline production, as a mark of efficiency of the process, is currently unknown (Yackee, 2021). Here a value was given based on intrinsic changes to the time series; however, an objective reference is needed to better understand the effects of when guidances are issued and when the affected industry actors react, as it may add much needed context to the effect of regulation on the day-to-day operations of the firm and potential impact on innovativeness.
- Beyond medical products, FDA guidance documents are also issued for veterinary products, tobacco, and foods. In principle, therefore, novel metrics may be constructed to consider innovation in these other sectors. Similarly, as other governmental agencies also use guidance documents, opportunities arise to use such regulated documentation to better understand industry response times and/or capabilities to drive new product development.
- Medical product subsector analysis was not performed for this study; however, it would be straightforward to consider medical devices and medicines separately, as the industry actors are different.
- Analytically, predicting rates of medical product development is critical to estimating future work force size, resource needs, and so on (Daizadeh, 2021e). Thus, this work may establish additional input variables to consider when developing such a model; predicting the number of guidance documents may also be of interest given the importance of considering regulatory complexity in medical product design.

The above present some ideas; there are probably many others, all in the hope of better understand the industrial regulatory economics.

**4.3 Conclusion**

In conclusion, this is a first work that investigates and potentially empirically identifies a regulation-innovation nexus in US medical product development. Irrespective of the residual uncertainty in the data and analysis, considering the elements that may affect innovation is imperative, if we are to optimize the rate of discovery, development, delivery, and ultimately the life-cycle processes for securing quality, safe, and effective medical products, which is critical to all.

**Disclosures**

The author is an employee of Takeda Pharmaceuticals; however, this work was completed independently of his employment. The views expressed in this article may not represent those of Takeda Pharmaceuticals. See Appendix for all data and methods to replicate (and hopefully extend) the results presented herein.

**Appendix**

```
citation()
R Core Team (2022). R: A language and environment for
  statistical computing. R Foundation for Statistical
  Computing, Vienna, Austria. URL
  https://www.R-project.org/.

version
               _
platform       x86_64-w64-mingw32
arch           x86_64
os             mingw32
crt            ucrt
system         x86_64, mingw32
status
major          4
minor          2.1
year           2022
month          06
day            23
svn rev        82513
language       R
version.string R version 4.2.1 (2022-06-23 ucrt)
nickname       Funny-Looking Kid

# Read data
Ucurve<-0;
Ucurve<-read.csv("510k-Ucurve-Final.csv", header=TRUE)

#Create MPI (sum of FDA Registered MD and Med) and MPI Guides (sum of FDA Registered MD and
#Med Guidelines)
MPI<-Ucurve$MDReg+Ucurve$MedReg
MPIGuides<-Ucurve$MDGuides+Ucurve$MedGuides

#Create Cumulative sum of MPI and MPI Guides
CumMPI<-cumsum(MPI);
CumMPIGuides<- cumsum(MPIGuides)

Ucurve<-cbind(Ucurve, MPI, MPIGuides, CumMPI, CumMPIGuides)
Ucurve
   MDReg MedReg MDGuides MedGuides  MPI MPIGuides CumMPI CumMPIGuides
1      4    251        0         0  255         0    255            0
2      2    185        0         0  187         0    442            0
3    130    149        0         0  279         0    721            0
4    171    124        2         0  295         2   1016            2
5    114    250        0         0  364         0   1380            2
6    169    124        2         0  293         2   1673            4
7    250    145        1         0  395         1   2068            5
```

| | | | | | | | | |
|---|---|---|---|---|---|---|---|---|
| 8 | 253 | 128 | 0 | 0 | 381 | 0 | 2449 | 5 |
| 9 | 199 | 121 | 1 | 0 | 320 | 1 | 2769 | 6 |
| 10 | 157 | 164 | 0 | 0 | 321 | 0 | 3090 | 6 |
| 11 | 180 | 81 | 1 | 0 | 261 | 1 | 3351 | 7 |
| 12 | 183 | 139 | 0 | 0 | 322 | 0 | 3673 | 7 |
| 13 | 106 | 158 | 0 | 0 | 264 | 0 | 3937 | 7 |
| 14 | 193 | 169 | 0 | 0 | 362 | 0 | 4299 | 7 |
| 15 | 180 | 150 | 0 | 0 | 330 | 0 | 4629 | 7 |
| 16 | 217 | 158 | 0 | 0 | 375 | 0 | 5004 | 7 |
| 17 | 198 | 131 | 0 | 2 | 329 | 2 | 5333 | 9 |
| 18 | 179 | 120 | 0 | 0 | 299 | 0 | 5632 | 9 |
| 19 | 215 | 68 | 1 | 0 | 283 | 1 | 5915 | 10 |
| 20 | 111 | 185 | 0 | 0 | 296 | 0 | 6211 | 10 |
| 21 | 205 | 164 | 0 | 0 | 369 | 0 | 6580 | 10 |
| 22 | 155 | 193 | 0 | 0 | 348 | 0 | 6928 | 10 |
| 23 | 141 | 224 | 0 | 0 | 365 | 0 | 7293 | 10 |
| 24 | 158 | 170 | 0 | 0 | 328 | 0 | 7621 | 10 |
| 25 | 147 | 144 | 1 | 0 | 291 | 1 | 7912 | 11 |
| 26 | 185 | 190 | 0 | 0 | 375 | 0 | 8287 | 11 |
| 27 | 123 | 242 | 0 | 0 | 365 | 0 | 8652 | 11 |
| 28 | 211 | 223 | 0 | 0 | 434 | 0 | 9086 | 11 |
| 29 | 151 | 172 | 0 | 0 | 323 | 0 | 9409 | 11 |
| 30 | 176 | 230 | 0 | 1 | 406 | 1 | 9815 | 12 |
| 31 | 118 | 234 | 0 | 1 | 352 | 1 | 10167 | 13 |
| 32 | 208 | 116 | 0 | 0 | 324 | 0 | 10491 | 13 |
| 33 | 152 | 145 | 0 | 0 | 297 | 0 | 10788 | 13 |
| 34 | 225 | 213 | 0 | 0 | 438 | 0 | 11226 | 13 |
| 35 | 154 | 153 | 0 | 0 | 307 | 0 | 11533 | 13 |
| 36 | 189 | 203 | 0 | 0 | 392 | 0 | 11925 | 13 |
| 37 | 183 | 164 | 0 | 0 | 347 | 0 | 12272 | 13 |
| 38 | 175 | 213 | 0 | 0 | 388 | 0 | 12660 | 13 |
| 39 | 147 | 208 | 0 | 0 | 355 | 0 | 13015 | 13 |
| 40 | 224 | 252 | 1 | 2 | 476 | 3 | 13491 | 16 |
| 41 | 276 | 163 | 1 | 0 | 439 | 1 | 13930 | 17 |
| 42 | 282 | 295 | 0 | 0 | 577 | 0 | 14507 | 17 |
| 43 | 221 | 168 | 0 | 0 | 389 | 0 | 14896 | 17 |
| 44 | 280 | 180 | 0 | 0 | 460 | 0 | 15356 | 17 |
| 45 | 273 | 254 | 0 | 0 | 527 | 0 | 15883 | 17 |
| 46 | 217 | 275 | 0 | 0 | 492 | 0 | 16375 | 17 |
| 47 | 211 | 135 | 1 | 0 | 346 | 1 | 16721 | 18 |
| 48 | 246 | 179 | 0 | 0 | 425 | 0 | 17146 | 18 |
| 49 | 311 | 290 | 1 | 0 | 601 | 1 | 17747 | 19 |
| 50 | 210 | 462 | 0 | 0 | 672 | 0 | 18419 | 19 |
| 51 | 191 | 293 | 0 | 0 | 484 | 0 | 18903 | 19 |
| 52 | 256 | 310 | 1 | 0 | 566 | 1 | 19469 | 20 |
| 53 | 230 | 219 | 3 | 4 | 449 | 7 | 19918 | 27 |
| 54 | 299 | 191 | 0 | 9 | 490 | 9 | 20408 | 36 |
| 55 | 200 | 119 | 0 | 0 | 319 | 0 | 20727 | 36 |
| 56 | 343 | 178 | 0 | 0 | 521 | 0 | 21248 | 36 |
| 57 | 313 | 331 | 0 | 1 | 644 | 1 | 21892 | 37 |
| 58 | 251 | 163 | 0 | 0 | 414 | 0 | 22306 | 37 |
| 59 | 340 | 238 | 0 | 0 | 578 | 0 | 22884 | 37 |
| 60 | 266 | 292 | 0 | 0 | 558 | 0 | 23442 | 37 |
| 61 | 284 | 243 | 1 | 0 | 527 | 1 | 23969 | 38 |
| 62 | 296 | 297 | 0 | 0 | 593 | 0 | 24562 | 38 |
| 63 | 376 | 158 | 0 | 1 | 534 | 1 | 25096 | 39 |
| 64 | 266 | 222 | 0 | 0 | 488 | 0 | 25584 | 39 |
| 65 | 323 | 195 | 1 | 0 | 518 | 1 | 26102 | 40 |
| 66 | 220 | 329 | 0 | 0 | 549 | 0 | 26651 | 40 |
| 67 | 282 | 309 | 0 | 0 | 591 | 0 | 27242 | 40 |
| 68 | 318 | 273 | 0 | 0 | 591 | 0 | 27833 | 40 |
| 69 | 243 | 183 | 0 | 0 | 426 | 0 | 28259 | 40 |
| 70 | 225 | 323 | 0 | 0 | 548 | 0 | 28807 | 40 |
| 71 | 379 | 356 | 0 | 0 | 735 | 0 | 29542 | 40 |
| 72 | 329 | 391 | 0 | 1 | 720 | 1 | 30262 | 41 |
| 73 | 328 | 328 | 0 | 1 | 656 | 1 | 30918 | 42 |
| 74 | 345 | 536 | 0 | 0 | 881 | 0 | 31799 | 42 |
| 75 | 391 | 449 | 0 | 0 | 840 | 0 | 32639 | 42 |
| 76 | 284 | 247 | 0 | 0 | 531 | 0 | 33170 | 42 |
| 77 | 288 | 267 | 0 | 0 | 555 | 0 | 33725 | 42 |
| 78 | 242 | 180 | 0 | 0 | 422 | 0 | 34147 | 42 |
| 79 | 312 | 224 | 0 | 0 | 536 | 0 | 34683 | 42 |
| 80 | 316 | 312 | 0 | 0 | 628 | 0 | 35311 | 42 |
| 81 | 341 | 261 | 0 | 0 | 602 | 0 | 35913 | 42 |
| 82 | 193 | 218 | 0 | 0 | 411 | 0 | 36324 | 42 |
| 83 | 385 | 246 | 0 | 1 | 631 | 1 | 36955 | 43 |
| 84 | 307 | 201 | 0 | 0 | 508 | 0 | 37463 | 43 |
| 85 | 253 | 180 | 0 | 0 | 433 | 0 | 37896 | 43 |
| 86 | 385 | 263 | 0 | 0 | 648 | 0 | 38544 | 43 |
| 87 | 287 | 356 | 1 | 0 | 643 | 1 | 39187 | 44 |
| 88 | 325 | 210 | 0 | 0 | 535 | 0 | 39722 | 44 |
| 89 | 344 | 170 | 0 | 0 | 514 | 0 | 40236 | 44 |
| 90 | 378 | 176 | 0 | 0 | 554 | 0 | 40790 | 44 |
| 91 | 306 | 256 | 0 | 0 | 562 | 0 | 41352 | 44 |
| 92 | 319 | 223 | 0 | 0 | 542 | 0 | 41894 | 44 |
| 93 | 425 | 274 | 0 | 1 | 699 | 1 | 42593 | 45 |
| 94 | 365 | 322 | 0 | 0 | 687 | 0 | 43280 | 45 |
| 95 | 486 | 439 | 0 | 0 | 925 | 0 | 44205 | 45 |
| 96 | 468 | 247 | 0 | 0 | 715 | 0 | 44920 | 45 |
| 97 | 452 | 217 | 0 | 2 | 669 | 2 | 45589 | 47 |
| 98 | 326 | 272 | 0 | 0 | 598 | 0 | 46187 | 47 |
| 99 | 341 | 226 | 0 | 0 | 567 | 0 | 46754 | 47 |
| 100 | 467 | 249 | 0 | 0 | 716 | 0 | 47470 | 47 |
| 101 | 330 | 359 | 0 | 0 | 689 | 0 | 48159 | 47 |
| 102 | 414 | 463 | 0 | 0 | 877 | 0 | 49036 | 47 |
| 103 | 323 | 270 | 0 | 0 | 593 | 0 | 49629 | 47 |
| 104 | 260 | 211 | 0 | 1 | 471 | 1 | 50100 | 48 |
| 105 | 380 | 362 | 1 | 0 | 742 | 1 | 50842 | 49 |
| 106 | 355 | 190 | 1 | 0 | 545 | 1 | 51387 | 50 |
| 107 | 467 | 276 | 0 | 1 | 743 | 1 | 52130 | 51 |
| 108 | 439 | 498 | 0 | 0 | 937 | 0 | 53067 | 51 |
| 109 | 501 | 408 | 0 | 0 | 909 | 0 | 53976 | 51 |
| 110 | 457 | 570 | 0 | 0 | 1027 | 0 | 55003 | 51 |
| 111 | 490 | 438 | 0 | 0 | 928 | 0 | 55931 | 51 |
| 112 | 413 | 503 | 0 | 0 | 916 | 0 | 56847 | 51 |
| 113 | 393 | 344 | 0 | 0 | 737 | 0 | 57584 | 51 |
| 114 | 341 | 530 | 1 | 0 | 871 | 1 | 58455 | 52 |
| 115 | 374 | 347 | 1 | 0 | 721 | 1 | 59176 | 53 |
| 116 | 337 | 344 | 0 | 0 | 681 | 0 | 59857 | 53 |
| 117 | 358 | 509 | 0 | 0 | 867 | 0 | 60724 | 53 |
| 118 | 436 | 421 | 0 | 0 | 857 | 0 | 61581 | 53 |
| 119 | 478 | 238 | 0 | 0 | 716 | 0 | 62297 | 53 |
| 120 | 425 | 303 | 0 | 0 | 728 | 0 | 63025 | 53 |
| 121 | 440 | 328 | 0 | 0 | 768 | 0 | 63793 | 53 |
| 122 | 395 | 326 | 0 | 0 | 721 | 0 | 64514 | 53 |
| 123 | 418 | 353 | 0 | 0 | 771 | 0 | 65285 | 53 |
| 124 | 426 | 369 | 0 | 0 | 795 | 0 | 66080 | 53 |
| 125 | 434 | 314 | 1 | 0 | 748 | 1 | 66828 | 54 |
| 126 | 398 | 354 | 0 | 0 | 752 | 0 | 67580 | 54 |
| 127 | 338 | 359 | 0 | 0 | 697 | 0 | 68277 | 54 |
| 128 | 337 | 292 | 0 | 0 | 629 | 0 | 68906 | 54 |
| 129 | 299 | 289 | 0 | 0 | 588 | 0 | 69494 | 54 |
| 130 | 437 | 290 | 0 | 5 | 727 | 5 | 70221 | 59 |
| 131 | 524 | 378 | 0 | 0 | 902 | 0 | 71123 | 59 |
| 132 | 369 | 408 | 0 | 0 | 777 | 0 | 71900 | 59 |
| 133 | 370 | 375 | 0 | 5 | 745 | 5 | 72645 | 64 |
| 134 | 427 | 291 | 2 | 0 | 718 | 2 | 73363 | 66 |
| 135 | 344 | 565 | 0 | 0 | 909 | 0 | 74272 | 66 |
| 136 | 394 | 287 | 1 | 0 | 681 | 1 | 74953 | 67 |
| 137 | 392 | 256 | 8 | 5 | 648 | 13 | 75601 | 80 |
| 138 | 423 | 271 | 1 | 1 | 694 | 2 | 76295 | 82 |
| 139 | 368 | 260 | 0 | 0 | 628 | 0 | 76923 | 82 |
| 140 | 359 | 310 | 0 | 1 | 669 | 1 | 77592 | 83 |
| 141 | 331 | 290 | 0 | 0 | 621 | 0 | 78213 | 83 |
| 142 | 444 | 446 | 0 | 0 | 890 | 0 | 79103 | 83 |
| 143 | 505 | 522 | 3 | 0 | 1027 | 3 | 80130 | 86 |
| 144 | 543 | 459 | 1 | 0 | 1002 | 1 | 81132 | 87 |
| 145 | 383 | 399 | 0 | 0 | 782 | 0 | 81914 | 87 |
| 146 | 378 | 498 | 0 | 0 | 876 | 0 | 82790 | 87 |
| 147 | 419 | 344 | 0 | 2 | 763 | 2 | 83553 | 89 |
| 148 | 503 | 482 | 1 | 0 | 985 | 1 | 84538 | 90 |
| 149 | 397 | 303 | 0 | 0 | 700 | 0 | 85238 | 90 |
| 150 | 370 | 360 | 0 | 0 | 730 | 0 | 85968 | 90 |
| 151 | 330 | 511 | 0 | 0 | 841 | 0 | 86809 | 90 |
| 152 | 336 | 498 | 0 | 0 | 834 | 0 | 87643 | 90 |
| 153 | 370 | 463 | 0 | 0 | 833 | 0 | 88476 | 90 |
| 154 | 431 | 434 | 3 | 0 | 865 | 3 | 89341 | 93 |
| 155 | 489 | 422 | 3 | 0 | 911 | 3 | 90252 | 96 |
| 156 | 487 | 379 | 0 | 0 | 866 | 0 | 91118 | 96 |
| 157 | 491 | 518 | 0 | 0 | 1009 | 0 | 92127 | 96 |
| 158 | 540 | 397 | 0 | 1 | 937 | 1 | 93064 | 97 |
| 159 | 362 | 551 | 0 | 0 | 913 | 0 | 93977 | 97 |
| 160 | 617 | 262 | 10 | 2 | 879 | 12 | 94856 | 109 |
| 161 | 543 | 236 | 1 | 0 | 779 | 1 | 95635 | 110 |
| 162 | 540 | 341 | 0 | 0 | 881 | 0 | 96516 | 110 |
| 163 | 371 | 301 | 0 | 3 | 672 | 3 | 97188 | 113 |
| 164 | 414 | 231 | 0 | 0 | 645 | 0 | 97833 | 113 |
| 165 | 491 | 441 | 1 | 1 | 932 | 2 | 98765 | 115 |
| 166 | 449 | 323 | 0 | 0 | 772 | 0 | 99537 | 115 |
| 167 | 493 | 269 | 0 | 0 | 762 | 0 | 100299 | 115 |
| 168 | 461 | 303 | 0 | 0 | 764 | 0 | 101063 | 115 |
| 169 | 410 | 245 | 0 | 0 | 655 | 0 | 101718 | 115 |
| 170 | 403 | 290 | 0 | 1 | 693 | 1 | 102411 | 116 |
| 171 | 467 | 203 | 0 | 0 | 670 | 0 | 103081 | 116 |
| 172 | 491 | 222 | 0 | 0 | 713 | 0 | 103794 | 116 |
| 173 | 468 | 410 | 0 | 0 | 878 | 0 | 104672 | 116 |
| 174 | 482 | 214 | 0 | 0 | 696 | 0 | 105368 | 116 |
| 175 | 337 | 302 | 0 | 0 | 639 | 0 | 106007 | 116 |
| 176 | 424 | 241 | 0 | 0 | 665 | 0 | 106672 | 116 |
| 177 | 365 | 363 | 0 | 1 | 728 | 1 | 107400 | 117 |
| 178 | 345 | 491 | 1 | 0 | 836 | 1 | 108236 | 118 |
| 179 | 431 | 490 | 1 | 0 | 921 | 1 | 109157 | 119 |
| 180 | 430 | 404 | 1 | 0 | 834 | 1 | 109991 | 120 |
| 181 | 369 | 339 | 1 | 0 | 708 | 1 | 110699 | 121 |
| 182 | 365 | 229 | 1 | 1 | 594 | 2 | 111293 | 123 |
| 183 | 395 | 294 | 0 | 0 | 689 | 0 | 111982 | 123 |
| 184 | 405 | 490 | 0 | 0 | 895 | 0 | 112877 | 123 |
| 185 | 428 | 375 | 0 | 1 | 803 | 1 | 113680 | 124 |
| 186 | 450 | 274 | 0 | 0 | 724 | 0 | 114404 | 124 |
| 187 | 340 | 300 | 0 | 0 | 640 | 0 | 115044 | 124 |
| 188 | 393 | 395 | 0 | 0 | 788 | 0 | 115832 | 124 |
| 189 | 409 | 399 | 0 | 0 | 808 | 0 | 116640 | 124 |
| 190 | 328 | 367 | 0 | 0 | 695 | 0 | 117335 | 124 |
| 191 | 396 | 283 | 0 | 1 | 679 | 1 | 118014 | 125 |
| 192 | 309 | 649 | 0 | 1 | 958 | 1 | 118972 | 126 |
| 193 | 260 | 395 | 1 | 3 | 655 | 4 | 119627 | 130 |
| 194 | 279 | 326 | 1 | 0 | 605 | 1 | 120232 | 131 |
| 195 | 334 | 294 | 0 | 0 | 628 | 0 | 120860 | 131 |
| 196 | 289 | 343 | 0 | 0 | 632 | 0 | 121492 | 131 |
| 197 | 388 | 389 | 0 | 0 | 777 | 0 | 122269 | 131 |
| 198 | 328 | 241 | 0 | 0 | 569 | 0 | 122838 | 131 |
| 199 | 346 | 283 | 0 | 2 | 629 | 2 | 123467 | 133 |
| 200 | 342 | 412 | 0 | 0 | 754 | 0 | 124221 | 133 |
| 201 | 296 | 221 | 0 | 0 | 517 | 0 | 124738 | 133 |
| 202 | 333 | 278 | 3 | 0 | 611 | 3 | 125349 | 136 |
| 203 | 446 | 361 | 3 | 0 | 807 | 3 | 126156 | 139 |
| 204 | 332 | 413 | 0 | 0 | 745 | 0 | 126901 | 139 |
| 205 | 363 | 324 | 1 | 0 | 687 | 1 | 127588 | 140 |
| 206 | 387 | 365 | 1 | 0 | 752 | 1 | 128340 | 141 |
| 207 | 494 | 413 | 0 | 2 | 907 | 2 | 129247 | 143 |
| 208 | 389 | 360 | 2 | 1 | 749 | 3 | 129996 | 146 |
| 209 | 362 | 258 | 1 | 0 | 620 | 1 | 130616 | 147 |
| 210 | 391 | 335 | 2 | 0 | 726 | 2 | 131342 | 149 |
| 211 | 364 | 290 | 0 | 0 | 654 | 0 | 131996 | 149 |
| 212 | 376 | 359 | 0 | 0 | 735 | 0 | 132731 | 149 |
| 213 | 533 | 458 | 0 | 0 | 991 | 0 | 133722 | 149 |
| 214 | 620 | 394 | 2 | 0 | 1014 | 2 | 134736 | 151 |
| 215 | 700 | 371 | 0 | 0 | 1071 | 0 | 135807 | 151 |
| 216 | 501 | 282 | 1 | 0 | 783 | 1 | 136590 | 152 |
| 217 | 521 | 323 | 0 | 0 | 844 | 0 | 137434 | 152 |
| 218 | 468 | 370 | 1 | 0 | 838 | 1 | 138272 | 153 |
| 219 | 490 | 325 | 1 | 0 | 815 | 1 | 139087 | 154 |
| 220 | 508 | 273 | 1 | 2 | 781 | 3 | 139868 | 157 |
| 221 | 463 | 480 | 2 | 2 | 943 | 4 | 140811 | 161 |
| 222 | 344 | 327 | 0 | 0 | 671 | 0 | 141482 | 161 |
| 223 | 500 | 237 | 1 | 2 | 737 | 3 | 142219 | 164 |
| 224 | 546 | 491 | 0 | 0 | 1037 | 0 | 143256 | 164 |
| 225 | 433 | 349 | 1 | 0 | 782 | 1 | 144038 | 165 |
| 226 | 516 | 409 | 12 | 3 | 925 | 15 | 144963 | 180 |
| 227 | 582 | 355 | 9 | 19 | 937 | 28 | 145900 | 208 |
| 228 | 451 | 246 | 1 | 0 | 697 | 1 | 146597 | 209 |
| 229 | 614 | 356 | 1 | 0 | 970 | 1 | 147567 | 210 |
| 230 | 519 | 379 | 2 | 3 | 898 | 5 | 148465 | 215 |
| 231 | 496 | 252 | 4 | 1 | 748 | 5 | 149213 | 220 |
| 232 | 528 | 501 | 1 | 1 | 1029 | 2 | 150242 | 222 |
| 233 | 508 | 294 | 1 | 0 | 802 | 1 | 151044 | 223 |
| 234 | 518 | 392 | 0 | 0 | 910 | 0 | 151954 | 223 |
| 235 | 417 | 356 | 0 | 3 | 773 | 3 | 152727 | 226 |
| 236 | 401 | 321 | 1 | 1 | 722 | 2 | 153449 | 228 |
| 237 | 358 | 277 | 0 | 0 | 635 | 0 | 154084 | 228 |
| 238 | 439 | 377 | 1 | 1 | 816 | 2 | 154900 | 230 |
| 239 | 488 | 584 | 3 | 6 | 1072 | 9 | 155972 | 239 |
| 240 | 421 | 512 | 1 | 2 | 933 | 3 | 156905 | 242 |
| 241 | 411 | 451 | 1 | 1 | 862 | 2 | 157767 | 244 |
| 242 | 345 | 362 | 1 | 0 | 707 | 1 | 158474 | 245 |
| 243 | 407 | 456 | 3 | 5 | 863 | 8 | 159337 | 253 |
| 244 | 409 | 824 | 3 | 3 | 1233 | 6 | 160570 | 259 |
| 245 | 361 | 346 | 2 | 0 | 707 | 2 | 161277 | 261 |
| 246 | 449 | 401 | 2 | 0 | 850 | 2 | 162127 | 263 |
| 247 | 456 | 254 | 0 | 0 | 710 | 0 | 162837 | 263 |
| 248 | 347 | 405 | 1 | 2 | 752 | 3 | 163589 | 266 |
| 249 | 344 | 452 | 2 | 1 | 796 | 3 | 164385 | 269 |
| 250 | 412 | 292 | 2 | 4 | 704 | 6 | 165089 | 275 |
| 251 | 426 | 366 | 2 | 0 | 792 | 2 | 165881 | 277 |
| 252 | 411 | 554 | 0 | 1 | 965 | 1 | 166846 | 278 |
| 253 | 395 | 395 | 1 | 5 | 790 | 6 | 167636 | 284 |
| 254 | 410 | 462 | 1 | 0 | 872 | 1 | 168508 | 285 |
| 255 | 428 | 473 | 0 | 4 | 901 | 4 | 169409 | 289 |
| 256 | 406 | 425 | 0 | 3 | 831 | 3 | 170240 | 292 |
| 257 | 356 | 359 | 0 | 1 | 715 | 1 | 170955 | 293 |
| 258 | 399 | 391 | 0 | 1 | 790 | 1 | 171745 | 294 |
| 259 | 367 | 386 | 2 | 3 | 753 | 5 | 172498 | 299 |
| 260 | 394 | 436 | 1 | 3 | 830 | 4 | 173328 | 303 |
| 261 | 345 | 334 | 10 | 4 | 679 | 14 | 174007 | 317 |
| 262 | 352 | 593 | 2 | 2 | 945 | 4 | 174952 | 321 |
| 263 | 345 | 425 | 0 | 17 | 770 | 17 | 175722 | 338 |
| 264 | 327 | 435 | 1 | 2 | 762 | 3 | 176484 | 341 |
| 265 | 361 | 411 | 1 | 4 | 772 | 5 | 177256 | 346 |
| 266 | 340 | 369 | 1 | 3 | 709 | 4 | 177965 | 350 |
| 267 | 365 | 332 | 3 | 3 | 697 | 6 | 178662 | 356 |
| 268 | 337 | 468 | 4 | 0 | 805 | 4 | 179467 | 360 |
| 269 | 364 | 520 | 1 | 7 | 884 | 8 | 180351 | 368 |
| 270 | 329 | 476 | 2 | 3 | 805 | 5 | 181156 | 373 |
| 271 | 397 | 418 | 9 | 0 | 815 | 9 | 181971 | 382 |
| 272 | 418 | 670 | 1 | 3 | 1088 | 4 | 183059 | 386 |
| 273 | 299 | 265 | 1 | 3 | 564 | 4 | 183623 | 390 |
| 274 | 347 | 601 | 0 | 3 | 948 | 3 | 184571 | 393 |
| 275 | 340 | 607 | 4 | 1 | 947 | 5 | 185518 | 398 |
| 276 | 314 | 426 | 0 | 1 | 740 | 1 | 186258 | 399 |
| 277 | 302 | 581 | 2 | 2 | 883 | 4 | 187141 | 403 |
| 278 | 365 | 373 | 1 | 2 | 738 | 3 | 187879 | 406 |
| 279 | 317 | 361 | 0 | 2 | 678 | 2 | 188557 | 408 |
| 280 | 363 | 377 | 0 | 5 | 740 | 5 | 189297 | 413 |
| 281 | 385 | 339 | 0 | 2 | 724 | 2 | 190021 | 415 |
| 282 | 310 | 385 | 0 | 0 | 695 | 0 | 190716 | 415 |
| 283 | 378 | 617 | 1 | 2 | 995 | 3 | 191711 | 418 |
| 284 | 381 | 439 | 0 | 2 | 820 | 2 | 192531 | 420 |
| 285 | 320 | 406 | 5 | 0 | 726 | 5 | 193257 | 425 |
| 286 | 381 | 468 | 1 | 1 | 849 | 2 | 194106 | 427 |
| 287 | 390 | 516 | 1 | 1 | 906 | 2 | 195012 | 429 |
| 288 | 367 | 642 | 2 | 2 | 1009 | 4 | 196021 | 433 |
| 289 | 369 | 550 | 0 | 0 | 919 | 0 | 196940 | 433 |
| 290 | 356 | 383 | 1 | 0 | 739 | 1 | 197679 | 434 |
| 291 | 287 | 513 | 2 | 0 | 800 | 2 | 198479 | 436 |
| 292 | 356 | 437 | 3 | 1 | 793 | 4 | 199272 | 440 |
| 293 | 256 | 416 | 0 | 0 | 672 | 0 | 199944 | 440 |
| 294 | 319 | 485 | 0 | 4 | 804 | 4 | 200748 | 444 |
| 295 | 312 | 410 | 1 | 1 | 722 | 2 | 201470 | 446 |
| 296 | 310 | 480 | 0 | 6 | 790 | 6 | 202260 | 452 |
| 297 | 276 | 366 | 1 | 2 | 642 | 3 | 202902 | 455 |
| 298 | 313 | 428 | 1 | 3 | 741 | 4 | 203643 | 459 |
| 299 | 344 | 607 | 0 | 5 | 951 | 5 | 204594 | 464 |
| 300 | 300 | 638 | 2 | 1 | 938 | 3 | 205532 | 467 |
| 301 | 385 | 452 | 2 | 5 | 837 | 7 | 206369 | 474 |
| 302 | 382 | 398 | 1 | 0 | 780 | 1 | 207149 | 475 |
| 303 | 325 | 383 | 2 | 7 | 708 | 9 | 207857 | 484 |
| 304 | 415 | 704 | 0 | 5 | 1119 | 5 | 208976 | 489 |
| 305 | 327 | 383 | 0 | 3 | 710 | 3 | 209686 | 492 |
| 306 | 366 | 475 | 0 | 3 | 841 | 3 | 210527 | 495 |
| 307 | 362 | 459 | 1 | 3 | 821 | 4 | 211348 | 499 |
| 308 | 355 | 333 | 0 | 1 | 688 | 1 | 212036 | 500 |
| 309 | 351 | 479 | 2 | 1 | 830 | 3 | 212866 | 503 |
| 310 | 353 | 457 | 0 | 0 | 810 | 0 | 213676 | 503 |
| 311 | 358 | 613 | 0 | 1 | 971 | 1 | 214647 | 504 |
| 312 | 323 | 555 | 0 | 2 | 878 | 2 | 215525 | 506 |
| 313 | 387 | 477 | 0 | 3 | 864 | 3 | 216389 | 509 |
| 314 | 357 | 400 | 1 | 0 | 757 | 1 | 217146 | 510 |
| 315 | 369 | 693 | 1 | 1 | 1062 | 2 | 218208 | 512 |
| 316 | 371 | 506 | 2 | 0 | 877 | 2 | 219085 | 514 |
| 317 | 350 | 434 | 0 | 0 | 784 | 0 | 219869 | 514 |
| 318 | 388 | 569 | 1 | 1 | 957 | 2 | 220826 | 516 |
| 319 | 344 | 607 | 2 | 1 | 951 | 3 | 221777 | 519 |
| 320 | 389 | 674 | 1 | 3 | 1063 | 4 | 222840 | 523 |
| 321 | 349 | 255 | 0 | 3 | 604 | 3 | 223444 | 526 |
| 322 | 307 | 153 | 1 | 3 | 460 | 4 | 223904 | 530 |
| 323 | 369 | 179 | 0 | 1 | 548 | 1 | 224452 | 531 |
| 324 | 306 | 214 | 1 | 4 | 520 | 5 | 224972 | 536 |
| 325 | 337 | 242 | 0 | 5 | 579 | 5 | 225551 | 541 |
| 326 | 306 | 227 | 1 | 1 | 533 | 2 | 226084 | 543 |
| 327 | 353 | 147 | 2 | 2 | 500 | 4 | 226584 | 547 |
| 328 | 383 | 150 | 0 | 0 | 533 | 0 | 227117 | 547 |
| 329 | 335 | 176 | 2 | 4 | 511 | 6 | 227628 | 553 |
| 330 | 440 | 192 | 0 | 0 | 632 | 0 | 228260 | 553 |
| 331 | 324 | 177 | 1 | 2 | 501 | 3 | 228761 | 556 |
| 332 | 347 | 189 | 2 | 0 | 536 | 2 | 229297 | 558 |
| 333 | 269 | 164 | 2 | 2 | 433 | 4 | 229730 | 562 |
| 334 | 316 | 177 | 1 | 0 | 493 | 1 | 230223 | 563 |
| 335 | 401 | 261 | 1 | 2 | 662 | 3 | 230885 | 566 |
| 336 | 333 | 232 | 0 | 1 | 565 | 1 | 231450 | 567 |
| 337 | 356 | 204 | 3 | 1 | 560 | 4 | 232010 | 571 |
| 338 | 343 | 247 | 1 | 10 | 590 | 11 | 232600 | 582 |
| 339 | 338 | 232 | 1 | 2 | 570 | 3 | 233170 | 585 |
| 340 | 310 | 163 | 0 | 3 | 473 | 3 | 233643 | 588 |
| 341 | 332 | 286 | 1 | 1 | 618 | 2 | 234261 | 590 |
| 342 | 311 | 226 | 1 | 5 | 537 | 6 | 234798 | 596 |
| 343 | 342 | 179 | 2 | 3 | 521 | 5 | 235319 | 601 |
| 344 | 346 | 201 | 1 | 4 | 547 | 5 | 235866 | 606 |
| 345 | 250 | 136 | 1 | 3 | 386 | 4 | 236252 | 610 |
| 346 | 320 | 133 | 6 | 0 | 453 | 6 | 236705 | 616 |
| 347 | 355 | 154 | 3 | 12 | 509 | 15 | 237214 | 631 |
| 348 | 302 | 227 | 3 | 6 | 529 | 9 | 237743 | 640 |
| 349 | 318 | 235 | 2 | 1 | 553 | 3 | 238296 | 643 |
| 350 | 327 | 245 | 0 | 3 | 572 | 3 | 238868 | 646 |
| 351 | 331 | 205 | 2 | 0 | 536 | 2 | 239404 | 648 |
| 352 | 333 | 378 | 1 | 0 | 711 | 1 | 240115 | 649 |
| 353 | 354 | 187 | 2 | 2 | 541 | 4 | 240656 | 653 |
| 354 | 310 | 146 | 1 | 2 | 456 | 3 | 241112 | 656 |
| 355 | 316 | 183 | 2 | 0 | 499 | 2 | 241611 | 658 |
| 356 | 383 | 186 | 0 | 0 | 569 | 0 | 242180 | 658 |
| 357 | 287 | 160 | 4 | 9 | 447 | 13 | 242627 | 671 |
| 358 | 293 | 173 | 3 | 6 | 466 | 9 | 243093 | 680 |
| 359 | 427 | 235 | 4 | 7 | 662 | 11 | 243755 | 691 |
| 360 | 340 | 172 | 1 | 3 | 512 | 4 | 244267 | 695 |
| 361 | 388 | 158 | 2 | 0 | 546 | 2 | 244813 | 697 |
| 362 | 373 | 176 | 0 | 8 | 549 | 8 | 245362 | 705 |
| 363 | 370 | 216 | 0 | 2 | 586 | 2 | 245948 | 707 |
| 364 | 361 | 221 | 0 | 3 | 582 | 3 | 246530 | 710 |
| 365 | 359 | 178 | 2 | 2 | 537 | 4 | 247067 | 714 |
| 366 | 362 | 204 | 2 | 5 | 566 | 7 | 247633 | 721 |
| 367 | 358 | 129 | 1 | 1 | 487 | 2 | 248120 | 723 |
| 368 | 373 | 209 | 1 | 0 | 582 | 1 | 248702 | 724 |
| 369 | 322 | 181 | 0 | 0 | 503 | 0 | 249205 | 724 |
| 370 | 290 | 240 | 2 | 4 | 530 | 6 | 249735 | 730 |
| 371 | 374 | 212 | 2 | 1 | 586 | 3 | 250321 | 733 |
| 372 | 305 | 240 | 0 | 0 | 545 | 0 | 250866 | 733 |
| 373 | 392 | 209 | 2 | 5 | 601 | 7 | 251467 | 740 |
| 374 | 322 | 279 | 5 | 2 | 601 | 7 | 252068 | 747 |
| 375 | 302 | 207 | 3 | 2 | 509 | 5 | 252577 | 752 |
| 376 | 379 | 256 | 0 | 3 | 635 | 3 | 253212 | 755 |
| 377 | 312 | 310 | 4 | 3 | 622 | 7 | 253834 | 762 |
| 378 | 369 | 311 | 3 | 1 | 680 | 4 | 254514 | 766 |
| 379 | 353 | 245 | 0 | 2 | 598 | 2 | 255112 | 768 |
| 380 | 383 | 469 | 0 | 1 | 852 | 1 | 255964 | 769 |
| 381 | 317 | 257 | 2 | 0 | 574 | 2 | 256538 | 771 |
| 382 | 348 | 286 | 2 | 6 | 634 | 8 | 257172 | 779 |
| 383 | 329 | 257 | 2 | 3 | 586 | 5 | 257758 | 784 |
| 384 | 383 | 248 | 3 | 4 | 631 | 7 | 258389 | 791 |
| 385 | 419 | 266 | 0 | 0 | 685 | 0 | 259074 | 791 |
| 386 | 343 | 257 | 0 | 5 | 600 | 5 | 259674 | 796 |
| 387 | 410 | 202 | 3 | 2 | 612 | 5 | 260286 | 801 |
| 388 | 348 | 234 | 3 | 0 | 582 | 3 | 260868 | 804 |
| 389 | 400 | 236 | 0 | 3 | 636 | 3 | 261504 | 807 |
| 390 | 425 | 308 | 1 | 1 | 733 | 2 | 262237 | 809 |
| 391 | 362 | 211 | 0 | 2 | 573 | 2 | 262810 | 811 |
| 392 | 460 | 250 | 3 | 1 | 710 | 4 | 263520 | 815 |
| 393 | 357 | 217 | 1 | 5 | 574 | 6 | 264094 | 821 |
| 394 | 373 | 221 | 1 | 1 | 594 | 2 | 264688 | 823 |
| 395 | 433 | 604 | 0 | 0 | 1037 | 0 | 265725 | 823 |
| 396 | 399 | 278 | 0 | 10 | 677 | 10 | 266402 | 833 |
| 397 | 366 | 250 | 0 | 3 | 616 | 3 | 267018 | 836 |
| 398 | 369 | 210 | 2 | 2 | 579 | 4 | 267597 | 840 |
| 399 | 433 | 264 | 2 | 5 | 697 | 7 | 268294 | 847 |
| 400 | 294 | 235 | 1 | 1 | 529 | 2 | 268823 | 849 |
| 401 | 339 | 233 | 0 | 3 | 572 | 3 | 269395 | 852 |
| 402 | 376 | 156 | 1 | 7 | 532 | 8 | 269927 | 860 |
| 403 | 338 | 250 | 0 | 4 | 588 | 4 | 270515 | 864 |
| 404 | 423 | 199 | 1 | 8 | 622 | 9 | 271137 | 873 |
| 405 | 341 | 224 | 1 | 4 | 565 | 5 | 271702 | 878 |
| 406 | 309 | 202 | 2 | 5 | 511 | 7 | 272213 | 885 |
| 407 | 430 | 255 | 1 | 6 | 685 | 7 | 272898 | 892 |
| 408 | 390 | 307 | 1 | 2 | 697 | 3 | 273595 | 895 |
| 409 | 348 | 234 | 0 | 1 | 582 | 1 | 274177 | 896 |
| 410 | 416 | 264 | 2 | 6 | 680 | 8 | 274857 | 904 |
| 411 | 400 | 216 | 1 | 0 | 616 | 1 | 275473 | 905 |
| 412 | 415 | 221 | 1 | 4 | 636 | 5 | 276109 | 910 |
| 413 | 374 | 211 | 2 | 6 | 585 | 8 | 276694 | 918 |
| 414 | 395 | 243 | 0 | 1 | 638 | 1 | 277332 | 919 |
| 415 | 377 | 259 | 2 | 3 | 636 | 5 | 277968 | 924 |
| 416 | 457 | 331 | 1 | 1 | 788 | 2 | 278756 | 926 |
| 417 | 417 | 235 | 0 | 3 | 652 | 3 | 279408 | 929 |
| 418 | 374 | 255 | 0 | 1 | 629 | 1 | 280037 | 930 |
| 419 | 484 | 330 | 1 | 1 | 814 | 2 | 280851 | 932 |
| 420 | 409 | 246 | 1 | 2 | 655 | 3 | 281506 | 935 |
| 421 | 520 | 306 | 1 | 2 | 826 | 3 | 282332 | 938 |
| 422 | 432 | 407 | 0 | 4 | 839 | 4 | 283171 | 942 |
| 423 | 439 | 282 | 1 | 0 | 721 | 1 | 283892 | 943 |
| 424 | 479 | 297 | 1 | 8 | 776 | 9 | 284668 | 952 |
| 425 | 444 | 301 | 0 | 7 | 745 | 7 | 285413 | 959 |
| 426 | 416 | 292 | 0 | 4 | 708 | 4 | 286121 | 963 |
| 427 | 421 | 300 | 0 | 6 | 721 | 6 | 286842 | 969 |
| 428 | 515 | 309 | 0 | 5 | 824 | 5 | 287666 | 974 |
| 429 | 352 | 331 | 0 | 0 | 683 | 0 | 288349 | 974 |
| 430 | 392 | 303 | 2 | 2 | 695 | 4 | 289044 | 978 |
| 431 | 499 | 346 | 0 | 4 | 845 | 4 | 289889 | 982 |
| 432 | 495 | 301 | 0 | 1 | 796 | 1 | 290685 | 983 |
| 433 | 491 | 280 | 0 | 2 | 771 | 2 | 291456 | 985 |
| 434 | 437 | 204 | 0 | 6 | 641 | 6 | 292097 | 991 |
| 435 | 451 | 302 | 1 | 1 | 753 | 2 | 292850 | 993 |
| 436 | 457 | 319 | 0 | 3 | 776 | 3 | 293626 | 996 |
| 437 | 455 | 302 | 0 | 3 | 757 | 3 | 294383 | 999 |
| 438 | 488 | 265 | 0 | 5 | 753 | 5 | 295136 | 1004 |
| 439 | 462 | 350 | 1 | 4 | 812 | 5 | 295948 | 1009 |
| 440 | 569 | 413 | 0 | 8 | 982 | 8 | 296930 | 1017 |
| 441 | 355 | 417 | 1 | 4 | 772 | 5 | 297702 | 1022 |
| 442 | 421 | 488 | 3 | 5 | 909 | 8 | 298611 | 1030 |
| 443 | 471 | 460 | 4 | 5 | 931 | 9 | 299542 | 1039 |
| 444 | 431 | 549 | 3 | 7 | 980 | 10 | 300522 | 1049 |
| 445 | 458 | 533 | 1 | 7 | 991 | 8 | 301513 | 1057 |
| 446 | 454 | 377 | 1 | 7 | 831 | 8 | 302344 | 1065 |
| 447 | 464 | 468 | 0 | 5 | 932 | 5 | 303276 | 1070 |
| 448 | 514 | 431 | 3 | 4 | 945 | 7 | 304221 | 1077 |
| 449 | 411 | 396 | 1 | 4 | 807 | 5 | 305028 | 1082 |
| 450 | 521 | 425 | 1 | 3 | 946 | 4 | 305974 | 1086 |
| 451 | 382 | 299 | 3 | 3 | 681 | 6 | 306655 | 1092 |
| 452 | 513 | 343 | 0 | 1 | 856 | 1 | 307511 | 1093 |
| 453 | 433 | 400 | 0 | 4 | 833 | 4 | 308344 | 1097 |
| 454 | 438 | 346 | 2 | 7 | 784 | 9 | 309128 | 1106 |
| 455 | 470 | 376 | 4 | 7 | 846 | 11 | 309974 | 1117 |
| 456 | 513 | 361 | 1 | 1 | 874 | 2 | 310848 | 1119 |
| 457 | 525 | 255 | 2 | 5 | 780 | 7 | 311628 | 1126 |
| 458 | 441 | 509 | 3 | 14 | 950 | 17 | 312578 | 1143 |
| 459 | 539 | 514 | 2 | 6 | 1053 | 8 | 313631 | 1151 |
| 460 | 474 | 537 | 6 | 5 | 1011 | 11 | 314642 | 1162 |
| 461 | 407 | 858 | 1 | 0 | 1265 | 1 | 315907 | 1163 |
| 462 | 547 | 334 | 4 | 5 | 881 | 9 | 316788 | 1172 |
| 463 | 376 | 507 | 2 | 5 | 883 | 7 | 317671 | 1179 |
| 464 | 523 | 675 | 4 | 10 | 1198 | 14 | 318869 | 1193 |
| 465 | 411 | 456 | 0 | 2 | 867 | 2 | 319736 | 1195 |
| 466 | 393 | 490 | 1 | 1 | 883 | 2 | 320619 | 1197 |
| 467 | 463 | 535 | 1 | 4 | 998 | 5 | 321617 | 1202 |
| 468 | 492 | 662 | 1 | 7 | 1154 | 8 | 322771 | 1210 |
| 469 | 409 | 451 | 0 | 2 | 860 | 2 | 323631 | 1212 |
| 470 | 444 | 545 | 1 | 8 | 989 | 9 | 324620 | 1221 |
| 471 | 496 | 560 | 1 | 3 | 1056 | 4 | 325676 | 1225 |
| 472 | 388 | 454 | 3 | 8 | 842 | 11 | 326518 | 1236 |
| 473 | 386 | 641 | 1 | 2 | 1027 | 3 | 327545 | 1239 |
| 474 | 458 | 524 | 1 | 4 | 982 | 5 | 328527 | 1244 |
| 475 | 408 | 592 | 0 | 7 | 1000 | 7 | 329527 | 1251 |
| 476 | 445 | 589 | 2 | 2 | 1034 | 4 | 330561 | 1255 |
| 477 | 387 | 556 | 2 | 2 | 943 | 4 | 331504 | 1259 |
| 478 | 434 | 488 | 4 | 6 | 922 | 10 | 332426 | 1269 |
| 479 | 510 | 470 | 2 | 4 | 980 | 6 | 333406 | 1275 |
| 480 | 477 | 496 | 2 | 8 | 973 | 10 | 334379 | 1285 |
| 481 | 398 | 610 | 2 | 1 | 1008 | 3 | 335387 | 1288 |
| 482 | 452 | 378 | 2 | 9 | 830 | 11 | 336217 | 1299 |
| 483 | 455 | 350 | 4 | 7 | 805 | 11 | 337022 | 1310 |
| 484 | 500 | 329 | 3 | 4 | 829 | 7 | 337851 | 1317 |
| 485 | 512 | 402 | 2 | 8 | 914 | 10 | 338765 | 1327 |
| 486 | 439 | 313 | 2 | 8 | 752 | 10 | 339517 | 1337 |
| 487 | 422 | 384 | 4 | 13 | 806 | 17 | 340323 | 1354 |
| 488 | 534 | 821 | 5 | 12 | 1355 | 17 | 341678 | 1371 |
| 489 | 355 | 449 | 3 | 18 | 804 | 21 | 342482 | 1392 |
| 490 | 441 | 405 | 0 | 1 | 846 | 1 | 343328 | 1393 |
| 491 | 491 | 345 | 0 | 0 | 836 | 0 | 344164 | 1393 |
| 492 | 454 | 411 | 0 | 0 | 865 | 0 | 345029 | 1393 |
| 493 | 559 | 455 | 0 | 0 | 1014 | 0 | 346043 | 1393 |
| 494 | 564 | 306 | 2 | 5 | 870 | 7 | 346913 | 1400 |
| 495 | 430 | 280 | 1 | 8 | 710 | 9 | 347623 | 1409 |
| 496 | 587 | 497 | 2 | 5 | 1084 | 7 | 348707 | 1416 |
| 497 | 443 | 307 | 7 | 9 | 750 | 16 | 349457 | 1432 |
| 498 | 526 | 299 | 11 | 20 | 825 | 31 | 350282 | 1463 |
| 499 | 512 | 285 | 2 | 11 | 797 | 13 | 351079 | 1476 |
| 500 | 563 | 342 | 6 | 20 | 905 | 26 | 351984 | 1502 |
| 501 | 394 | 186 | 1 | 8 | 580 | 9 | 352564 | 1511 |
| 502 | 452 | 233 | 2 | 11 | 685 | 13 | 353249 | 1524 |
| 503 | 581 | 281 | 1 | 12 | 862 | 13 | 354111 | 1537 |
| 504 | 567 | 303 | 3 | 12 | 870 | 15 | 354981 | 1552 |
| 505 | 497 | 310 | 1 | 17 | 807 | 18 | 355788 | 1570 |
| 506 | 515 | 337 | 3 | 10 | 852 | 13 | 356640 | 1583 |
| 507 | 513 | 341 | 0 | 22 | 854 | 22 | 357494 | 1605 |
| 508 | 464 | 246 | 1 | 11 | 710 | 12 | 358204 | 1617 |
| 509 | 377 | 773 | 3 | 16 | 1150 | 19 | 359354 | 1636 |
| 510 | 491 | 374 | 3 | 21 | 865 | 24 | 360219 | 1660 |
| 511 | 527 | 444 | 0 | 6 | 971 | 6 | 361190 | 1666 |
| 512 | 479 | 356 | 4 | 15 | 835 | 19 | 362025 | 1685 |
| 513 | 376 | 342 | 0 | 0 | 718 | 0 | 362743 | 1685 |
| 514 | 302 | 338 | 7 | 25 | 640 | 32 | 363383 | 1717 |
| 515 | 576 | 266 | 3 | 20 | 842 | 23 | 364225 | 1740 |
| 516 | 441 | 441 | 5 | 11 | 882 | 16 | 365107 | 1756 |
| 517 | 526 | 483 | 5 | 17 | 1009 | 22 | 366116 | 1778 |
| 518 | 473 | 229 | 1 | 12 | 702 | 13 | 366818 | 1791 |
| 519 | 507 | 419 | 5 | 19 | 926 | 24 | 367744 | 1815 |
| 520 | 436 | 453 | 4 | 15 | 889 | 19 | 368633 | 1834 |
| 521 | 438 | 488 | 17 | 22 | 926 | 39 | 369559 | 1873 |
| 522 | 502 | 840 | 2 | 15 | 1342 | 17 | 370901 | 1890 |
| 523 | 438 | 776 | 0 | 6 | 1214 | 6 | 372115 | 1896 |
| 524 | 540 | 524 | 7 | 21 | 1064 | 28 | 373179 | 1924 |
| 525 | 376 | 413 | 3 | 17 | 789 | 20 | 373968 | 1944 |
| 526 | 358 | 531 | 3 | 13 | 889 | 16 | 374857 | 1960 |
| 527 | 477 | 431 | 13 | 22 | 908 | 35 | 375765 | 1995 |
| 528 | 522 | 548 | 13 | 15 | 1070 | 28 | 376835 | 2023 |
| 529 | 442 | 498 | 5 | 13 | 940 | 18 | 377775 | 2041 |
| 530 | 409 | 419 | 2 | 10 | 828 | 12 | 378603 | 2053 |
| 531 | 459 | 483 | 10 | 22 | 942 | 32 | 379545 | 2085 |
| 532 | 444 | 513 | 4 | 19 | 957 | 23 | 380502 | 2108 |
| 533 | 430 | 456 | 9 | 19 | 886 | 28 | 381388 | 2136 |
| 534 | 504 | 469 | 6 | 18 | 973 | 24 | 382361 | 2160 |
| 535 | 437 | 461 | 6 | 18 | 898 | 24 | 383259 | 2184 |

```
536   572    377       7       18  949       25 384208        2209

str(Ucurve)
'data.frame':   536 obs. of  8 variables:
 $ MDReg      : int  4 2 130 171 114 169 250 253 199 157 ...
 $ MedReg     : int  251 185 149 124 250 124 145 128 121 164 ...
 $ MDGuides   : int  0 0 0 2 0 2 1 0 1 0 ...
 $ MedGuides  : int  0 0 0 0 0 0 0 0 0 0 ...
 $ MPI        : int  255 187 279 295 364 293 395 381 320 321 ...
 $ MPIGuides  : int  0 0 0 2 0 2 1 0 1 0 ...
 $ CumMPI     : int  255 442 721 1016 1380 1673 2068 2449 2769 3090 ...
 $ CumMPIGuides: int  0 0 0 2 2 4 5 5 6 6 ...

library(psych); citation("psych")
Revelle, W. (2022) psych: Procedures for Personality and
  Psychological Research, Northwestern University,
  Evanston, Illinois, USA,
  https://CRAN.R-project.org/package=psych Version = 2.2.9.

psych::describe(Ucurve)
             vars   n     mean       sd   median   trimmed      mad min    max  range  skew kurtosis      se
MDReg           1 536    377.71     98.84    376.5    382.16     84.51   2    700    698 -0.41     0.74    4.27
MedReg          2 536    339.09    136.15    321.5    328.94    135.66  68    858    790  0.78     0.63    5.88
MDGuides        3 536      1.23      2.04      1.0      0.80      1.48   0     17     17  3.29    14.81    0.09
MedGuides       4 536      2.89      4.65      1.0      1.78      1.48   0     25     25  2.36     5.58    0.20
MPI             5 536    716.81    197.80    721.0    720.55    206.08 187   1355   1168 -0.06    -0.09    8.54
MPIGuides       6 536      4.12      6.09      2.0      2.74      2.97   0     39     39  2.55     7.42    0.26
CumMPI          7 536 177557.10 113659.38 179909.0 175762.11 144213.98 255 384208 383953  0.05    -
1.23 4909.34
CumMPIGuides    8 536    526.80    536.86    364.0    447.12    474.43   0   2209   2209  1.07     0.35   23.19


#Normality
library(nortest);citation("nortest")
 Gross J, Ligges U (2015). _nortest: Tests for Normality_.
  R package version 1.0-4,
  <https://CRAN.R-project.org/package=nortest>.

library(tseries); citation("tseries")
Adrian Trapletti and Kurt Hornik (2022). tseries: Time
  Series Analysis and Computational Finance. R package
  version 0.10-52.

i=0; for (i in 1:8) { print(i); print (jarque.bera.test (Ucurve[,i]) ) }
[1] 1
        Jarque Bera Test
data:  Ucurve[, i]
X-squared = 27.361, df = 2, p-value = 1.145e-06
```

```
[1] 2
	Jarque Bera Test

data:  Ucurve[, i]
X-squared = 63.264, df = 2, p-value = 1.832e-14
[1] 3
	Jarque Bera Test

data:  Ucurve[, i]
X-squared = 5912.7, df = 2, p-value < 2.2e-16

[1] 4
	Jarque Bera Test
data:  Ucurve[, i]
X-squared = 1201.9, df = 2, p-value < 2.2e-16

[1] 5
	Jarque Bera Test
data:  Ucurve[, i]
X-squared = 0.45603, df = 2, p-value = 0.7961

[1] 6
	Jarque Bera Test
data:  Ucurve[, i]
X-squared = 1826, df = 2, p-value < 2.2e-16

[1] 7
	Jarque Bera Test
data:  Ucurve[, i]
X-squared = 33.42, df = 2, p-value = 5.532e-08

[1] 8
	Jarque Bera Test
data:  Ucurve[, i]
X-squared = 105.52, df = 2, p-value < 2.2e-16

i=0; for (i in 1:8) { c=0; c <- ad.test (Ucurve[,i]); print (c$p.value) }
[1] 1.248561e-06
[1] 3.844127e-11
[1] 3.7e-24
[1] 3.7e-24
[1] 0.09462634
[1] 3.7e-24
[1] 3.334211e-17
[1] 3.7e-24

i=0; for (i in 1:8) { c=0; c <- cvm.test (Ucurve[,i]); print (c$p.value) }
[1] 3.285281e-05
```

```
[1] 6.232428e-08
[1] 7.37e-10
[1] 7.37e-10
[1] 0.2449725
[1] 7.37e-10
[1] 1.760278e-09
[1] 7.37e-10
Warning messages:
1: In cvm.test(Ucurve[, i]) :
  p-value is smaller than 7.37e-10, cannot be computed more accurately
2: In cvm.test(Ucurve[, i]) :
  p-value is smaller than 7.37e-10, cannot be computed more accurately
3: In cvm.test(Ucurve[, i]) :
  p-value is smaller than 7.37e-10, cannot be computed more accurately
4: In cvm.test(Ucurve[, i]) :
  p-value is smaller than 7.37e-10, cannot be computed more accurately


#Stationarity
library(aTSA);citation("aTSA")
Qiu D (2015). _aTSA: Alternative Time Series Analysis_. R
  package version 3.1.2,
  <https://CRAN.R-project.org/package=aTSA>.

i=0; for (i in 1:8) { print(i); aTSA::stationary.test(Ucurve[,i], method="kpss") }
[1] 1
KPSS Unit Root Test
alternative: nonstationary

Type 1: no drift no trend
 lag stat p.value
   5 5.36    0.01
-----
 Type 2: with drift no trend
 lag stat p.value
   5 2.53    0.01
-----
 Type 1: with drift and trend
 lag  stat p.value
   5 0.641    0.01
-----------
Note: p.value = 0.01 means p.value <= 0.01
    : p.value = 0.10 means p.value >= 0.10
[1] 2
KPSS Unit Root Test
alternative: nonstationary

Type 1: no drift no trend
```

```
 lag stat p.value
   5 10.3    0.01
-----
 Type 2: with drift no trend
 lag stat p.value
   5  1.2    0.01
-----
 Type 1: with drift and trend
 lag  stat p.value
   5 0.447    0.01
-----------
Note: p.value = 0.01 means p.value <= 0.01
    : p.value = 0.10 means p.value >= 0.10
[1] 3
KPSS Unit Root Test
alternative: nonstationary

Type 1: no drift no trend
 lag stat p.value
   5 5.84    0.01
-----
 Type 2: with drift no trend
 lag stat p.value
   5 3.04    0.01
-----
 Type 1: with drift and trend
 lag stat p.value
   5  0.3    0.01
-----------
Note: p.value = 0.01 means p.value <= 0.01
    : p.value = 0.10 means p.value >= 0.10
[1] 4
KPSS Unit Root Test
alternative: nonstationary

Type 1: no drift no trend
 lag stat p.value
   5  2.5   0.016
-----
 Type 2: with drift no trend
 lag stat p.value
   5  4.2    0.01
-----
 Type 1: with drift and trend
 lag  stat p.value
   5 0.845    0.01
-----------
Note: p.value = 0.01 means p.value <= 0.01
```

```
    : p.value = 0.10 means p.value >= 0.10
[1] 5
KPSS Unit Root Test
alternative: nonstationary

Type 1: no drift no trend
 lag stat p.value
   5 5.69    0.01
-----
 Type 2: with drift no trend
 lag stat p.value
   5  2.1    0.01
-----
 Type 1: with drift and trend
 lag  stat p.value
   5 0.644    0.01
-----------
Note: p.value = 0.01 means p.value <= 0.01
    : p.value = 0.10 means p.value >= 0.10
[1] 6
KPSS Unit Root Test
alternative: nonstationary

Type 1: no drift no trend
 lag stat p.value
   5 3.18    0.01
-----
 Type 2: with drift no trend
 lag stat p.value
   5 4.08    0.01
-----
 Type 1: with drift and trend
 lag  stat p.value
   5 0.722    0.01
-----------
Note: p.value = 0.01 means p.value <= 0.01
    : p.value = 0.10 means p.value >= 0.10
[1] 7
KPSS Unit Root Test
alternative: nonstationary

Type 1: no drift no trend
 lag stat p.value
   5 22.7    0.01
-----
 Type 2: with drift no trend
 lag stat p.value
   5 0.77    0.01
```

```
-----
 Type 1: with drift and trend
 lag  stat p.value
   5 0.719    0.01
-----------
Note: p.value = 0.01 means p.value <= 0.01
    : p.value = 0.10 means p.value >= 0.10
[1] 8
KPSS Unit Root Test
alternative: nonstationary

Type 1: no drift no trend
 lag  stat p.value
   5 0.594     0.1
-----
 Type 2: with drift no trend
 lag  stat p.value
   5 0.734  0.0105
-----
 Type 1: with drift and trend
 lag  stat p.value
   5 0.516    0.01
-----------
Note: p.value = 0.01 means p.value <= 0.01
    : p.value = 0.10 means p.value >= 0.10

library(nonlinearTseries); citation("nonlinearTseries")
  Garcia C (2022). _nonlinearTseries: Nonlinear Time Series Analysis_. R package version 0.2.12,
  <https://CRAN.R-project.org/package=nonlinearTseries>.

i=0; for (i in 1:8) { print(i); lag.length=0; lag.length=timeLag(Ucurve[,i]); print
(Box.test(Ucurve[i],lag=lag.length, type="Ljung-Box") ) }
[1] 1

        Box-Ljung test

data:  Ucurve[i]
X-squared = 5173.4, df = 30, p-value < 2.2e-16

[1] 2

        Box-Ljung test

data:  Ucurve[i]
X-squared = 1944, df = 16, p-value < 2.2e-16

[1] 3
```

```
        Box-Ljung test

data:  Ucurve[i]
X-squared = 127.83, df = 2, p-value < 2.2e-16

[1] 4

        Box-Ljung test

data:  Ucurve[i]
X-squared = 4155.7, df = 28, p-value < 2.2e-16

[1] 5

        Box-Ljung test

data:  Ucurve[i]
X-squared = 5153.7, df = 33, p-value < 2.2e-16

[1] 6

        Box-Ljung test

data:  Ucurve[i]
X-squared = 3488.5, df = 25, p-value < 2.2e-16

[1] 7

        Box-Ljung test

data:  Ucurve[i]
X-squared = 33868, df = 118, p-value < 2.2e-16

[1] 8

        Box-Ljung test

data:  Ucurve[i]
X-squared = 26320, df = 105, p-value < 2.2e-16


library(forecast); citation("forecast")

To cite the forecast package in publications, please use:

Hyndman R, Athanasopoulos G, Bergmeir C, Caceres G, Chhay
  L, O'Hara-Wild M, Petropoulos F, Razbash S, Wang E,
  Yasmeen F (2022). _forecast: Forecasting functions for
```

```
  time series and linear models_. R package version 8.18,
  <https://pkg.robjhyndman.com/forecast/>.

  Hyndman RJ, Khandakar Y (2008). "Automatic time series
  forecasting: the forecast package for R." _Journal of
  Statistical Software_, *26*(3), 1-22.
  doi:10.18637/jss.v027.i03
  <https://doi.org/10.18637/jss.v027.i03>.

#nonstationary with diff = 1 or 2
i=0; for (i in 1:8) { print( ndiffs( Ucurve[,i] ) ) }
[1] 1
[1] 1
[1] 1
[1] 1
[1] 1
[1] 1
[1] 2
[1] 2

#Long-memory
#Determine Time lag
i=0; for (i in 1:8) { print (timeLag(Ucurve[,i]) )}
[1] 30
[1] 16
[1] 2
[1] 28
[1] 33
[1] 25
[1] 118
[1] 105

library(LongMemoryTS); citation("LongMemoryTS")
  Leschinski C (2019). _LongMemoryTS: Long Memory Time
  Series_. R package version 0.1.0,
  <https://CRAN.R-project.org/package=LongMemoryTS>.

m<-floor(1+536^0.75)
# Qu test
Qu.test(diff(Ucurve),m)
i=0; for (i in 1:8) { print(i); print ( Qu.test (diff(Ucurve[,i]) , m ) ) }
[1] 1
$W.stat
[1] 1.474096

$CriticalValues
         eps=.02 eps=.05
alpha=.1    1.118   1.022
```

```
alpha=.05    1.252   1.155
alpha=.025   1.374   1.277
alpha=.01    1.517   1.426

[1] 2
$W.stat
[1] 0.8295874

$CriticalValues
         eps=.02 eps=.05
alpha=.1     1.118   1.022
alpha=.05    1.252   1.155
alpha=.025   1.374   1.277
alpha=.01    1.517   1.426

[1] 3
$W.stat
[1] 3.796412

$CriticalValues
         eps=.02 eps=.05
alpha=.1     1.118   1.022
alpha=.05    1.252   1.155
alpha=.025   1.374   1.277
alpha=.01    1.517   1.426

[1] 4
$W.stat
[1] 1.939415

$CriticalValues
         eps=.02 eps=.05
alpha=.1     1.118   1.022
alpha=.05    1.252   1.155
alpha=.025   1.374   1.277
alpha=.01    1.517   1.426

[1] 5
$W.stat
[1] 1.941642

$CriticalValues
         eps=.02 eps=.05
alpha=.1     1.118   1.022
alpha=.05    1.252   1.155
alpha=.025   1.374   1.277
alpha=.01    1.517   1.426
```

```
[1] 6
$W.stat
[1] 1.915778

$CriticalValues
          eps=.02 eps=.05
alpha=.1     1.118   1.022
alpha=.05    1.252   1.155
alpha=.025   1.374   1.277
alpha=.01    1.517   1.426

[1] 7
$W.stat
[1] 1.334307

$CriticalValues
          eps=.02 eps=.05
alpha=.1     1.118   1.022
alpha=.05    1.252   1.155
alpha=.025   1.374   1.277
alpha=.01    1.517   1.426

[1] 8
$W.stat
[1] 1.473759

$CriticalValues
          eps=.02 eps=.05
alpha=.1     1.118   1.022
alpha=.05    1.252   1.155
alpha=.025   1.374   1.277
alpha=.01    1.517   1.426


#Multivariate local Whittle Score
i=0; for (i in 1:8) { print(i); print ( MLWS (diff(Ucurve[,i]) , m=m ) ) }
[1] 1
$B
     [,1]
[1,]    1

$d
[1] -0.3406924

$W.stat
[1] 1.58504

$CriticalValues
```

```
  alpha=.1  alpha=.05 alpha=.025  alpha=.01
    1.118     1.252     1.374     1.517

[1] 2
$B
    [,1]
[1,]    1

$d
[1] -0.4985455

$W.stat
[1] 0.9129707

$CriticalValues
  alpha=.1  alpha=.05 alpha=.025  alpha=.01
    1.118     1.252     1.374     1.517

[1] 3
$B
    [,1]
[1,]    1

$d
[1] -0.4999415

$W.stat
[1] 3.7631

$CriticalValues
  alpha=.1  alpha=.05 alpha=.025  alpha=.01
    1.118     1.252     1.374     1.517

[1] 4
$B
    [,1]
[1,]    1

$d
[1] -0.4999415

$W.stat
[1] 2.102016

$CriticalValues
  alpha=.1  alpha=.05 alpha=.025  alpha=.01
    1.118     1.252     1.374     1.517
```

```
[1] 5
$B
     [,1]
[1,]    1

$d
[1] -0.4719694

$W.stat
[1] 2.14884

$CriticalValues
  alpha=.1  alpha=.05 alpha=.025  alpha=.01
     1.118      1.252      1.374      1.517

[1] 6
$B
     [,1]
[1,]    1

$d
[1] -0.4999415

$W.stat
[1] 2.076575

$CriticalValues
  alpha=.1  alpha=.05 alpha=.025  alpha=.01
     1.118      1.252      1.374      1.517

[1] 7
$B
     [,1]
[1,]    1

$d
[1] 0.5687095

$W.stat
[1] 1.344365

$CriticalValues
  alpha=.1  alpha=.05 alpha=.025  alpha=.01
     1.118      1.252      1.374      1.517

[1] 8
$B
     [,1]
```

```
[1,]    1

$d
[1] 0.5135193

$W.stat
[1] 1.393347

$CriticalValues
  alpha=.1  alpha=.05 alpha=.025  alpha=.01
     1.118      1.252      1.374      1.517

#Seasonality
library(seastests); citation("seastests")
  Ollech D (2021). _seastests: Seasonality Tests_. R package version 0.15.4,
  <https://CRAN.R-project.org/package=seastests>.

i=0; for (i in 1:8) {
print(i);
print(isSeasonal(ts (Ucurve[,i], frequency=12, start=c(1976,5)),  test = "combined"));
}
#Webel, K. and Ollech, D. (2019). An overall seasonality test. Deutsche Bundesbank's Discussion Paper series.

[1] 1
[1] TRUE
[1] 2
[1] FALSE
[1] 3
[1] FALSE
[1] 4
[1] FALSE
[1] 5
[1] TRUE
[1] 6
[1] TRUE
[1] 7
[1] TRUE
[1] 8
[1] TRUE


#Nonlinearity tests
i=0; for (i in 1:8) { print (i) ; nonlinearityTest( diff (Ucurve[,i]), verbose = TRUE)}
[1] 1
        ** Teraesvirta's neural network test  **
        Null hypothesis: Linearity in "mean"
        X-squared =  1.90914  df =  2  p-value =  0.3849777
```

** White neural network test **
Null hypothesis: Linearity in "mean"
X-squared = 2.529153 df = 2 p-value = 0.2823588

** Keenan's one-degree test for nonlinearity **
Null hypothesis: The time series follows some AR process
F-stat = 4.682326 p-value = 0.03094262

** McLeod-Li test **
Null hypothesis: The time series follows some ARIMA process
Maximum p-value = 0.0003988775

** Tsay's Test for nonlinearity **
Null hypothesis: The time series follows some AR process
F-stat = 1.35379 p-value = 0.0207586

** Likelihood ratio test for threshold nonlinearity **
Null hypothesis: The time series follows some AR process
Alternativce hypothesis: The time series follows some TAR process
X-squared = 42.2632 p-value = 0.004565301

[1] 2

** Teraesvirta's neural network test **
Null hypothesis: Linearity in "mean"
X-squared = 40.78587 df = 2 p-value = 1.391427e-09

** White neural network test **
Null hypothesis: Linearity in "mean"
X-squared = 44.01912 df = 2 p-value = 2.762928e-10

** Keenan's one-degree test for nonlinearity **
Null hypothesis: The time series follows some AR process
F-stat = 17.4128 p-value = 3.535055e-05

** McLeod-Li test **
Null hypothesis: The time series follows some ARIMA process
Maximum p-value = 1.905055e-07

** Tsay's Test for nonlinearity **
Null hypothesis: The time series follows some AR process
F-stat = 1.511547 p-value = 0.008831497

** Likelihood ratio test for threshold nonlinearity **
Null hypothesis: The time series follows some AR process
Alternativce hypothesis: The time series follows some TAR process
X-squared = 34.85769 p-value = 0.00957431

[1] 3

** Teraesvirta's neural network test **
Null hypothesis: Linearity in "mean"
X-squared = 63.50306 df = 2 p-value = 1.620926e-14

** White neural network test **
Null hypothesis: Linearity in "mean"
X-squared = 55.54954 df = 2 p-value = 8.66085e-13

** Keenan's one-degree test for nonlinearity **
Null hypothesis: The time series follows some AR process
F-stat = 0.4898334 p-value = 0.4843265

** McLeod-Li test **
Null hypothesis: The time series follows some ARIMA process
Maximum p-value = 0

** Tsay's Test for nonlinearity **
Null hypothesis: The time series follows some AR process
F-stat = 2.486122 p-value = 5.397464e-12

** Likelihood ratio test for threshold nonlinearity **
Null hypothesis: The time series follows some AR process
Alternativce hypothesis: The time series follows some TAR process
X-squared = 41.56346 p-value = 0.0144255

[1] 4

** Teraesvirta's neural network test **
Null hypothesis: Linearity in "mean"
X-squared = 20.71877 df = 2 p-value = 3.1694e-05

** White neural network test **
Null hypothesis: Linearity in "mean"
X-squared = 16.5868 df = 2 p-value = 0.0002501625

** Keenan's one-degree test for nonlinearity **
Null hypothesis: The time series follows some AR process
F-stat = 1.948734 p-value = 0.1633491

** McLeod-Li test **
Null hypothesis: The time series follows some ARIMA process
Maximum p-value = 0

** Tsay's Test for nonlinearity **
Null hypothesis: The time series follows some AR process
F-stat = 2.554092 p-value = 1.263933e-13

** Likelihood ratio test for threshold nonlinearity **

Null hypothesis: The time series follows some AR process
Alternativce hypothesis: The time series follows some TAR process
X-squared = 47.4056 p-value = 0.01026376

[1] 5

** Teraesvirta's neural network test **
Null hypothesis: Linearity in "mean"
X-squared = 29.83743 df = 2 p-value = 3.318064e-07

** White neural network test **
Null hypothesis: Linearity in "mean"
X-squared = 26.88544 df = 2 p-value = 1.451782e-06

** Keenan's one-degree test for nonlinearity **
Null hypothesis: The time series follows some AR process
F-stat = 7.832092 p-value = 0.005326578

** McLeod-Li test **
Null hypothesis: The time series follows some ARIMA process
Maximum p-value = 8.643369e-08

** Tsay's Test for nonlinearity **
Null hypothesis: The time series follows some AR process
F-stat = 1.597208 p-value = 0.003461193

** Likelihood ratio test for threshold nonlinearity **
Null hypothesis: The time series follows some AR process
Alternativce hypothesis: The time series follows some TAR process
X-squared = 41.95389 p-value = 0.0009159348

[1] 6

** Teraesvirta's neural network test **
Null hypothesis: Linearity in "mean"
X-squared = 28.52837 df = 2 p-value = 6.384755e-07

** White neural network test **
Null hypothesis: Linearity in "mean"
X-squared = 22.77942 df = 2 p-value = 1.131126e-05

** Keenan's one-degree test for nonlinearity **
Null hypothesis: The time series follows some AR process
F-stat = 2.342567 p-value = 0.1265131

** McLeod-Li test **
Null hypothesis: The time series follows some ARIMA process
Maximum p-value = 0

** Tsay's Test for nonlinearity **

Null hypothesis: The time series follows some AR process
F-stat = 3.542435 p-value = 6.332389e-22

** Likelihood ratio test for threshold nonlinearity **
Null hypothesis: The time series follows some AR process
Alternativce hypothesis: The time series follows some TAR process
X-squared = 47.22592 p-value = 0.002742857

[1] 7

** Teraesvirta's neural network test **
Null hypothesis: Linearity in "mean"
X-squared = 59.30496 df = 2 p-value = 1.324496e-13

** White neural network test **
Null hypothesis: Linearity in "mean"
X-squared = 61.16617 df = 2 p-value = 5.218048e-14

** Keenan's one-degree test for nonlinearity **
Null hypothesis: The time series follows some AR process
F-stat = 2.307904 p-value = 0.1293384

** McLeod-Li test **
Null hypothesis: The time series follows some ARIMA process
Maximum p-value = 0

** Tsay's Test for nonlinearity **
Null hypothesis: The time series follows some AR process
F-stat = 1.41285 p-value = 0.01770818

** Likelihood ratio test for threshold nonlinearity **
Null hypothesis: The time series follows some AR process
Alternativce hypothesis: The time series follows some TAR process
X-squared = 24.75322 p-value = 0.1856432

[1] 8

** Teraesvirta's neural network test **
Null hypothesis: Linearity in "mean"
X-squared = 10.68095 df = 2 p-value = 0.004793589

** White neural network test **
Null hypothesis: Linearity in "mean"
X-squared = 5.957434 df = 2 p-value = 0.05085804

** Keenan's one-degree test for nonlinearity **
Null hypothesis: The time series follows some AR process
F-stat = 0.9434776 p-value = 0.3318483

** McLeod-Li test **

```
Null hypothesis: The time series follows some ARIMA process
Maximum p-value =  0

** Tsay's Test for nonlinearity **
Null hypothesis: The time series follows some AR process
F-stat =  4.374126  p-value =  3.136824e-25

** Likelihood ratio test for threshold nonlinearity **
Null hypothesis: The time series follows some AR process
Alternativce hypothesis: The time series follows some TAR process
X-squared =  36.40977  p-value =  0.01618581
```

```
#Determine correlation
 cor(Ucurve, method="spearman")
             MDReg    MedReg  MDGuides MedGuides       MPI MPIGuides    CumMPI CumMPIGuides
MDReg       1.0000000 0.4361785 0.2495336 0.3439078 0.7463243 0.3509028 0.5071342    0.5073940
MedReg      0.4361785 1.0000000 0.1762174 0.2121413 0.9061992 0.2198800 0.3082031    0.3081392
MDGuides    0.2495336 0.1762174 1.0000000 0.4901232 0.2545086 0.7471684 0.5068244    0.5077051
MedGuides   0.3439078 0.2121413 0.4901232 1.0000000 0.3123561 0.9175882 0.7289273    0.7295685
MPI         0.7463243 0.9061992 0.2545086 0.3123561 1.0000000 0.3266415 0.4473291    0.4474296
MPIGuides   0.3509028 0.2198800 0.7471684 0.9175882 0.3266415 1.0000000 0.7344358    0.7354322
CumMPI      0.5071342 0.3082031 0.5068244 0.7289273 0.4473291 0.7344358 1.0000000    0.9999789
CumMPIGuides 0.5073940 0.3081392 0.5077051 0.7295685 0.4474296 0.7354322 0.9999789
1.0000000

library(corrplot); citation("corrplot")
Taiyun Wei and Viliam Simko (2021). R package 'corrplot':
  Visualization of a Correlation Matrix (Version 0.92).
  Available from https://github.com/taiyun/corrplot

corrplot(cor(Ucurve, method="spearman"))

#Normalization
#Formula for Min-Max renormalization follows from https://www.geeksforgeeks.org/how-to-normalize-
and-standardize-data-in-r/
minMax <- function(x) {
  (x - min(x)) / (max(x) - min(x))
}

NormUcurve<- as.data.frame(lapply(Ucurve, minMax))

#Determine Causation
library(VLTimeCausality); citation("VLTimeCausality")

 Chainarong Amornbunchornvej, Elena Zheleva, and Tanya
  Berger-Wolf. 2021. Variable-lag Granger Causality and
  Transfer Entropy for Time Series Analysis. ACM
```

```
#Variable-Lag Transfer Entropy
#Note: The flag is true if X (VL-)Transfer-Entropy-causes Y using Transfer Entropy
#ratio where TEratio >1 if X causes Y.
#Additionally, if nboot>1, the flag is true only when pval<=alpha

i=0; for (i in 1:8) {
  j=0; for (j in 1:8) {
out=0;
out<-VLTransferEntropy(Ucurve[,j], Ucurve[,i], maxLag=1, nboot=500, VLflag=TRUE)
print( paste ( "TE ratio=", format(out$TEratio,
digits=4),names(Ucurve[i]),out$XgCsY_trns,names(Ucurve[j]), "p value", format(out$pval,digits=10) ))
  }
}


[1] "TE ratio= 5.444 MDReg TRUE MDReg p value 0"
[1] "TE ratio= 1.297 MDReg TRUE MedReg p value 0.004"
[1] "TE ratio= 0.7526 MDReg FALSE MDGuides p value 0.002"
[1] "TE ratio= 0.6022 MDReg FALSE MedGuides p value 0.032"
[1] "TE ratio= 1.978 MDReg TRUE MPI p value 0"
[1] "TE ratio= 0.741 MDReg FALSE MPIGuides p value 0.002"
[1] "TE ratio= 0.04626 MDReg FALSE CumMPI p value 0.992"
[1] "TE ratio= 0.01632 MDReg FALSE CumMPIGuides p value 1"

[1] "TE ratio= 0.5626 MedReg FALSE MDReg p value 0.002"
[1] "TE ratio= 23.14 MedReg TRUE MedReg p value 0"
[1] "TE ratio= NaN MedReg FALSE MDGuides p value 0"
[1] "TE ratio= 0.935 MedReg FALSE MedGuides p value 0.018"
[1] "TE ratio= 3.809 MedReg TRUE MPI p value 0"
[1] "TE ratio= 2.166 MedReg TRUE MPIGuides p value 0"
[1] "TE ratio= 0.3122 MedReg FALSE CumMPI p value 0.18"
[1] "TE ratio= 0.259 MedReg FALSE CumMPIGuides p value 0.284"

[1] "TE ratio= 2.336 MDGuides TRUE MDReg p value 0"
[1] "TE ratio= 2.669 MDGuides FALSE MedReg p value 0.054"
[1] "TE ratio= 19.64 MDGuides TRUE MDGuides p value 0"
[1] "TE ratio= 1.417 MDGuides TRUE MedGuides p value 0"
[1] "TE ratio= 2.866 MDGuides TRUE MPI p value 0.012"
[1] "TE ratio= 4.517 MDGuides TRUE MPIGuides p value 0"
[1] "TE ratio= 0.05025 MDGuides FALSE CumMPI p value 0.998"
[1] "TE ratio= 0.04843 MDGuides FALSE CumMPIGuides p value 1"

[1] "TE ratio= 1.436 MedGuides TRUE MDReg p value 0"
[1] "TE ratio= NaN MedGuides FALSE MedReg p value 0"
[1] "TE ratio= 1.936 MedGuides TRUE MDGuides p value 0"
```

```
[1] "TE ratio= 9.322 MedGuides TRUE MedGuides p value 0"
[1] "TE ratio= 1.252 MedGuides TRUE MPI p value 0.042"
[1] "TE ratio= 4.362 MedGuides TRUE MPIGuides p value 0"
[1] "TE ratio= 0.1191 MedGuides FALSE CumMPI p value 0.53"
[1] "TE ratio= 0.2228 MedGuides FALSE CumMPIGuides p value 0.148"

[1] "TE ratio= 1.327 MPI TRUE MDReg p value 0"
[1] "TE ratio= 4.987 MPI TRUE MedReg p value 0"
[1] "TE ratio= NaN MPI FALSE MDGuides p value 0"
[1] "TE ratio= 0.6051 MPI FALSE MedGuides p value 0.026"
[1] "TE ratio= 5.992 MPI TRUE MPI p value 0"
[1] "TE ratio= NaN MPI FALSE MPIGuides p value 0"
[1] "TE ratio= 0.03936 MPI FALSE CumMPI p value 0.986"
[1] "TE ratio= 0.03403 MPI FALSE CumMPIGuides p value 0.992"

[1] "TE ratio= 1.313 MPIGuides TRUE MDReg p value 0"
[1] "TE ratio= NaN MPIGuides FALSE MedReg p value 0"
[1] "TE ratio= 7.454 MPIGuides TRUE MDGuides p value 0"
[1] "TE ratio= 4.308 MPIGuides TRUE MedGuides p value 0"
[1] "TE ratio= 1.127 MPIGuides TRUE MPI p value 0.004"
[1] "TE ratio= 7.189 MPIGuides TRUE MPIGuides p value 0"
[1] "TE ratio= 0.1173 MPIGuides FALSE CumMPI p value 0.428"
[1] "TE ratio= 0.1452 MPIGuides FALSE CumMPIGuides p value 0.298"

[1] "TE ratio= 38.22 CumMPI TRUE MDReg p value 0"
[1] "TE ratio= NaN CumMPI FALSE MedReg p value 0"
[1] "TE ratio= 8.582 CumMPI TRUE MDGuides p value 0"
[1] "TE ratio= 35.64 CumMPI TRUE MedGuides p value 0"
[1] "TE ratio= 16.81 CumMPI TRUE MPI p value 0"
[1] "TE ratio= 35.54 CumMPI TRUE MPIGuides p value 0"
[1] "TE ratio= 282.5 CumMPI TRUE CumMPI p value 0"
[1] "TE ratio= 215 CumMPI TRUE CumMPIGuides p value 0"

[1] "TE ratio= 18.67 CumMPIGuides TRUE MDReg p value 0"
[1] "TE ratio= NaN CumMPIGuides FALSE MedReg p value 0"
[1] "TE ratio= 8.76 CumMPIGuides TRUE MDGuides p value 0"
[1] "TE ratio= 36.58 CumMPIGuides TRUE MedGuides p value 0"
[1] "TE ratio= 20.84 CumMPIGuides TRUE MPI p value 0"
[1] "TE ratio= 36.84 CumMPIGuides TRUE MPIGuides p value 0"
[1] "TE ratio= 1.649 CumMPIGuides TRUE CumMPI p value 0.004"
[1] "TE ratio= 284.1 CumMPIGuides TRUE CumMPIGuides p value 0"
```

```
library(Hmisc);citation("Hmisc")
   Harrell Jr F (2022). _Hmisc: Harrell Miscellaneous_. R package version
  4.7-1, <https://CRAN.R-project.org/package=Hmisc>.
```

```
library(tseries);citation("tseries")
 Adrian Trapletti and Kurt Hornik (2022). tseries: Time Series Analysis
```

Figure 1:
plot(ts (Ucurve$MDReg, frequency=12, start=c(1976,05)), ylab="Monthly FDA-Registered Medical Devices" );minor.tick(10)
plot(ts (Ucurve$MedReg, frequency=12, start=c(1976,05)), ylab="Monthly FDA-Registered Medicines" );minor.tick(10)

Figure 2:
plot(ts (Ucurve$MDGuides, frequency=12, start=c(1976,05)), ylab="Monthly FDA-Issued Medical Device Guidelines" );minor.tick(10)
plot(ts (Ucurve$MedGuides, frequency=12, start=c(1976,05)), ylab="Monthly FDA-Issued Medicines Guidelines" );minor.tick(10)

Figure 3:
plot(ts (Ucurve$MPI, frequency=12, start=c(1976,05)), ylab="Monthly FDA-Registered Medical Products" );minor.tick(10)
plot(ts (Ucurve$MPIGuides, frequency=12, start=c(1976,05)), ylab="Monthly FDA-Issued Medical Products Guidelines" );minor.tick(10)

Figure 4:
plot(ts (Ucurve$CumMPI, frequency=12, start=c(1976,05)), ylab="Cumulative FDA-Registered Medical Products" );minor.tick(10)
plot(ts (Ucurve$CumMPIGuides, frequency=12, start=c(1976,05)), ylab="Cumulative FDA-Issued Medical Products Guidelines" );minor.tick(10)

library(regr.easy); citation("regr.easy")
Martins dos Santos W (2022). _regr.easy: Easy Linear, Quadratic and Cubic Regression Models_. R package version 1.0.1,
<https://CRAN.R-project.org/package=regr.easy>.

regr_easy_calc(CumMPIGuides, CumMPI, model="all")

```
Observed and estimated - Linear Model
==========================================
     x     y      y_est      Resid
-----------------------------------------
1    0    255   73,288.780  -73,033.780
2    0    442   73,288.780  -72,846.780
3    0    721   73,288.780  -72,567.780
4    2  1,016  73,684.630  -72,668.630
5    2  1,380  73,684.630  -72,304.630
6    4  1,673  74,080.490  -72,407.490
7    5  2,068  74,278.410  -72,210.410
8    5  2,449  74,278.410  -71,829.410
9    6  2,769  74,476.340  -71,707.340
```

| | | | | |
|---|---|---|---|---|
| 10 | 6 | 3,090 | 74,476.340 | -71,386.340 |
| 11 | 7 | 3,351 | 74,674.270 | -71,323.270 |
| 12 | 7 | 3,673 | 74,674.270 | -71,001.270 |
| 13 | 7 | 3,937 | 74,674.270 | -70,737.270 |
| 14 | 7 | 4,299 | 74,674.270 | -70,375.270 |
| 15 | 7 | 4,629 | 74,674.270 | -70,045.270 |
| 16 | 7 | 5,004 | 74,674.270 | -69,670.270 |
| 17 | 9 | 5,333 | 75,070.120 | -69,737.120 |
| 18 | 9 | 5,632 | 75,070.120 | -69,438.120 |
| 19 | 10 | 5,915 | 75,268.050 | -69,353.050 |
| 20 | 10 | 6,211 | 75,268.050 | -69,057.050 |
| 21 | 10 | 6,580 | 75,268.050 | -68,688.050 |
| 22 | 10 | 6,928 | 75,268.050 | -68,340.050 |
| 23 | 10 | 7,293 | 75,268.050 | -67,975.050 |
| 24 | 10 | 7,621 | 75,268.050 | -67,647.050 |
| 25 | 11 | 7,912 | 75,465.980 | -67,553.980 |
| 26 | 11 | 8,287 | 75,465.980 | -67,178.980 |
| 27 | 11 | 8,652 | 75,465.980 | -66,813.980 |
| 28 | 11 | 9,086 | 75,465.980 | -66,379.980 |
| 29 | 11 | 9,409 | 75,465.980 | -66,056.980 |
| 30 | 12 | 9,815 | 75,663.900 | -65,848.900 |
| 31 | 13 | 10,167 | 75,861.830 | -65,694.830 |
| 32 | 13 | 10,491 | 75,861.830 | -65,370.830 |
| 33 | 13 | 10,788 | 75,861.830 | -65,073.830 |
| 34 | 13 | 11,226 | 75,861.830 | -64,635.830 |
| 35 | 13 | 11,533 | 75,861.830 | -64,328.830 |
| 36 | 13 | 11,925 | 75,861.830 | -63,936.830 |
| 37 | 13 | 12,272 | 75,861.830 | -63,589.830 |
| 38 | 13 | 12,660 | 75,861.830 | -63,201.830 |
| 39 | 13 | 13,015 | 75,861.830 | -62,846.830 |
| 40 | 16 | 13,491 | 76,455.610 | -62,964.610 |
| 41 | 17 | 13,930 | 76,653.540 | -62,723.540 |
| 42 | 17 | 14,507 | 76,653.540 | -62,146.540 |
| 43 | 17 | 14,896 | 76,653.540 | -61,757.540 |
| 44 | 17 | 15,356 | 76,653.540 | -61,297.540 |
| 45 | 17 | 15,883 | 76,653.540 | -60,770.540 |
| 46 | 17 | 16,375 | 76,653.540 | -60,278.540 |
| 47 | 18 | 16,721 | 76,851.460 | -60,130.460 |
| 48 | 18 | 17,146 | 76,851.460 | -59,705.460 |
| 49 | 19 | 17,747 | 77,049.390 | -59,302.390 |
| 50 | 19 | 18,419 | 77,049.390 | -58,630.390 |
| 51 | 19 | 18,903 | 77,049.390 | -58,146.390 |
| 52 | 20 | 19,469 | 77,247.320 | -57,778.320 |
| 53 | 27 | 19,918 | 78,632.810 | -58,714.810 |
| 54 | 36 | 20,408 | 80,414.150 | -60,006.150 |
| 55 | 36 | 20,727 | 80,414.150 | -59,687.150 |
| 56 | 36 | 21,248 | 80,414.150 | -59,166.150 |
| 57 | 37 | 21,892 | 80,612.070 | -58,720.070 |
| 58 | 37 | 22,306 | 80,612.070 | -58,306.070 |
| 59 | 37 | 22,884 | 80,612.070 | -57,728.070 |
| 60 | 37 | 23,442 | 80,612.070 | -57,170.070 |
| 61 | 38 | 23,969 | 80,810.000 | -56,841.000 |
| 62 | 38 | 24,562 | 80,810.000 | -56,248.000 |
| 63 | 39 | 25,096 | 81,007.930 | -55,911.930 |
| 64 | 39 | 25,584 | 81,007.930 | -55,423.930 |
| 65 | 40 | 26,102 | 81,205.850 | -55,103.860 |
| 66 | 40 | 26,651 | 81,205.850 | -54,554.860 |
| 67 | 40 | 27,242 | 81,205.850 | -53,963.860 |
| 68 | 40 | 27,833 | 81,205.850 | -53,372.860 |
| 69 | 40 | 28,259 | 81,205.850 | -52,946.860 |
| 70 | 40 | 28,807 | 81,205.850 | -52,398.860 |
| 71 | 40 | 29,542 | 81,205.850 | -51,663.860 |
| 72 | 41 | 30,262 | 81,403.780 | -51,141.780 |
| 73 | 42 | 30,918 | 81,601.710 | -50,683.710 |
| 74 | 42 | 31,799 | 81,601.710 | -49,802.710 |
| 75 | 42 | 32,639 | 81,601.710 | -48,962.710 |
| 76 | 42 | 33,170 | 81,601.710 | -48,431.710 |
| 77 | 42 | 33,725 | 81,601.710 | -47,876.710 |
| 78 | 42 | 34,147 | 81,601.710 | -47,454.710 |
| 79 | 42 | 34,683 | 81,601.710 | -46,918.710 |
| 80 | 42 | 35,311 | 81,601.710 | -46,290.710 |
| 81 | 42 | 35,913 | 81,601.710 | -45,688.710 |
| 82 | 42 | 36,324 | 81,601.710 | -45,277.710 |
| 83 | 43 | 36,955 | 81,799.640 | -44,844.640 |
| 84 | 43 | 37,463 | 81,799.640 | -44,336.640 |
| 85 | 43 | 37,896 | 81,799.640 | -43,903.640 |
| 86 | 43 | 38,544 | 81,799.640 | -43,255.640 |
| 87 | 44 | 39,187 | 81,997.560 | -42,810.560 |
| 88 | 44 | 39,722 | 81,997.560 | -42,275.560 |
| 89 | 44 | 40,236 | 81,997.560 | -41,761.560 |
| 90 | 44 | 40,790 | 81,997.560 | -41,207.560 |
| 91 | 44 | 41,352 | 81,997.560 | -40,645.560 |
| 92 | 44 | 41,894 | 81,997.560 | -40,103.560 |
| 93 | 45 | 42,593 | 82,195.490 | -39,602.490 |
| 94 | 45 | 43,280 | 82,195.490 | -38,915.490 |
| 95 | 45 | 44,205 | 82,195.490 | -37,990.490 |
| 96 | 45 | 44,920 | 82,195.490 | -37,275.490 |
| 97 | 47 | 45,589 | 82,591.340 | -37,002.340 |
| 98 | 47 | 46,187 | 82,591.340 | -36,404.340 |
| 99 | 47 | 46,754 | 82,591.340 | -35,837.340 |
| 100 | 47 | 47,470 | 82,591.340 | -35,121.340 |
| 101 | 47 | 48,159 | 82,591.340 | -34,432.340 |
| 102 | 47 | 49,036 | 82,591.340 | -33,555.340 |
| 103 | 47 | 49,629 | 82,591.340 | -32,962.340 |
| 104 | 48 | 50,100 | 82,789.270 | -32,689.270 |
| 105 | 49 | 50,842 | 82,987.200 | -32,145.200 |
| 106 | 50 | 51,387 | 83,185.120 | -31,798.120 |
| 107 | 51 | 52,130 | 83,383.050 | -31,253.050 |
| 108 | 51 | 53,067 | 83,383.050 | -30,316.050 |
| 109 | 51 | 53,976 | 83,383.050 | -29,407.050 |
| 110 | 51 | 55,003 | 83,383.050 | -28,380.050 |
| 111 | 51 | 55,931 | 83,383.050 | -27,452.050 |
| 112 | 51 | 56,847 | 83,383.050 | -26,536.050 |
| 113 | 51 | 57,584 | 83,383.050 | -25,799.050 |
| 114 | 52 | 58,455 | 83,580.980 | -25,125.980 |
| 115 | 53 | 59,176 | 83,778.900 | -24,602.900 |
| 116 | 53 | 59,857 | 83,778.900 | -23,921.900 |
| 117 | 53 | 60,724 | 83,778.900 | -23,054.900 |
| 118 | 53 | 61,581 | 83,778.900 | -22,197.900 |
| 119 | 53 | 62,297 | 83,778.900 | -21,481.900 |
| 120 | 53 | 63,025 | 83,778.900 | -20,753.900 |
| 121 | 53 | 63,793 | 83,778.900 | -19,985.900 |
| 122 | 53 | 64,514 | 83,778.900 | -19,264.900 |
| 123 | 53 | 65,285 | 83,778.900 | -18,493.900 |
| 124 | 53 | 66,080 | 83,778.900 | -17,698.900 |
| 125 | 54 | 66,828 | 83,976.830 | -17,148.830 |
| 126 | 54 | 67,580 | 83,976.830 | -16,396.830 |
| 127 | 54 | 68,277 | 83,976.830 | -15,699.830 |
| 128 | 54 | 68,906 | 83,976.830 | -15,070.830 |
| 129 | 54 | 69,494 | 83,976.830 | -14,482.830 |
| 130 | 59 | 70,221 | 84,966.470 | -14,745.470 |
| 131 | 59 | 71,123 | 84,966.470 | -13,843.470 |
| 132 | 59 | 71,900 | 84,966.470 | -13,066.470 |
| 133 | 64 | 72,645 | 85,956.100 | -13,311.100 |
| 134 | 66 | 73,363 | 86,351.950 | -12,988.950 |
| 135 | 66 | 74,272 | 86,351.950 | -12,079.950 |
| 136 | 67 | 74,953 | 86,549.880 | -11,596.880 |
| 137 | 80 | 75,601 | 89,122.930 | -13,521.930 |
| 138 | 82 | 76,295 | 89,518.780 | -13,223.780 |
| 139 | 82 | 76,923 | 89,518.780 | -12,595.780 |
| 140 | 83 | 77,592 | 89,716.710 | -12,124.710 |
| 141 | 83 | 78,213 | 89,716.710 | -11,503.710 |
| 142 | 83 | 79,103 | 89,716.710 | -10,613.710 |
| 143 | 86 | 80,130 | 90,310.490 | -10,180.490 |
| 144 | 87 | 81,132 | 90,508.420 | -9,376.419 |
| 145 | 87 | 81,914 | 90,508.420 | -8,594.419 |
| 146 | 87 | 82,790 | 90,508.420 | -7,718.419 |
| 147 | 89 | 83,553 | 90,904.270 | -7,351.272 |
| 148 | 90 | 84,538 | 91,102.200 | -6,564.199 |
| 149 | 90 | 85,238 | 91,102.200 | -5,864.199 |
| 150 | 90 | 85,968 | 91,102.200 | -5,134.199 |
| 151 | 90 | 86,809 | 91,102.200 | -4,293.199 |
| 152 | 90 | 87,643 | 91,102.200 | -3,459.199 |
| 153 | 90 | 88,476 | 91,102.200 | -2,626.199 |
| 154 | 93 | 89,341 | 91,695.980 | -2,354.980 |
| 155 | 96 | 90,252 | 92,289.760 | -2,037.761 |
| 156 | 96 | 91,118 | 92,289.760 | -1,171.761 |
| 157 | 96 | 92,127 | 92,289.760 | -162.761 |
| 158 | 97 | 93,064 | 92,487.690 | 576.313 |
| 159 | 97 | 93,977 | 92,487.690 | 1,489.313 |
| 160 | 109 | 94,856 | 94,862.810 | -6.810 |
| 161 | 110 | 95,635 | 95,060.740 | 574.263 |
| 162 | 110 | 96,516 | 95,060.740 | 1,455.263 |
| 163 | 113 | 97,188 | 95,654.520 | 1,533.482 |
| 164 | 113 | 97,833 | 95,654.520 | 2,178.482 |
| 165 | 115 | 98,765 | 96,050.370 | 2,714.629 |
| 166 | 115 | 99,537 | 96,050.370 | 3,486.629 |
| 167 | 115 | 100,299 | 96,050.370 | 4,248.629 |
| 168 | 115 | 101,063 | 96,050.370 | 5,012.629 |
| 169 | 115 | 101,718 | 96,050.370 | 5,667.629 |
| 170 | 116 | 102,411 | 96,248.300 | 6,162.702 |
| 171 | 116 | 103,081 | 96,248.300 | 6,832.702 |
| 172 | 116 | 103,794 | 96,248.300 | 7,545.702 |
| 173 | 116 | 104,672 | 96,248.300 | 8,423.702 |
| 174 | 116 | 105,368 | 96,248.300 | 9,119.702 |
| 175 | 116 | 106,007 | 96,248.300 | 9,758.702 |
| 176 | 116 | 106,672 | 96,248.300 | 10,423.700 |
| 177 | 117 | 107,400 | 96,446.230 | 10,953.770 |
| 178 | 118 | 108,236 | 96,644.150 | 11,591.850 |
| 179 | 119 | 109,157 | 96,842.080 | 12,314.920 |
| 180 | 120 | 109,991 | 97,040.010 | 12,950.990 |
| 181 | 121 | 110,699 | 97,237.930 | 13,461.070 |
| 182 | 123 | 111,293 | 97,633.790 | 13,659.210 |
| 183 | 123 | 111,982 | 97,633.790 | 14,348.210 |
| 184 | 123 | 112,877 | 97,633.790 | 15,243.210 |
| 185 | 124 | 113,680 | 97,831.710 | 15,848.290 |
| 186 | 124 | 114,404 | 97,831.710 | 16,572.290 |
| 187 | 124 | 115,044 | 97,831.710 | 17,212.290 |
| 188 | 124 | 115,832 | 97,831.710 | 18,000.290 |
| 189 | 124 | 116,640 | 97,831.710 | 18,808.290 |
| 190 | 124 | 117,335 | 97,831.710 | 19,503.290 |
| 191 | 125 | 118,014 | 98,029.640 | 19,984.360 |
| 192 | 126 | 118,972 | 98,227.570 | 20,744.430 |
| 193 | 130 | 119,627 | 99,019.270 | 20,607.730 |
| 194 | 131 | 120,232 | 99,217.200 | 21,014.800 |
| 195 | 131 | 120,860 | 99,217.200 | 21,642.800 |
| 196 | 131 | 121,492 | 99,217.200 | 22,274.800 |
| 197 | 131 | 122,269 | 99,217.200 | 23,051.800 |
| 198 | 131 | 122,838 | 99,217.200 | 23,620.800 |
| 199 | 133 | 123,467 | 99,613.050 | 23,853.940 |
| 200 | 133 | 124,221 | 99,613.050 | 24,607.940 |
| 201 | 133 | 124,738 | 99,613.050 | 25,124.940 |
| 202 | 136 | 125,349 | 100,206.800 | 25,142.160 |
| 203 | 139 | 126,156 | 100,800.600 | 25,355.380 |
| 204 | 139 | 126,901 | 100,800.600 | 26,100.380 |
| 205 | 140 | 127,588 | 100,998.500 | 26,589.460 |
| 206 | 141 | 128,340 | 101,196.500 | 27,143.530 |
| 207 | 143 | 129,247 | 101,592.300 | 27,654.680 |
| 208 | 146 | 129,996 | 102,186.100 | 27,809.900 |
| 209 | 147 | 130,616 | 102,384.000 | 28,231.970 |
| 210 | 149 | 131,342 | 102,779.900 | 28,562.120 |
| 211 | 149 | 131,996 | 102,779.900 | 29,216.120 |
| 212 | 149 | 132,731 | 102,779.900 | 29,951.120 |
| 213 | 149 | 133,722 | 102,779.900 | 30,942.120 |
| 214 | 151 | 134,736 | 103,175.700 | 31,560.260 |
| 215 | 151 | 135,807 | 103,175.700 | 32,631.260 |
| 216 | 152 | 136,590 | 103,373.700 | 33,216.330 |
| 217 | 152 | 137,434 | 103,373.700 | 34,060.330 |
| 218 | 153 | 138,272 | 103,571.600 | 34,700.410 |
| 219 | 154 | 139,087 | 103,769.500 | 35,317.480 |
| 220 | 157 | 139,868 | 104,363.300 | 35,504.700 |
| 221 | 161 | 140,811 | 105,155.000 | 35,655.990 |
| 222 | 161 | 141,482 | 105,155.000 | 36,326.990 |
| 223 | 164 | 142,219 | 105,748.800 | 36,470.210 |
| 224 | 164 | 143,256 | 105,748.800 | 37,507.210 |
| 225 | 165 | 144,038 | 105,946.700 | 38,091.290 |
| 226 | 180 | 144,963 | 108,915.600 | 36,047.380 |
| 227 | 208 | 145,900 | 114,457.600 | 31,442.430 |
| 228 | 209 | 146,597 | 114,655.500 | 31,941.500 |
| 229 | 210 | 147,567 | 114,853.400 | 32,713.580 |
| 230 | 215 | 148,465 | 115,843.100 | 32,621.940 |
| 231 | 220 | 149,213 | 116,832.700 | 32,380.310 |
| 232 | 222 | 150,242 | 117,228.500 | 33,013.450 |
| 233 | 223 | 151,044 | 117,426.500 | 33,617.530 |
| 234 | 223 | 151,954 | 117,426.500 | 34,527.530 |
| 235 | 226 | 152,727 | 118,020.300 | 34,706.750 |
| 236 | 228 | 153,449 | 118,416.100 | 35,032.890 |
| 237 | 228 | 154,084 | 118,416.100 | 35,667.890 |
| 238 | 230 | 154,900 | 118,812.000 | 36,088.040 |
| 239 | 239 | 155,972 | 120,593.300 | 35,378.700 |
| 240 | 242 | 156,905 | 121,187.100 | 35,717.920 |
| 241 | 244 | 157,767 | 121,582.900 | 36,184.060 |
| 242 | 245 | 158,474 | 121,780.900 | 36,693.140 |
| 243 | 253 | 159,337 | 123,364.300 | 35,972.720 |
| 244 | 259 | 160,570 | 124,551.800 | 36,018.160 |
| 245 | 261 | 161,277 | 124,947.700 | 36,329.310 |
| 246 | 263 | 162,127 | 125,343.500 | 36,783.450 |
| 247 | 263 | 162,837 | 125,343.500 | 37,493.450 |
| 248 | 266 | 163,589 | 125,937.300 | 37,651.670 |
| 249 | 269 | 164,385 | 126,531.100 | 37,853.890 |
| 250 | 275 | 165,089 | 127,718.700 | 37,370.330 |
| 251 | 277 | 165,881 | 128,114.500 | 37,766.470 |
| 252 | 278 | 166,846 | 128,312.500 | 38,533.550 |
| 253 | 284 | 167,636 | 129,500.000 | 38,135.990 |
| 254 | 285 | 168,508 | 129,697.900 | 38,810.060 |
| 255 | 289 | 169,409 | 130,489.600 | 38,919.350 |
| 256 | 292 | 170,240 | 131,083.400 | 39,156.570 |
| 257 | 293 | 170,955 | 131,281.400 | 39,673.640 |
| 258 | 294 | 171,745 | 131,479.300 | 40,265.720 |
| 259 | 299 | 172,498 | 132,468.900 | 40,029.080 |
| 260 | 303 | 173,328 | 133,260.600 | 40,067.380 |
| 261 | 317 | 174,007 | 136,031.600 | 37,975.400 |
| 262 | 321 | 174,952 | 136,823.300 | 38,128.690 |
| 263 | 338 | 175,722 | 140,188.100 | 35,533.930 |
| 264 | 341 | 176,484 | 140,781.800 | 35,702.150 |
| 265 | 346 | 177,256 | 141,771.500 | 35,484.520 |
| 266 | 350 | 177,965 | 142,563.200 | 35,401.810 |
| 267 | 356 | 178,662 | 143,750.700 | 34,911.250 |
| 268 | 360 | 179,467 | 144,542.500 | 34,924.540 |
| 269 | 368 | 180,351 | 146,125.900 | 34,225.130 |
| 270 | 373 | 181,156 | 147,115.500 | 34,040.490 |
| 271 | 382 | 181,971 | 148,896.800 | 33,074.150 |
| 272 | 386 | 183,059 | 149,688.600 | 33,370.440 |
| 273 | 390 | 183,623 | 150,480.300 | 33,142.740 |
| 274 | 393 | 184,571 | 151,074.000 | 33,496.960 |
| 275 | 398 | 185,518 | 152,063.700 | 33,454.320 |
| 276 | 399 | 186,258 | 152,261.600 | 33,996.390 |
| 277 | 403 | 187,141 | 153,053.300 | 34,087.690 |
| 278 | 406 | 187,879 | 153,647.100 | 34,231.910 |
| 279 | 408 | 188,557 | 154,042.900 | 34,514.050 |
| 280 | 413 | 189,297 | 155,032.600 | 34,264.420 |
| 281 | 415 | 190,021 | 155,428.400 | 34,592.570 |
| 282 | 415 | 190,716 | 155,428.400 | 35,287.570 |
| 283 | 418 | 191,711 | 156,022.200 | 35,688.790 |
| 284 | 420 | 192,531 | 156,418.100 | 36,112.930 |
| 285 | 425 | 193,257 | 157,407.700 | 35,849.300 |
| 286 | 427 | 194,106 | 157,803.600 | 36,302.440 |
| 287 | 429 | 195,012 | 158,199.400 | 36,812.590 |
| 288 | 433 | 196,021 | 158,991.100 | 37,029.880 |
| 289 | 433 | 196,940 | 158,991.100 | 37,948.880 |
| 290 | 434 | 197,679 | 159,189.000 | 38,489.960 |
| 291 | 436 | 198,479 | 159,584.900 | 38,894.100 |
| 292 | 440 | 199,272 | 160,376.600 | 38,895.390 |
| 293 | 440 | 199,944 | 160,376.600 | 39,567.390 |
| 294 | 444 | 200,748 | 161,168.300 | 39,579.690 |
| 295 | 446 | 201,470 | 161,564.200 | 39,905.830 |
| 296 | 452 | 202,260 | 162,751.700 | 39,508.270 |
| 297 | 455 | 202,902 | 163,345.500 | 39,556.490 |
| 298 | 459 | 203,643 | 164,137.200 | 39,505.780 |
| 299 | 464 | 204,594 | 165,126.900 | 39,467.150 |
| 300 | 467 | 205,532 | 165,720.600 | 39,811.370 |
| 301 | 474 | 206,369 | 167,106.100 | 39,262.880 |
| 302 | 475 | 207,149 | 167,304.000 | 39,844.950 |
| 303 | 484 | 207,857 | 169,085.400 | 38,771.610 |
| 304 | 489 | 208,976 | 170,075.000 | 38,900.980 |
| 305 | 492 | 209,686 | 170,668.800 | 39,017.200 |
| 306 | 495 | 210,527 | 171,262.600 | 39,264.420 |
| 307 | 499 | 211,348 | 172,054.300 | 39,293.710 |
| 308 | 500 | 212,036 | 172,252.200 | 39,783.780 |
| 309 | 503 | 212,866 | 172,846.000 | 40,020.000 |
| 310 | 503 | 213,676 | 172,846.000 | 40,830.000 |
| 311 | 504 | 214,647 | 173,043.900 | 41,603.070 |
| 312 | 506 | 215,525 | 173,439.800 | 42,085.220 |
| 313 | 509 | 216,389 | 174,033.600 | 42,355.440 |
| 314 | 510 | 217,146 | 174,231.500 | 42,914.510 |
| 315 | 512 | 218,208 | 174,627.300 | 43,580.660 |
| 316 | 514 | 219,085 | 175,023.200 | 44,061.810 |
| 317 | 514 | 219,869 | 175,023.200 | 44,845.810 |
| 318 | 516 | 220,826 | 175,419.000 | 45,406.950 |
| 319 | 519 | 221,777 | 176,012.800 | 45,764.170 |
| 320 | 523 | 222,840 | 176,804.500 | 46,035.460 |
| 321 | 526 | 223,444 | 177,398.300 | 46,045.680 |
| 322 | 530 | 223,904 | 178,190.000 | 45,713.970 |
| 323 | 531 | 224,452 | 178,388.000 | 46,064.050 |
| 324 | 536 | 224,972 | 179,377.600 | 45,594.410 |
| 325 | 541 | 225,551 | 180,367.200 | 45,183.780 |
| 326 | 543 | 226,084 | 180,763.100 | 45,320.930 |
| 327 | 547 | 226,584 | 181,554.800 | 45,029.220 |
| 328 | 547 | 227,117 | 181,554.800 | 45,562.220 |
| 329 | 553 | 227,628 | 182,742.300 | 44,885.660 |
| 330 | 553 | 228,260 | 182,742.300 | 45,517.660 |
| 331 | 556 | 228,761 | 183,336.100 | 45,424.880 |
| 332 | 558 | 229,297 | 183,732.000 | 45,565.020 |
| 333 | 562 | 229,730 | 184,523.700 | 45,206.310 |
| 334 | 563 | 230,223 | 184,721.600 | 45,501.390 |
| 335 | 566 | 230,885 | 185,315.400 | 45,569.610 |
| 336 | 567 | 231,450 | 185,513.300 | 45,936.680 |
| 337 | 571 | 232,010 | 186,305.000 | 45,704.970 |
| 338 | 582 | 232,600 | 188,482.200 | 44,117.780 |
| 339 | 585 | 233,170 | 189,076.000 | 44,094.000 |
| 340 | 588 | 233,643 | 189,669.800 | 43,973.220 |
| 341 | 590 | 234,261 | 190,065.600 | 44,195.360 |
| 342 | 596 | 234,798 | 191,253.200 | 43,544.800 |
| 343 | 601 | 235,319 | 192,242.800 | 43,076.170 |
| 344 | 606 | 235,866 | 193,232.500 | 42,633.530 |
| 345 | 610 | 236,252 | 194,024.200 | 42,227.820 |
| 346 | 616 | 236,705 | 195,211.700 | 41,493.260 |
| 347 | 631 | 237,214 | 198,180.600 | 39,033.360 |
| 348 | 640 | 237,743 | 199,962.000 | 37,781.020 |
| 349 | 643 | 238,296 | 200,555.800 | 37,740.240 |
| 350 | 646 | 238,868 | 201,149.500 | 37,718.460 |
| 351 | 648 | 239,404 | 201,545.400 | 37,858.600 |
| 352 | 649 | 240,115 | 201,743.300 | 38,371.680 |
| 353 | 653 | 240,656 | 202,535.000 | 38,120.970 |
| 354 | 656 | 241,112 | 203,128.800 | 37,983.190 |
| 355 | 658 | 241,611 | 203,524.700 | 38,086.330 |
| 356 | 658 | 242,180 | 203,524.700 | 38,655.330 |
| 357 | 671 | 242,627 | 206,097.700 | 36,529.290 |
| 358 | 680 | 243,093 | 207,879.100 | 35,213.940 |
| 359 | 691 | 243,755 | 210,056.300 | 33,698.750 |
| 360 | 695 | 244,267 | 210,848.000 | 33,419.040 |
| 361 | 697 | 244,813 | 211,243.800 | 33,569.190 |
| 362 | 705 | 245,362 | 212,827.200 | 32,534.770 |
| 363 | 707 | 245,948 | 213,223.100 | 32,724.920 |
| 364 | 710 | 246,530 | 213,816.900 | 32,713.140 |
| 365 | 714 | 247,067 | 214,608.600 | 32,458.430 |
| 366 | 721 | 247,633 | 215,994.100 | 31,638.940 |
| 367 | 723 | 248,120 | 216,389.900 | 31,730.090 |
| 368 | 724 | 248,702 | 216,587.800 | 32,114.160 |
| 369 | 724 | 249,205 | 216,587.800 | 32,617.160 |
| 370 | 730 | 249,735 | 217,775.400 | 31,959.600 |
| 371 | 733 | 250,321 | 218,369.200 | 31,951.820 |
| 372 | 733 | 250,866 | 218,369.200 | 32,496.820 |
| 373 | 740 | 251,467 | 219,754.700 | 31,712.330 |
| 374 | 747 | 252,068 | 221,140.200 | 30,927.840 |
| 375 | 752 | 252,577 | 222,129.800 | 30,447.210 |
| 376 | 755 | 253,212 | 222,723.600 | 30,488.430 |
| 377 | 762 | 253,834 | 224,109.100 | 29,724.940 |
| 378 | 766 | 254,514 | 224,900.800 | 29,613.230 |
| 379 | 768 | 255,112 | 225,296.600 | 29,815.380 |
| 380 | 769 | 255,964 | 225,494.500 | 30,469.450 |
| 381 | 771 | 256,538 | 225,890.400 | 30,647.600 |
| 382 | 779 | 257,172 | 227,473.800 | 29,698.180 |
| 383 | 784 | 257,758 | 228,463.500 | 29,294.550 |
| 384 | 791 | 258,389 | 229,848.900 | 28,540.060 |
| 385 | 791 | 259,074 | 229,848.900 | 29,225.060 |
| 386 | 796 | 259,674 | 230,838.600 | 28,835.420 |
| 387 | 801 | 260,286 | 231,828.200 | 28,457.790 |
| 388 | 804 | 260,868 | 232,422.000 | 28,446.010 |
| 389 | 807 | 261,504 | 233,015.800 | 28,488.230 |
| 390 | 809 | 262,237 | 233,411.600 | 28,825.380 |
| 391 | 811 | 262,810 | 233,807.500 | 29,002.520 |
| 392 | 815 | 263,520 | 234,599.200 | 28,920.810 |
| 393 | 821 | 264,094 | 235,786.700 | 28,307.250 |
| 394 | 823 | 264,688 | 236,182.600 | 28,505.400 |
| 395 | 823 | 265,725 | 236,182.600 | 29,542.400 |
| 396 | 833 | 266,402 | 238,161.900 | 28,240.130 |
| 397 | 836 | 267,018 | 238,755.600 | 28,262.350 |
| 398 | 840 | 267,597 | 239,547.400 | 28,049.640 |
| 399 | 847 | 268,294 | 240,932.800 | 27,361.150 |
| 400 | 849 | 268,823 | 241,328.700 | 27,494.300 |
| 401 | 852 | 269,395 | 241,922.500 | 27,472.520 |
| 402 | 860 | 269,927 | 243,505.900 | 26,421.100 |
| 403 | 864 | 270,515 | 244,297.600 | 26,217.400 |
| 404 | 873 | 271,137 | 246,078.900 | 25,058.060 |
| 405 | 878 | 271,702 | 247,068.600 | 24,633.420 |
| 406 | 885 | 272,213 | 248,454.100 | 23,758.930 |
| 407 | 892 | 272,898 | 249,839.600 | 23,058.440 |
| 408 | 895 | 273,595 | 250,433.300 | 23,161.660 |
| 409 | 896 | 274,177 | 250,631.300 | 23,545.740 |
| 410 | 904 | 274,857 | 252,214.700 | 22,642.320 |
| 411 | 905 | 275,473 | 252,412.600 | 23,060.400 |
| 412 | 910 | 276,109 | 253,402.200 | 22,706.760 |
| 413 | 918 | 276,694 | 254,985.700 | 21,708.350 |
| 414 | 919 | 277,332 | 255,183.600 | 22,148.420 |
| 415 | 924 | 277,968 | 256,173.200 | 21,794.780 |
| 416 | 926 | 278,756 | 256,569.100 | 22,186.930 |
| 417 | 929 | 279,408 | 257,162.900 | 22,245.150 |
| 418 | 930 | 280,037 | 257,360.800 | 22,676.220 |
| 419 | 932 | 280,851 | 257,756.600 | 23,094.370 |
| 420 | 935 | 281,506 | 258,350.400 | 23,155.590 |
| 421 | 938 | 282,332 | 258,944.200 | 23,387.810 |
| 422 | 942 | 283,171 | 259,735.900 | 23,435.100 |
| 423 | 943 | 283,892 | 259,933.800 | 23,958.170 |
| 424 | 952 | 284,668 | 261,715.200 | 22,952.830 |
| 425 | 959 | 285,413 | 263,100.700 | 22,312.340 |
| 426 | 963 | 286,121 | 263,892.400 | 22,228.640 |
| 427 | 969 | 286,842 | 265,079.900 | 21,762.080 |
| 428 | 974 | 287,666 | 266,069.600 | 21,596.440 |
| 429 | 974 | 288,349 | 266,069.600 | 22,279.440 |
| 430 | 978 | 289,044 | 266,861.300 | 22,182.730 |
| 431 | 982 | 289,889 | 267,653.000 | 22,236.030 |
| 432 | 983 | 290,685 | 267,850.900 | 22,834.100 |
| 433 | 985 | 291,456 | 268,246.800 | 23,209.240 |
| 434 | 991 | 292,097 | 269,434.300 | 22,662.680 |
| 435 | 993 | 292,850 | 269,830.200 | 23,019.830 |
| 436 | 996 | 293,626 | 270,424.000 | 23,202.050 |
| 437 | 999 | 294,383 | 271,017.700 | 23,365.270 |
| 438 | 1,004 | 295,136 | 272,007.400 | 23,128.630 |
| 439 | 1,009 | 295,948 | 272,997.000 | 22,951.000 |
| 440 | 1,017 | 296,930 | 274,580.400 | 22,349.580 |
| 441 | 1,022 | 297,702 | 275,570.000 | 22,131.950 |
| 442 | 1,030 | 298,611 | 277,153.500 | 21,457.540 |
| 443 | 1,039 | 299,542 | 278,934.800 | 20,607.190 |
| 444 | 1,049 | 300,522 | 280,914.100 | 19,607.920 |
| 445 | 1,057 | 301,513 | 282,497.500 | 19,015.510 |
| 446 | 1,065 | 302,344 | 284,080.900 | 18,263.100 |
| 447 | 1,070 | 303,276 | 285,070.500 | 18,205.460 |
| 448 | 1,077 | 304,221 | 286,456.000 | 17,764.970 |
| 449 | 1,082 | 305,028 | 287,445.700 | 17,582.340 |
| 450 | 1,086 | 305,974 | 288,237.400 | 17,736.630 |
| 451 | 1,092 | 306,655 | 289,424.900 | 17,230.070 |
| 452 | 1,093 | 307,511 | 289,622.900 | 17,888.140 |
| 453 | 1,097 | 308,344 | 290,414.600 | 17,929.440 |
| 454 | 1,106 | 309,128 | 292,195.900 | 16,932.090 |
| 455 | 1,117 | 309,974 | 294,373.100 | 15,600.900 |
| 456 | 1,119 | 310,848 | 294,769.000 | 16,079.040 |
| 457 | 1,126 | 311,628 | 296,154.400 | 15,473.560 |
| 458 | 1,143 | 312,578 | 299,519.200 | 13,058.800 |
| 459 | 1,151 | 313,631 | 301,102.600 | 12,528.380 |
| 460 | 1,162 | 314,642 | 303,279.800 | 11,362.190 |
| 461 | 1,163 | 315,907 | 303,477.700 | 12,429.260 |
| 462 | 1,172 | 316,788 | 305,259.100 | 11,528.920 |
| 463 | 1,179 | 317,671 | 306,644.600 | 11,026.430 |
| 464 | 1,193 | 318,869 | 309,415.500 | 9,453.455 |
| 465 | 1,195 | 319,736 | 309,811.400 | 9,924.601 |
| 466 | 1,197 | 320,619 | 310,207.300 | 10,411.750 |
| 467 | 1,202 | 321,617 | 311,196.900 | 10,420.110 |
| 468 | 1,210 | 322,771 | 312,780.300 | 9,990.698 |
| 469 | 1,212 | 323,631 | 313,176.200 | 10,454.840 |
| 470 | 1,221 | 324,620 | 314,957.500 | 9,662.502 |
| 471 | 1,225 | 325,676 | 315,749.200 | 9,926.795 |
| 472 | 1,236 | 326,518 | 317,926.400 | 8,591.599 |
| 473 | 1,239 | 327,545 | 318,520.200 | 9,024.818 |
| 474 | 1,244 | 328,527 | 319,509.800 | 9,017.184 |
| 475 | 1,251 | 329,527 | 320,895.300 | 8,631.696 |
| 476 | 1,255 | 330,561 | 321,687.000 | 8,873.988 |
| 477 | 1,259 | 331,504 | 322,478.700 | 9,025.281 |
| 478 | 1,269 | 332,426 | 324,458.000 | 7,968.012 |
| 479 | 1,275 | 333,406 | 325,645.500 | 7,760.451 |
| 480 | 1,285 | 334,379 | 327,624.800 | 6,754.182 |
| 481 | 1,288 | 335,387 | 328,218.600 | 7,168.401 |
| 482 | 1,299 | 336,217 | 330,395.800 | 5,821.206 |
| 483 | 1,310 | 337,022 | 332,573.000 | 4,449.010 |
| 484 | 1,317 | 337,851 | 333,958.500 | 3,892.522 |
| 485 | 1,327 | 338,765 | 335,937.700 | 2,827.253 |
| 486 | 1,337 | 339,517 | 337,917.000 | 1,599.984 |
| 487 | 1,354 | 340,323 | 341,281.800 | -958.773 |
| 488 | 1,371 | 341,678 | 344,646.500 | -2,968.530 |
| 489 | 1,392 | 342,482 | 348,803.000 | -6,320.994 |

490 1,393 343,328 349,000.900 -5,672.921
491 1,393 344,164 349,000.900 -4,836.921
492 1,393 345,029 349,000.900 -3,971.921
493 1,393 346,043 349,000.900 -2,957.921
494 1,400 346,913 350,386.400 -3,473.409
495 1,409 347,623 352,167.800 -4,544.751
496 1,416 348,707 353,553.200 -4,846.239
497 1,432 349,457 356,720.100 -7,263.069
498 1,463 350,282 362,855.800 -12,573.800
499 1,476 351,079 365,428.900 -14,349.850
500 1,502 351,984 370,575.000 -18,590.950
501 1,511 352,564 372,356.300 -19,792.290
502 1,524 353,249 374,929.300 -21,680.340
503 1,537 354,111 377,502.400 -23,391.390
504 1,552 354,981 380,471.300 -25,490.290
505 1,570 355,788 384,034.000 -28,245.980
506 1,583 356,640 386,607.000 -29,967.030
507 1,605 357,494 390,961.400 -33,467.420
508 1,617 358,204 393,336.500 -35,132.540
509 1,636 359,354 397,097.200 -37,743.150
510 1,660 360,219 401,847.400 -41,628.400
511 1,666 361,190 403,035.000 -41,844.960
512 1,685 362,025 406,795.600 -44,770.570
513 1,685 362,743 406,795.600 -44,052.570
514 1,717 363,383 413,129.200 -49,746.230
515 1,740 364,225 417,681.500 -53,456.550
516 1,756 365,107 420,848.400 -55,741.380
517 1,778 366,116 425,202.800 -59,086.770
518 1,791 366,818 427,775.800 -60,957.820
519 1,815 367,744 432,526.100 -64,782.060
520 1,834 368,633 436,286.700 -67,653.670
521 1,873 369,559 444,005.800 -74,446.820
522 1,890 370,901 447,370.600 -76,469.580
523 1,896 372,115 448,558.100 -76,443.140
524 1,924 373,179 454,100.100 -80,921.090
525 1,944 373,968 458,058.600 -84,090.630
526 1,960 374,857 461,225.500 -86,368.460
527 1,995 375,765 468,152.900 -92,387.900
528 2,023 376,835 473,694.900 -96,859.850
529 2,041 377,775 477,257.500 -99,482.540
530 2,053 378,603 479,632.700 -101,029.700
531 2,085 379,545 485,966.300 -106,421.300
532 2,108 380,502 490,518.600 -110,016.600
533 2,136 381,388 496,060.600 -114,672.600
534 2,160 382,361 500,810.800 -118,449.800
535 2,184 383,259 505,561.100 -122,302.100
536 2,209 384,208 510,509.300 -126,301.300
----------------------------------------

Analysis of Variance Table - Linear Model

===================================================================================

| | Df | Sum Sq | Mean Sq | F value | Pr(> F) |
|---|---|---|---|---|---|
| Linear effect | 1 | 6,040,593,804,231.000000 | 6,040,593,804,231.000000 | 3,704.357000 | 0 |
| Residuals | 534 | 870,779,264,417.000000 | 1,630,672,780.000000 | | |

Linear Model

=====================================================

| | Estimate | Std. Error | t value | Pr(> \| t\| ) |
|---|---|---|---|---|
| b0 | 73,288.780000 | 2,444.829000 | 29.977050 | 0 |
| b1 | 197.926900 | 3.251984 | 60.863430 | 0 |

R-square - Linear Model

========

0.874008

--------

Equation - Linear Model

======================

73288.7804+197.926878x

----------------------

Observed and estimated - Quadratic Model

=========================================

| | x | y | y_est | Resid |
|---|---|---|---|---|
| 1 | 0 | 255 | 41,900.900 | -41,645.900 |
| 2 | 0 | 442 | 41,900.900 | -41,458.900 |
| 3 | 0 | 721 | 41,900.900 | -41,179.900 |
| 4 | 2 | 1,016 | 42,645.670 | -41,629.670 |
| 5 | 2 | 1,380 | 42,645.670 | -41,265.670 |
| 6 | 4 | 1,673 | 43,389.590 | -41,716.590 |
| 7 | 5 | 2,068 | 43,761.220 | -41,693.220 |
| 8 | 5 | 2,449 | 43,761.220 | -41,312.220 |
| 9 | 6 | 2,769 | 44,132.640 | -41,363.640 |
| 10 | 6 | 3,090 | 44,132.640 | -41,042.640 |
| 11 | 7 | 3,351 | 44,503.840 | -41,152.840 |
| 12 | 7 | 3,673 | 44,503.840 | -40,830.840 |
| 13 | 7 | 3,937 | 44,503.840 | -40,566.840 |
| 14 | 7 | 4,299 | 44,503.840 | -40,204.840 |
| 15 | 7 | 4,629 | 44,503.840 | -39,874.840 |
| 16 | 7 | 5,004 | 44,503.840 | -39,499.840 |
| 17 | 9 | 5,333 | 45,245.610 | -39,912.610 |
| 18 | 9 | 5,632 | 45,245.610 | -39,613.610 |
| 19 | 10 | 5,915 | 45,616.170 | -39,701.170 |
| 20 | 10 | 6,211 | 45,616.170 | -39,405.170 |
| 21 | 10 | 6,580 | 45,616.170 | -39,036.170 |
| 22 | 10 | 6,928 | 45,616.170 | -38,688.170 |
| 23 | 10 | 7,293 | 45,616.170 | -38,323.170 |
| 24 | 10 | 7,621 | 45,616.170 | -37,995.170 |
| 25 | 11 | 7,912 | 45,986.520 | -38,074.520 |
| 26 | 11 | 8,287 | 45,986.520 | -37,699.520 |
| 27 | 11 | 8,652 | 45,986.520 | -37,334.520 |
| 28 | 11 | 9,086 | 45,986.520 | -36,900.520 |
| 29 | 11 | 9,409 | 45,986.520 | -36,577.520 |
| 30 | 12 | 9,815 | 46,356.650 | -36,541.650 |
| 31 | 13 | 10,167 | 46,726.570 | -36,559.570 |
| 32 | 13 | 10,491 | 46,726.570 | -36,235.570 |
| 33 | 13 | 10,788 | 46,726.570 | -35,938.570 |
| 34 | 13 | 11,226 | 46,726.570 | -35,500.570 |
| 35 | 13 | 11,533 | 46,726.570 | -35,193.570 |
| 36 | 13 | 11,925 | 46,726.570 | -34,801.570 |
| 37 | 13 | 12,272 | 46,726.570 | -34,454.570 |
| 38 | 13 | 12,660 | 46,726.570 | -34,066.570 |
| 39 | 13 | 13,015 | 46,726.570 | -33,711.570 |
| 40 | 16 | 13,491 | 47,835.030 | -34,344.030 |
| 41 | 17 | 13,930 | 48,204.090 | -34,274.090 |
| 42 | 17 | 14,507 | 48,204.090 | -33,697.090 |
| 43 | 17 | 14,896 | 48,204.090 | -33,308.090 |
| 44 | 17 | 15,356 | 48,204.090 | -32,848.090 |
| 45 | 17 | 15,883 | 48,204.090 | -32,321.090 |
| 46 | 17 | 16,375 | 48,204.090 | -31,829.090 |
| 47 | 18 | 16,721 | 48,572.940 | -31,851.940 |
| 48 | 18 | 17,146 | 48,572.940 | -31,426.940 |
| 49 | 19 | 17,747 | 48,941.570 | -31,194.570 |
| 50 | 19 | 18,419 | 48,941.570 | -30,522.570 |
| 51 | 19 | 18,903 | 48,941.570 | -30,038.570 |
| 52 | 20 | 19,469 | 49,309.980 | -29,840.980 |
| 53 | 27 | 19,918 | 51,882.890 | -31,964.890 |
| 54 | 36 | 20,408 | 55,175.460 | -34,767.460 |
| 55 | 36 | 20,727 | 55,175.460 | -34,448.460 |
| 56 | 36 | 21,248 | 55,175.460 | -33,927.460 |
| 57 | 37 | 21,892 | 55,540.230 | -33,648.230 |
| 58 | 37 | 22,306 | 55,540.230 | -33,234.230 |
| 59 | 37 | 22,884 | 55,540.230 | -32,656.230 |
| 60 | 37 | 23,442 | 55,540.230 | -32,098.230 |
| 61 | 38 | 23,969 | 55,904.780 | -31,935.780 |
| 62 | 38 | 24,562 | 55,904.780 | -31,342.780 |
| 63 | 39 | 25,096 | 56,269.120 | -31,173.120 |
| 64 | 39 | 25,584 | 56,269.120 | -30,685.120 |
| 65 | 40 | 26,102 | 56,633.250 | -30,531.250 |
| 66 | 40 | 26,651 | 56,633.250 | -29,982.250 |
| 67 | 40 | 27,242 | 56,633.250 | -29,391.250 |
| 68 | 40 | 27,833 | 56,633.250 | -28,800.250 |
| 69 | 40 | 28,259 | 56,633.250 | -28,374.250 |
| 70 | 40 | 28,807 | 56,633.250 | -27,826.250 |
| 71 | 40 | 29,542 | 56,633.250 | -27,091.250 |
| 72 | 41 | 30,262 | 56,997.160 | -26,735.160 |
| 73 | 42 | 30,918 | 57,360.850 | -26,442.850 |
| 74 | 42 | 31,799 | 57,360.850 | -25,561.850 |
| 75 | 42 | 32,639 | 57,360.850 | -24,721.850 |
| 76 | 42 | 33,170 | 57,360.850 | -24,190.850 |
| 77 | 42 | 33,725 | 57,360.850 | -23,635.850 |
| 78 | 42 | 34,147 | 57,360.850 | -23,213.850 |
| 79 | 42 | 34,683 | 57,360.850 | -22,677.850 |
| 80 | 42 | 35,311 | 57,360.850 | -22,049.850 |
| 81 | 42 | 35,913 | 57,360.850 | -21,447.850 |
| 82 | 42 | 36,324 | 57,360.850 | -21,036.850 |
| 83 | 43 | 36,955 | 57,724.330 | -20,769.330 |
| 84 | 43 | 37,463 | 57,724.330 | -20,261.330 |
| 85 | 43 | 37,896 | 57,724.330 | -19,828.330 |
| 86 | 43 | 38,544 | 57,724.330 | -19,180.330 |
| 87 | 44 | 39,187 | 58,087.600 | -18,900.600 |
| 88 | 44 | 39,722 | 58,087.600 | -18,365.600 |
| 89 | 44 | 40,236 | 58,087.600 | -17,851.600 |
| 90 | 44 | 40,790 | 58,087.600 | -17,297.600 |
| 91 | 44 | 41,352 | 58,087.600 | -16,735.600 |
| 92 | 44 | 41,894 | 58,087.600 | -16,193.600 |
| 93 | 45 | 42,593 | 58,450.650 | -15,857.650 |
| 94 | 45 | 43,280 | 58,450.650 | -15,170.650 |
| 95 | 45 | 44,205 | 58,450.650 | -14,245.650 |
| 96 | 45 | 44,920 | 58,450.650 | -13,530.650 |
| 97 | 47 | 45,589 | 59,176.110 | -13,587.110 |
| 98 | 47 | 46,187 | 59,176.110 | -12,989.110 |
| 99 | 47 | 46,754 | 59,176.110 | -12,422.110 |
| 100 | 47 | 47,470 | 59,176.110 | -11,706.110 |
| 101 | 47 | 48,159 | 59,176.110 | -11,017.110 |
| 102 | 47 | 49,036 | 59,176.110 | -10,140.110 |
| 103 | 47 | 49,629 | 59,176.110 | -9,547.114 |
| 104 | 48 | 50,100 | 59,538.520 | -9,438.522 |
| 105 | 49 | 50,842 | 59,900.720 | -9,058.716 |
| 106 | 50 | 51,387 | 60,262.700 | -8,875.696 |
| 107 | 51 | 52,130 | 60,624.460 | -8,494.461 |
| 108 | 51 | 53,067 | 60,624.460 | -7,557.461 |
| 109 | 51 | 53,976 | 60,624.460 | -6,648.461 |
| 110 | 51 | 55,003 | 60,624.460 | -5,621.461 |
| 111 | 51 | 55,931 | 60,624.460 | -4,693.461 |
| 112 | 51 | 56,847 | 60,624.460 | -3,777.461 |
| 113 | 51 | 57,584 | 60,624.460 | -3,040.461 |
| 114 | 52 | 58,455 | 60,986.010 | -2,531.011 |
| 115 | 53 | 59,176 | 61,347.350 | -2,171.347 |
| 116 | 53 | 59,857 | 61,347.350 | -1,490.347 |
| 117 | 53 | 60,724 | 61,347.350 | -623.347 |
| 118 | 53 | 61,581 | 61,347.350 | 233.653 |
| 119 | 53 | 62,297 | 61,347.350 | 949.653 |
| 120 | 53 | 63,025 | 61,347.350 | 1,677.653 |
| 121 | 53 | 63,793 | 61,347.350 | 2,445.653 |
| 122 | 53 | 64,514 | 61,347.350 | 3,166.653 |
| 123 | 53 | 65,285 | 61,347.350 | 3,937.653 |
| 124 | 53 | 66,080 | 61,347.350 | 4,732.653 |
| 125 | 54 | 66,828 | 61,708.470 | 5,119.532 |
| 126 | 54 | 67,580 | 61,708.470 | 5,871.532 |
| 127 | 54 | 68,277 | 61,708.470 | 6,568.532 |
| 128 | 54 | 68,906 | 61,708.470 | 7,197.532 |
| 129 | 54 | 69,494 | 61,708.470 | 7,785.532 |
| 130 | 59 | 70,221 | 63,510.860 | 6,710.143 |
| 131 | 59 | 71,123 | 63,510.860 | 7,612.143 |
| 132 | 59 | 71,900 | 63,510.860 | 8,389.143 |
| 133 | 64 | 72,645 | 65,307.880 | 7,337.119 |
| 134 | 66 | 73,363 | 66,025.190 | 7,337.811 |
| 135 | 66 | 74,272 | 66,025.190 | 8,246.811 |
| 136 | 67 | 74,953 | 66,383.520 | 8,569.478 |
| 137 | 80 | 75,601 | 71,022.320 | 4,578.683 |
| 138 | 82 | 76,295 | 71,732.760 | 4,562.240 |
| 139 | 82 | 76,923 | 71,732.760 | 5,190.240 |
| 140 | 83 | 77,592 | 72,087.660 | 5,504.341 |
| 141 | 83 | 78,213 | 72,087.660 | 6,125.341 |
| 142 | 83 | 79,103 | 72,087.660 | 7,015.341 |
| 143 | 86 | 80,130 | 73,151.070 | 6,978.929 |
| 144 | 87 | 81,132 | 73,505.110 | 7,626.888 |
| 145 | 87 | 81,914 | 73,505.110 | 8,408.888 |
| 146 | 87 | 82,790 | 73,505.110 | 9,284.888 |
| 147 | 89 | 83,553 | 74,212.550 | 9,340.449 |
| 148 | 90 | 84,538 | 74,565.950 | 9,972.051 |
| 149 | 90 | 85,238 | 74,565.950 | 10,672.050 |
| 150 | 90 | 85,968 | 74,565.950 | 11,402.050 |
| 151 | 90 | 86,809 | 74,565.950 | 12,243.050 |
| 152 | 90 | 87,643 | 74,565.950 | 13,077.050 |
| 153 | 90 | 88,476 | 74,565.950 | 13,910.050 |
| 154 | 93 | 89,341 | 75,624.850 | 13,716.150 |
| 155 | 96 | 90,252 | 76,681.830 | 13,570.170 |
| 156 | 96 | 91,118 | 76,681.830 | 14,436.170 |
| 157 | 96 | 92,127 | 76,681.830 | 15,445.170 |
| 158 | 97 | 93,064 | 77,033.730 | 16,030.270 |
| 159 | 97 | 93,977 | 77,033.730 | 16,943.280 |
| 160 | 109 | 94,856 | 81,239.740 | 13,616.260 |
| 161 | 110 | 95,635 | 81,588.850 | 14,046.150 |
| 162 | 110 | 96,516 | 81,588.850 | 14,927.150 |
| 163 | 113 | 97,188 | 82,634.880 | 14,553.120 |
| 164 | 113 | 97,833 | 82,634.880 | 15,198.120 |
| 165 | 115 | 98,765 | 83,331.160 | 15,433.840 |
| 166 | 115 | 99,537 | 83,331.160 | 16,205.840 |
| 167 | 115 | 100,299 | 83,331.160 | 16,967.840 |
| 168 | 115 | 101,063 | 83,331.160 | 17,731.840 |
| 169 | 115 | 101,718 | 83,331.160 | 18,386.840 |
| 170 | 116 | 102,411 | 83,678.980 | 18,732.020 |
| 171 | 116 | 103,081 | 83,678.980 | 19,402.020 |
| 172 | 116 | 103,794 | 83,678.980 | 20,115.020 |
| 173 | 116 | 104,672 | 83,678.980 | 20,993.020 |
| 174 | 116 | 105,368 | 83,678.980 | 21,689.020 |
| 175 | 116 | 106,007 | 83,678.980 | 22,328.020 |
| 176 | 116 | 106,672 | 83,678.980 | 22,993.020 |
| 177 | 117 | 107,400 | 84,026.590 | 23,373.410 |
| 178 | 118 | 108,236 | 84,373.980 | 23,862.020 |
| 179 | 119 | 109,157 | 84,721.150 | 24,435.850 |
| 180 | 120 | 109,991 | 85,068.120 | 24,922.880 |
| 181 | 121 | 110,699 | 85,414.860 | 25,284.140 |
| 182 | 123 | 111,293 | 86,107.710 | 25,185.290 |
| 183 | 123 | 111,982 | 86,107.710 | 25,874.290 |
| 184 | 123 | 112,877 | 86,107.710 | 26,769.290 |
| 185 | 124 | 113,680 | 86,453.820 | 27,226.180 |
| 186 | 124 | 114,404 | 86,453.820 | 27,950.180 |
| 187 | 124 | 115,044 | 86,453.820 | 28,590.180 |
| 188 | 124 | 115,832 | 86,453.820 | 29,378.180 |
| 189 | 124 | 116,640 | 86,453.820 | 30,186.180 |
| 190 | 124 | 117,335 | 86,453.820 | 30,881.180 |
| 191 | 125 | 118,014 | 86,799.710 | 31,214.290 |
| 192 | 126 | 118,972 | 87,145.380 | 31,826.620 |
| 193 | 130 | 119,627 | 88,525.930 | 31,101.070 |
| 194 | 131 | 120,232 | 88,870.530 | 31,361.470 |
| 195 | 131 | 120,860 | 88,870.530 | 31,989.470 |
| 196 | 131 | 121,492 | 88,870.530 | 32,621.470 |
| 197 | 131 | 122,269 | 88,870.530 | 33,398.470 |
| 198 | 131 | 122,838 | 88,870.530 | 33,967.470 |
| 199 | 133 | 123,467 | 89,559.090 | 33,907.910 |
| 200 | 133 | 124,221 | 89,559.090 | 34,661.910 |
| 201 | 133 | 124,738 | 89,559.090 | 35,178.910 |
| 202 | 136 | 125,349 | 90,590.320 | 34,758.680 |
| 203 | 139 | 126,156 | 91,619.620 | 34,536.380 |
| 204 | 139 | 126,901 | 91,619.620 | 35,281.380 |
| 205 | 140 | 127,588 | 91,962.290 | 35,625.710 |
| 206 | 141 | 128,340 | 92,304.750 | 36,035.250 |
| 207 | 143 | 129,247 | 92,989.010 | 36,257.990 |
| 208 | 146 | 129,996 | 94,013.810 | 35,982.190 |
| 209 | 147 | 130,616 | 94,354.980 | 36,261.020 |
| 210 | 149 | 131,342 | 95,036.670 | 36,305.330 |
| 211 | 149 | 131,996 | 95,036.670 | 36,959.330 |
| 212 | 149 | 132,731 | 95,036.670 | 37,694.330 |
| 213 | 149 | 133,722 | 95,036.670 | 38,685.330 |
| 214 | 151 | 134,736 | 95,717.510 | 39,018.500 |
| 215 | 151 | 135,807 | 95,717.510 | 40,089.500 |
| 216 | 152 | 136,590 | 96,057.600 | 40,532.400 |
| 217 | 152 | 137,434 | 96,057.600 | 41,376.400 |
| 218 | 153 | 138,272 | 96,397.480 | 41,874.520 |
| 219 | 154 | 139,087 | 96,737.150 | 42,349.850 |
| 220 | 157 | 139,868 | 97,754.860 | 42,113.140 |
| 221 | 161 | 140,811 | 99,108.810 | 41,702.190 |
| 222 | 161 | 141,482 | 99,108.810 | 42,373.190 |
| 223 | 164 | 142,219 | 100,122.000 | 42,096.980 |
| 224 | 164 | 143,256 | 100,122.000 | 43,133.980 |
| 225 | 165 | 144,038 | 100,459.300 | 43,578.680 |
| 226 | 180 | 144,963 | 105,493.200 | 39,469.820 |
| 227 | 208 | 145,900 | 114,760.600 | 31,139.440 |
| 228 | 209 | 146,597 | 115,088.400 | 31,508.570 |
| 229 | 210 | 147,567 | 115,416.100 | 32,150.920 |
| 230 | 215 | 148,465 | 117,051.100 | 31,413.880 |
| 231 | 220 | 149,213 | 118,680.800 | 30,532.200 |
| 232 | 222 | 150,242 | 119,331.200 | 30,910.830 |
| 233 | 223 | 151,044 | 119,656.000 | 31,387.960 |
| 234 | 223 | 151,954 | 119,656.000 | 32,297.960 |
| 235 | 226 | 152,727 | 120,629.300 | 32,097.660 |
| 236 | 228 | 153,449 | 121,277.100 | 32,171.860 |
| 237 | 228 | 154,084 | 121,277.100 | 32,806.860 |
| 238 | 230 | 154,900 | 121,924.100 | 32,975.930 |
| 239 | 239 | 155,972 | 124,824.700 | 31,147.330 |
| 240 | 242 | 156,905 | 125,787.700 | 31,117.320 |
| 241 | 244 | 157,767 | 126,428.600 | 31,338.390 |
| 242 | 245 | 158,474 | 126,748.800 | 31,725.250 |
| 243 | 253 | 159,337 | 129,302.200 | 30,034.830 |
| 244 | 259 | 160,570 | 131,208.200 | 29,361.770 |
| 245 | 261 | 161,277 | 131,841.900 | 29,435.130 |
| 246 | 263 | 162,127 | 132,474.600 | 29,652.360 |
| 247 | 263 | 162,837 | 132,474.600 | 30,362.360 |
| 248 | 266 | 163,589 | 133,422.200 | 30,166.800 |
| 249 | 269 | 164,385 | 134,367.800 | 30,017.170 |
| 250 | 275 | 165,089 | 136,253.300 | 28,835.710 |
| 251 | 277 | 165,881 | 136,880.100 | 29,000.940 |
| 252 | 278 | 166,846 | 137,193.100 | 29,652.880 |
| 253 | 284 | 167,636 | 139,067.000 | 28,569.010 |
| 254 | 285 | 168,508 | 139,378.600 | 29,129.440 |
| 255 | 289 | 169,409 | 140,622.700 | 28,786.340 |
| 256 | 292 | 170,240 | 141,553.500 | 28,686.520 |
| 257 | 293 | 170,955 | 141,863.300 | 29,091.670 |
| 258 | 294 | 171,745 | 142,173.000 | 29,572.040 |
| 259 | 299 | 172,498 | 143,717.900 | 28,780.110 |
| 260 | 303 | 173,328 | 144,950.000 | 28,378.020 |
| 261 | 317 | 174,007 | 149,235.200 | 24,771.750 |
| 262 | 321 | 174,952 | 150,451.900 | 24,500.120 |
| 263 | 338 | 175,722 | 155,584.300 | 20,137.690 |
| 264 | 341 | 176,484 | 156,483.600 | 20,000.410 |
| 265 | 346 | 177,256 | 157,978.100 | 19,277.890 |
| 266 | 350 | 177,965 | 159,169.900 | 18,795.140 |
| 267 | 356 | 178,662 | 160,951.000 | 17,710.950 |
| 268 | 360 | 179,467 | 162,134.200 | 17,332.780 |
| 269 | 368 | 180,351 | 164,490.300 | 15,860.740 |
| 270 | 373 | 181,156 | 165,955.800 | 15,200.180 |
| 271 | 382 | 181,971 | 168,580.300 | 13,390.700 |
| 272 | 386 | 183,059 | 169,741.200 | 13,317.850 |
| 273 | 390 | 183,623 | 170,898.600 | 12,724.420 |
| 274 | 393 | 184,571 | 171,764.400 | 12,806.610 |
| 275 | 398 | 185,518 | 173,203.100 | 12,314.870 |
| 276 | 399 | 186,258 | 173,490.200 | 12,767.770 |
| 277 | 403 | 187,141 | 174,636.500 | 12,504.500 |
| 278 | 406 | 187,879 | 175,494.000 | 12,385.050 |
| 279 | 408 | 188,557 | 176,064.500 | 12,492.490 |
| 280 | 413 | 189,297 | 177,487.200 | 11,809.850 |
| 281 | 415 | 190,021 | 178,054.700 | 11,966.290 |
| 282 | 415 | 190,716 | 178,054.700 | 12,661.290 |
| 283 | 418 | 191,711 | 178,904.400 | 12,806.570 |
| 284 | 420 | 192,531 | 179,469.800 | 13,061.150 |
| 285 | 425 | 193,257 | 180,879.600 | 12,377.380 |
| 286 | 427 | 194,106 | 181,442.000 | 12,663.980 |
| 287 | 429 | 195,012 | 182,003.600 | 13,008.430 |
| 288 | 433 | 196,021 | 183,124.100 | 12,896.900 |
| 289 | 433 | 196,940 | 183,124.100 | 13,815.900 |
| 290 | 434 | 197,679 | 183,403.700 | 14,275.310 |
| 291 | 436 | 198,479 | 183,962.200 | 14,516.760 |
| 292 | 440 | 199,272 | 185,076.800 | 14,195.250 |
| 293 | 440 | 199,944 | 185,076.800 | 14,867.250 |
| 294 | 444 | 200,748 | 186,187.800 | 14,560.170 |
| 295 | 446 | 201,470 | 186,742.100 | 14,727.910 |
| 296 | 452 | 202,260 | 188,399.700 | 13,860.300 |
| 297 | 455 | 202,902 | 189,225.600 | 13,676.390 |
| 298 | 459 | 203,643 | 190,323.800 | 13,319.170 |
| 299 | 464 | 204,594 | 191,691.800 | 12,902.240 |
| 300 | 467 | 205,532 | 192,509.900 | 13,022.050 |
| 301 | 474 | 206,369 | 194,411.500 | 11,957.460 |
| 302 | 475 | 207,149 | 194,682.300 | 12,466.660 |
| 303 | 484 | 207,857 | 197,109.900 | 10,747.130 |
| 304 | 489 | 208,976 | 198,451.000 | 10,525.010 |
| 305 | 492 | 209,686 | 199,253.100 | 10,432.920 |
| 306 | 495 | 210,527 | 200,053.200 | 10,473.750 |
| 307 | 499 | 211,348 | 201,117.100 | 10,230.870 |
| 308 | 500 | 212,036 | 201,382.600 | 10,653.430 |
| 309 | 503 | 212,866 | 202,177.600 | 10,688.420 |
| 310 | 503 | 213,676 | 202,177.600 | 11,498.420 |
| 311 | 504 | 214,647 | 202,442.200 | 12,204.840 |
| 312 | 506 | 215,525 | 202,970.700 | 12,554.330 |
| 313 | 509 | 216,389 | 203,761.800 | 12,627.180 |
| 314 | 510 | 217,146 | 204,025.100 | 13,120.890 |
| 315 | 512 | 218,208 | 204,551.000 | 13,656.960 |
| 316 | 514 | 219,085 | 205,076.100 | 14,008.880 |
| 317 | 514 | 219,869 | 205,076.100 | 14,792.880 |
| 318 | 516 | 220,826 | 205,600.300 | 15,225.660 |
| 319 | 519 | 221,777 | 206,385.100 | 15,391.940 |
| 320 | 523 | 222,840 | 207,428.300 | 15,411.660 |
| 321 | 526 | 223,444 | 208,208.600 | 15,235.440 |
| 322 | 530 | 223,904 | 209,245.800 | 14,658.160 |
| 323 | 531 | 224,452 | 209,504.600 | 14,947.380 |
| 324 | 536 | 224,972 | 210,795.300 | 14,176.680 |
| 325 | 541 | 225,551 | 212,080.700 | 13,470.340 |
| 326 | 543 | 226,084 | 212,593.300 | 13,490.710 |
| 327 | 547 | 226,584 | 213,616.000 | 12,968.020 |
| 328 | 547 | 227,117 | 213,616.000 | 13,501.020 |
| 329 | 553 | 227,628 | 215,143.600 | 12,484.420 |
| 330 | 553 | 228,260 | 215,143.600 | 13,116.420 |
| 331 | 556 | 228,761 | 215,904.500 | 12,856.520 |
| 332 | 558 | 229,297 | 216,410.700 | 12,886.320 |
| 333 | 562 | 229,730 | 217,420.500 | 12,309.500 |
| 334 | 563 | 230,223 | 217,672.400 | 12,550.580 |
| 335 | 566 | 230,885 | 218,426.900 | 12,458.120 |
| 336 | 567 | 231,450 | 218,677.900 | 12,772.060 |
| 337 | 571 | 232,010 | 219,680.000 | 12,329.960 |
| 338 | 582 | 232,600 | 222,418.100 | 10,181.900 |
| 339 | 585 | 233,170 | 223,160.300 | 10,009.670 |
| 340 | 588 | 233,643 | 223,900.600 | 9,742.358 |
| 341 | 590 | 234,261 | 224,393.100 | 9,867.892 |
| 342 | 596 | 234,798 | 225,865.400 | 8,932.645 |
| 343 | 601 | 235,319 | 227,086.300 | 8,232.672 |
| 344 | 606 | 235,866 | 228,301.900 | 7,564.063 |
| 345 | 610 | 236,252 | 229,270.600 | 6,981.437 |
| 346 | 616 | 236,705 | 230,717.100 | 5,987.935 |
| 347 | 631 | 237,214 | 234,299.500 | 2,914.470 |
| 348 | 640 | 237,743 | 236,425.800 | 1,317.163 |
| 349 | 643 | 238,296 | 237,130.700 | 1,165.255 |
| 350 | 646 | 238,868 | 237,833.700 | 1,034.278 |
| 351 | 648 | 239,404 | 238,301.300 | 1,102.700 |
| 352 | 649 | 240,115 | 238,534.800 | 1,580.233 |
| 353 | 653 | 240,656 | 239,466.500 | 1,189.509 |
| 354 | 656 | 241,112 | 240,163.000 | 948.969 |
| 355 | 658 | 241,611 | 240,626.300 | 984.681 |
| 356 | 658 | 242,180 | 240,626.300 | 1,553.681 |
| 357 | 671 | 242,627 | 243,616.800 | -989.769 |
| 358 | 680 | 243,093 | 245,665.800 | -2,572.841 |
| 359 | 691 | 243,755 | 248,146.700 | -4,391.662 |
| 360 | 695 | 244,267 | 249,042.300 | -4,775.343 |
| 361 | 697 | 244,813 | 249,488.900 | -4,675.896 |
| 362 | 705 | 245,362 | 251,266.500 | -5,904.526 |
| 363 | 707 | 245,948 | 251,708.800 | -5,760.788 |
| 364 | 710 | 246,530 | 252,370.600 | -5,840.572 |
| 365 | 714 | 247,067 | 253,249.900 | -6,182.947 |
| 366 | 721 | 247,633 | 254,780.600 | -7,147.594 |
| 367 | 723 | 248,120 | 255,216.000 | -7,095.990 |
| 368 | 724 | 248,702 | 255,433.400 | -6,731.367 |
| 369 | 724 | 249,205 | 255,433.400 | -6,228.367 |
| 370 | 730 | 249,735 | 256,733.100 | -6,998.120 |
| 371 | 733 | 250,321 | 257,380.100 | -7,059.101 |
| 372 | 733 | 250,866 | 257,380.100 | -6,514.101 |
| 373 | 740 | 251,467 | 258,882.200 | -7,415.213 |
| 374 | 747 | 252,068 | 260,373.800 | -8,305.812 |
| 375 | 752 | 252,577 | 261,432.800 | -8,855.804 |
| 376 | 755 | 253,212 | 262,065.600 | -8,853.625 |
| 377 | 762 | 253,834 | 263,534.700 | -9,700.697 |
| 378 | 766 | 254,514 | 264,369.400 | -9,855.447 |
| 379 | 768 | 255,112 | 264,785.500 | -9,673.535 |
| 380 | 769 | 255,964 | 264,993.300 | -9,029.257 |
| 381 | 771 | 256,538 | 265,408.100 | -8,870.057 |
| 382 | 779 | 257,172 | 267,058.700 | -9,886.677 |
| 383 | 784 | 257,758 | 268,083.300 | -10,325.340 |
| 384 | 791 | 258,389 | 269,508.900 | -11,119.860 |
| 385 | 791 | 259,074 | 269,508.900 | -10,434.860 |
| 386 | 796 | 259,674 | 270,520.700 | -10,846.650 |
| 387 | 801 | 260,286 | 271,527.100 | -11,241.080 |
| 388 | 804 | 260,868 | 272,128.400 | -11,260.360 |
| 389 | 807 | 261,504 | 272,727.700 | -11,223.720 |
| 390 | 809 | 262,237 | 273,126.200 | -10,889.210 |
| 391 | 811 | 262,810 | 273,523.800 | -10,713.850 |
| 392 | 815 | 263,520 | 274,316.500 | -10,796.550 |
| 393 | 821 | 264,094 | 275,499.200 | -11,405.160 |
| 394 | 823 | 264,688 | 275,891.600 | -11,203.650 |
| 395 | 823 | 265,725 | 275,891.600 | -10,166.650 |
| 396 | 833 | 266,402 | 277,841.200 | -11,439.210 |
| 397 | 836 | 267,018 | 278,421.900 | -11,403.900 |
| 398 | 840 | 267,597 | 279,193.100 | -11,596.140 |
| 399 | 847 | 268,294 | 280,534.600 | -12,240.560 |
| 400 | 849 | 268,823 | 280,915.900 | -12,092.890 |
| 401 | 852 | 269,395 | 281,486.300 | -12,091.280 |
| 402 | 860 | 269,927 | 282,997.900 | -13,070.880 |
| 403 | 864 | 270,515 | 283,748.500 | -13,233.520 |
| 404 | 873 | 271,137 | 285,424.900 | -14,287.930 |
| 405 | 878 | 271,702 | 286,348.800 | -14,646.760 |
| 406 | 885 | 272,213 | 287,633.100 | -15,420.110 |
| 407 | 892 | 272,898 | 288,906.900 | -16,008.950 |
| 408 | 895 | 273,595 | 289,449.700 | -15,854.660 |
| 409 | 896 | 274,177 | 289,630.100 | -15,453.140 |
| 410 | 904 | 274,857 | 291,066.200 | -16,209.210 |
| 411 | 905 | 275,473 | 291,244.800 | -15,771.750 |
| 412 | 910 | 276,109 | 292,134.300 | -16,025.260 |
| 413 | 918 | 276,694 | 293,546.300 | -16,852.300 |
| 414 | 919 | 277,332 | 293,721.800 | -16,389.840 |
| 415 | 924 | 277,968 | 294,596.300 | -16,628.330 |
| 416 | 926 | 278,756 | 294,944.600 | -16,188.620 |
| 417 | 929 | 279,408 | 295,465.400 | -16,057.450 |
| 418 | 930 | 280,037 | 295,638.600 | -15,601.630 |
| 419 | 932 | 280,851 | 295,984.300 | -15,133.340 |
| 420 | 935 | 281,506 | 296,501.300 | -14,995.310 |
| 421 | 938 | 282,332 | 297,016.300 | -14,684.350 |
| 422 | 942 | 283,171 | 297,700.100 | -14,529.060 |
| 423 | 943 | 283,892 | 297,870.400 | -13,978.450 |
| 424 | 952 | 284,668 | 299,394.300 | -14,726.320 |
| 425 | 959 | 285,413 | 300,567.500 | -15,154.530 |
| 426 | 963 | 286,121 | 301,233.200 | -15,112.220 |
| 427 | 969 | 286,842 | 302,225.300 | -15,383.320 |
| 428 | 974 | 287,666 | 303,046.200 | -15,380.160 |
| 429 | 974 | 288,349 | 303,046.200 | -14,697.160 |
| 430 | 978 | 289,044 | 303,699.000 | -14,654.980 |
| 431 | 982 | 289,889 | 304,348.400 | -14,459.360 |
| 432 | 983 | 290,685 | 304,510.200 | -13,825.170 |
| 433 | 985 | 291,456 | 304,833.100 | -13,377.150 |
| 434 | 991 | 292,097 | 305,796.900 | -13,699.920 |
| 435 | 993 | 292,850 | 306,116.500 | -13,266.470 |
| 436 | 996 | 293,626 | 306,594.200 | -12,968.170 |
| 437 | 999 | 294,383 | 307,069.900 | -12,686.950 |
| 438 | 1,004 | 295,136 | 307,858.600 | -12,722.610 |
| 439 | 1,009 | 295,948 | 308,641.900 | -12,693.910 |
| 440 | 1,017 | 296,930 | 309,884.000 | -12,954.040 |
| 441 | 1,022 | 297,702 | 310,653.400 | -12,951.400 |
| 442 | 1,030 | 298,611 | 311,873.200 | -13,262.210 |
| 443 | 1,039 | 299,542 | 313,229.100 | -13,687.090 |
| 444 | 1,049 | 300,522 | 314,715.200 | -14,193.240 |
| 445 | 1,057 | 301,513 | 315,888.700 | -14,375.720 |
| 446 | 1,065 | 302,344 | 317,048.500 | -14,704.460 |
| 447 | 1,070 | 303,276 | 317,766.300 | -14,490.320 |
| 448 | 1,077 | 304,221 | 318,762.300 | -14,541.320 |

449 1,082 305,028 319,467.300 -14,439.320
450 1,086 305,974 320,027.500 -14,053.450
451 1,092 306,655 320,861.200 -14,206.210
452 1,093 307,511 320,999.400 -13,488.420
453 1,097 308,344 321,550.100 -13,206.120
454 1,106 309,128 322,776.600 -13,648.620
455 1,117 309,974 324,252.100 -14,278.090
456 1,119 310,848 324,517.600 -13,669.570
457 1,126 311,628 325,440.000 -13,811.980
458 1,143 312,578 327,636.400 -15,058.360
459 1,151 313,631 328,648.500 -15,017.500
460 1,162 314,642 330,017.800 -15,375.760
461 1,163 315,907 330,141.000 -14,233.950
462 1,172 316,788 331,240.000 -14,452.020
463 1,179 317,671 332,082.800 -14,411.840
464 1,193 318,869 333,736.900 -14,867.930
465 1,195 319,736 333,969.800 -14,233.800
466 1,197 320,619 334,201.800 -13,582.810
467 1,202 321,617 334,778.100 -13,161.080
468 1,210 322,771 335,688.900 -12,917.950
469 1,212 323,631 335,914.500 -12,283.520
470 1,221 324,620 336,919.000 -12,298.970
471 1,225 325,676 337,359.800 -11,683.820
472 1,236 326,518 338,554.400 -12,036.450
473 1,239 327,545 338,875.800 -11,330.750
474 1,244 328,527 339,407.000 -10,879.960
475 1,251 329,527 340,141.600 -10,614.650
476 1,255 330,561 340,556.700 -9,995.749
477 1,259 331,504 340,968.400 -9,464.418
478 1,269 332,426 341,982.600 -9,556.570
479 1,275 333,406 342,580.800 -9,174.763
480 1,285 334,379 343,560.600 -9,181.588
481 1,288 335,387 343,850.400 -8,463.352
482 1,299 336,217 344,896.300 -8,679.300
483 1,310 337,022 345,916.300 -8,894.288
484 1,317 337,851 346,551.900 -8,700.855
485 1,327 338,765 347,441.600 -8,676.572
486 1,337 339,517 348,309.800 -8,792.834
487 1,354 340,323 349,736.600 -9,413.641
488 1,371 341,678 351,101.400 -9,423.446
489 1,392 342,482 352,701.800 -10,219.780
490 1,393 343,328 352,775.600 -9,447.623
491 1,393 344,164 352,775.600 -8,611.623
492 1,393 345,029 352,775.600 -7,746.623
493 1,393 346,043 352,775.600 -6,732.623
494 1,400 346,913 353,286.500 -6,373.540
495 1,409 347,623 353,928.000 -6,304.986
496 1,416 348,707 354,414.900 -5,707.873

497 1,432 349,457 355,488.300 -6,031.283
498 1,463 350,282 357,411.700 -7,129.719
499 1,476 351,079 358,157.000 -7,077.962
500 1,502 351,984 359,538.700 -7,554.673
501 1,511 352,564 359,983.200 -7,419.167
502 1,524 353,249 360,594.500 -7,345.533
503 1,537 354,111 361,169.600 -7,058.642
504 1,552 354,981 361,788.200 -6,807.175
505 1,570 355,788 362,466.700 -6,678.694
506 1,583 356,640 362,913.500 -6,273.505
507 1,605 357,494 363,587.000 -6,093.048
508 1,617 358,204 363,910.700 -5,706.667
509 1,636 359,354 364,359.900 -5,005.882
510 1,660 360,219 364,816.600 -4,597.605
511 1,666 361,190 364,911.500 -3,721.477
512 1,685 362,025 365,161.000 -3,135.951
513 1,685 362,743 365,161.000 -2,417.951
514 1,717 363,383 365,406.000 -2,023.049
515 1,740 364,225 365,446.500 -1,221.513
516 1,756 365,107 365,407.700 -300.725
517 1,778 366,116 365,264.700 851.289
518 1,791 366,818 365,131.400 1,686.606
519 1,815 367,744 364,790.000 2,953.988
520 1,834 368,633 364,432.100 4,200.890
521 1,873 369,559 363,454.800 6,104.180
522 1,890 370,901 362,926.700 7,974.302
523 1,896 372,115 362,725.500 9,389.501
524 1,924 373,179 361,684.400 11,494.560
525 1,944 373,968 360,837.900 13,130.150
526 1,960 374,857 360,098.800 14,758.210
527 1,995 375,765 358,290.600 17,474.390
528 2,023 376,835 356,654.800 20,180.160
529 2,041 377,775 355,514.500 22,260.550
530 2,053 378,603 354,715.600 23,887.420
531 2,085 379,545 352,434.200 27,110.800
532 2,108 380,502 350,658.800 29,843.240
533 2,136 381,388 348,344.200 33,043.830
534 2,160 382,361 346,226.400 36,134.640
535 2,184 383,259 343,985.000 39,274.030
536 2,209 384,208 341,518.800 42,689.210
----------------------------------------

Analysis of Variance Table - Quadratic Model
===============================================================================
=======
Df Sum Sq Mean Sq F value Pr(> F)
-------------------------------------------------------------------------------------------
Linear effect 1 6,040,593,804,231.000000 6,040,593,804,231.000000 12,865.400000 0

| | | | | | |
|---|---|---|---|---|---|
| Quadratic effect | 1 | 620,523,772,463.000000 | 620,523,772,463.000000 | 1,321.606000 | 0 |
| Residuals | 533 | 250,255,491,954.000000 | 469,522,499.000000 | | |

-----------------------------------------------------------------------------------------

Quadratic Model

=====================================================

| | Estimate | Std. Error | t value | Pr(> \| t\| ) |
|---|---|---|---|---|
| b0 | 41,900.900000 | 1,570.503000 | 26.679930 | 0 |
| b1 | 372.599400 | 5.111841 | 72.889480 | 0 |
| b2 | -0.107272 | 0.002951 | -36.353900 | 0 |

--------------------------------------------------

R-square - Quadratic Model

========

0.963791

--------

Equation - Quadratic Model

=================================

41900.9044+372.599437x-0.107272x2

---------------------------------

Observed and estimated - Cubic Model

==========================================

| | x | y | y_est | Resid |
|---|---|---|---|---|
| 1 | 0 | 255 | 28,329.280 | -28,074.280 |
| 2 | 0 | 442 | 28,329.280 | -27,887.280 |
| 3 | 0 | 721 | 28,329.280 | -27,608.280 |
| 4 | 2 | 1,016 | 29,377.160 | -28,361.160 |
| 5 | 2 | 1,380 | 29,377.160 | -27,997.160 |
| 6 | 4 | 1,673 | 30,422.390 | -28,749.390 |
| 7 | 5 | 2,068 | 30,944.010 | -28,876.010 |
| 8 | 5 | 2,449 | 30,944.010 | -28,495.010 |
| 9 | 6 | 2,769 | 31,464.970 | -28,695.970 |
| 10 | 6 | 3,090 | 31,464.970 | -28,374.970 |
| 11 | 7 | 3,351 | 31,985.260 | -28,634.260 |
| 12 | 7 | 3,673 | 31,985.260 | -28,312.260 |
| 13 | 7 | 3,937 | 31,985.260 | -28,048.260 |
| 14 | 7 | 4,299 | 31,985.260 | -27,686.260 |
| 15 | 7 | 4,629 | 31,985.260 | -27,356.260 |
| 16 | 7 | 5,004 | 31,985.260 | -26,981.260 |
| 17 | 9 | 5,333 | 33,023.880 | -27,690.880 |
| 18 | 9 | 5,632 | 33,023.880 | -27,391.880 |
| 19 | 10 | 5,915 | 33,542.190 | -27,627.190 |
| 20 | 10 | 6,211 | 33,542.190 | -27,331.190 |
| 21 | 10 | 6,580 | 33,542.190 | -26,962.190 |
| 22 | 10 | 6,928 | 33,542.190 | -26,614.190 |
| 23 | 10 | 7,293 | 33,542.190 | -26,249.190 |
| 24 | 10 | 7,621 | 33,542.190 | -25,921.190 |
| 25 | 11 | 7,912 | 34,059.850 | -26,147.850 |
| 26 | 11 | 8,287 | 34,059.850 | -25,772.850 |
| 27 | 11 | 8,652 | 34,059.850 | -25,407.850 |
| 28 | 11 | 9,086 | 34,059.850 | -24,973.850 |
| 29 | 11 | 9,409 | 34,059.850 | -24,650.850 |
| 30 | 12 | 9,815 | 34,576.850 | -24,761.850 |
| 31 | 13 | 10,167 | 35,093.190 | -24,926.190 |
| 32 | 13 | 10,491 | 35,093.190 | -24,602.190 |
| 33 | 13 | 10,788 | 35,093.190 | -24,305.190 |
| 34 | 13 | 11,226 | 35,093.190 | -23,867.190 |
| 35 | 13 | 11,533 | 35,093.190 | -23,560.190 |
| 36 | 13 | 11,925 | 35,093.190 | -23,168.190 |
| 37 | 13 | 12,272 | 35,093.190 | -22,821.190 |
| 38 | 13 | 12,660 | 35,093.190 | -22,433.190 |
| 39 | 13 | 13,015 | 35,093.190 | -22,078.190 |
| 40 | 16 | 13,491 | 36,638.270 | -23,147.270 |
| 41 | 17 | 13,930 | 37,151.980 | -23,221.980 |
| 42 | 17 | 14,507 | 37,151.980 | -22,644.980 |
| 43 | 17 | 14,896 | 37,151.980 | -22,255.980 |
| 44 | 17 | 15,356 | 37,151.980 | -21,795.980 |
| 45 | 17 | 15,883 | 37,151.980 | -21,268.980 |
| 46 | 17 | 16,375 | 37,151.980 | -20,776.980 |
| 47 | 18 | 16,721 | 37,665.030 | -20,944.030 |
| 48 | 18 | 17,146 | 37,665.030 | -20,519.030 |
| 49 | 19 | 17,747 | 38,177.430 | -20,430.430 |
| 50 | 19 | 18,419 | 38,177.430 | -19,758.430 |
| 51 | 19 | 18,903 | 38,177.430 | -19,274.430 |
| 52 | 20 | 19,469 | 38,689.180 | -19,220.180 |
| 53 | 27 | 19,918 | 42,253.090 | -22,335.090 |
| 54 | 36 | 20,408 | 46,788.400 | -26,380.400 |
| 55 | 36 | 20,727 | 46,788.400 | -26,061.400 |
| 56 | 36 | 21,248 | 46,788.400 | -25,540.400 |
| 57 | 37 | 21,892 | 47,289.080 | -25,397.080 |
| 58 | 37 | 22,306 | 47,289.080 | -24,983.080 |
| 59 | 37 | 22,884 | 47,289.080 | -24,405.080 |
| 60 | 37 | 23,442 | 47,289.080 | -23,847.080 |
| 61 | 38 | 23,969 | 47,789.120 | -23,820.120 |
| 62 | 38 | 24,562 | 47,789.120 | -23,227.120 |
| 63 | 39 | 25,096 | 48,288.510 | -23,192.510 |
| 64 | 39 | 25,584 | 48,288.510 | -22,704.510 |
| 65 | 40 | 26,102 | 48,787.250 | -22,685.250 |
| 66 | 40 | 26,651 | 48,787.250 | -22,136.250 |
| 67 | 40 | 27,242 | 48,787.250 | -21,545.250 |
| 68 | 40 | 27,833 | 48,787.250 | -20,954.250 |
| 69 | 40 | 28,259 | 48,787.250 | -20,528.250 |
| 70 | 40 | 28,807 | 48,787.250 | -19,980.250 |
| 71 | 40 | 29,542 | 48,787.250 | -19,245.250 |
| 72 | 41 | 30,262 | 49,285.350 | -19,023.350 |
| 73 | 42 | 30,918 | 49,782.800 | -18,864.800 |
| 74 | 42 | 31,799 | 49,782.800 | -17,983.800 |
| 75 | 42 | 32,639 | 49,782.800 | -17,143.800 |
| 76 | 42 | 33,170 | 49,782.800 | -16,612.800 |
| 77 | 42 | 33,725 | 49,782.800 | -16,057.800 |
| 78 | 42 | 34,147 | 49,782.800 | -15,635.800 |
| 79 | 42 | 34,683 | 49,782.800 | -15,099.800 |
| 80 | 42 | 35,311 | 49,782.800 | -14,471.800 |
| 81 | 42 | 35,913 | 49,782.800 | -13,869.800 |
| 82 | 42 | 36,324 | 49,782.800 | -13,458.800 |
| 83 | 43 | 36,955 | 50,279.610 | -13,324.610 |
| 84 | 43 | 37,463 | 50,279.610 | -12,816.610 |
| 85 | 43 | 37,896 | 50,279.610 | -12,383.610 |
| 86 | 43 | 38,544 | 50,279.610 | -11,735.610 |
| 87 | 44 | 39,187 | 50,775.780 | -11,588.780 |
| 88 | 44 | 39,722 | 50,775.780 | -11,053.780 |
| 89 | 44 | 40,236 | 50,775.780 | -10,539.780 |
| 90 | 44 | 40,790 | 50,775.780 | -9,985.780 |
| 91 | 44 | 41,352 | 50,775.780 | -9,423.780 |
| 92 | 44 | 41,894 | 50,775.780 | -8,881.780 |
| 93 | 45 | 42,593 | 51,271.300 | -8,678.303 |
| 94 | 45 | 43,280 | 51,271.300 | -7,991.303 |
| 95 | 45 | 44,205 | 51,271.300 | -7,066.303 |
| 96 | 45 | 44,920 | 51,271.300 | -6,351.303 |
| 97 | 47 | 45,589 | 52,260.420 | -6,671.421 |
| 98 | 47 | 46,187 | 52,260.420 | -6,073.421 |
| 99 | 47 | 46,754 | 52,260.420 | -5,506.421 |
| 100 | 47 | 47,470 | 52,260.420 | -4,790.421 |
| 101 | 47 | 48,159 | 52,260.420 | -4,101.421 |
| 102 | 47 | 49,036 | 52,260.420 | -3,224.421 |
| 103 | 47 | 49,629 | 52,260.420 | -2,631.421 |
| 104 | 48 | 50,100 | 52,754.020 | -2,654.017 |
| 105 | 49 | 50,842 | 53,246.970 | -2,404.972 |
| 106 | 50 | 51,387 | 53,739.290 | -2,352.286 |
| 107 | 51 | 52,130 | 54,230.960 | -2,100.960 |
| 108 | 51 | 53,067 | 54,230.960 | -1,163.960 |
| 109 | 51 | 53,976 | 54,230.960 | -254.960 |
| 110 | 51 | 55,003 | 54,230.960 | 772.040 |
| 111 | 51 | 55,931 | 54,230.960 | 1,700.040 |
| 112 | 51 | 56,847 | 54,230.960 | 2,616.040 |
| 113 | 51 | 57,584 | 54,230.960 | 3,353.040 |
| 114 | 52 | 58,455 | 54,721.990 | 3,733.007 |
| 115 | 53 | 59,176 | 55,212.390 | 3,963.612 |
| 116 | 53 | 59,857 | 55,212.390 | 4,644.612 |
| 117 | 53 | 60,724 | 55,212.390 | 5,511.612 |
| 118 | 53 | 61,581 | 55,212.390 | 6,368.612 |
| 119 | 53 | 62,297 | 55,212.390 | 7,084.612 |
| 120 | 53 | 63,025 | 55,212.390 | 7,812.612 |
| 121 | 53 | 63,793 | 55,212.390 | 8,580.612 |
| 122 | 53 | 64,514 | 55,212.390 | 9,301.612 |
| 123 | 53 | 65,285 | 55,212.390 | 10,072.610 |
| 124 | 53 | 66,080 | 55,212.390 | 10,867.610 |
| 125 | 54 | 66,828 | 55,702.140 | 11,125.860 |
| 126 | 54 | 67,580 | 55,702.140 | 11,877.860 |
| 127 | 54 | 68,277 | 55,702.140 | 12,574.860 |
| 128 | 54 | 68,906 | 55,702.140 | 13,203.860 |
| 129 | 54 | 69,494 | 55,702.140 | 13,791.860 |
| 130 | 59 | 70,221 | 58,141.350 | 12,079.650 |
| 131 | 59 | 71,123 | 58,141.350 | 12,981.650 |
| 132 | 59 | 71,900 | 58,141.350 | 13,758.650 |
| 133 | 64 | 72,645 | 60,564.660 | 12,080.340 |
| 134 | 66 | 73,363 | 61,529.550 | 11,833.450 |
| 135 | 66 | 74,272 | 61,529.550 | 12,742.450 |
| 136 | 67 | 74,953 | 62,011.040 | 12,941.960 |
| 137 | 80 | 75,601 | 68,213.070 | 7,387.925 |
| 138 | 82 | 76,295 | 69,157.820 | 7,137.184 |
| 139 | 82 | 76,923 | 69,157.820 | 7,765.184 |
| 140 | 83 | 77,592 | 69,629.250 | 7,962.751 |
| 141 | 83 | 78,213 | 69,629.250 | 8,583.751 |
| 142 | 83 | 79,103 | 69,629.250 | 9,473.751 |
| 143 | 86 | 80,130 | 71,039.800 | 9,090.199 |
| 144 | 87 | 81,132 | 71,508.740 | 9,623.263 |
| 145 | 87 | 81,914 | 71,508.740 | 10,405.260 |
| 146 | 87 | 82,790 | 71,508.740 | 11,281.260 |
| 147 | 89 | 83,553 | 72,444.740 | 11,108.260 |
| 148 | 90 | 84,538 | 72,911.810 | 11,626.190 |
| 149 | 90 | 85,238 | 72,911.810 | 12,326.190 |
| 150 | 90 | 85,968 | 72,911.810 | 13,056.190 |
| 151 | 90 | 86,809 | 72,911.810 | 13,897.190 |
| 152 | 90 | 87,643 | 72,911.810 | 14,731.190 |
| 153 | 90 | 88,476 | 72,911.810 | 15,564.190 |
| 154 | 93 | 89,341 | 74,309.290 | 15,031.700 |
| 155 | 96 | 90,252 | 75,701.200 | 14,550.800 |
| 156 | 96 | 91,118 | 75,701.200 | 15,416.800 |
| 157 | 96 | 92,127 | 75,701.200 | 16,425.800 |
| 158 | 97 | 93,064 | 76,163.930 | 16,900.070 |
| 159 | 97 | 93,977 | 76,163.930 | 17,813.070 |
| 160 | 109 | 94,856 | 81,668.610 | 13,187.390 |
| 161 | 110 | 95,635 | 82,123.340 | 13,511.660 |
| 162 | 110 | 96,516 | 82,123.340 | 14,392.660 |
| 163 | 113 | 97,188 | 83,483.870 | 13,704.130 |
| 164 | 113 | 97,833 | 83,483.870 | 14,349.130 |
| 165 | 115 | 98,765 | 84,387.830 | 14,377.170 |
| 166 | 115 | 99,537 | 84,387.830 | 15,149.170 |
| 167 | 115 | 100,299 | 84,387.830 | 15,911.170 |
| 168 | 115 | 101,063 | 84,387.830 | 16,675.170 |
| 169 | 115 | 101,718 | 84,387.830 | 17,330.170 |
| 170 | 116 | 102,411 | 84,838.900 | 17,572.100 |
| 171 | 116 | 103,081 | 84,838.900 | 18,242.100 |
| 172 | 116 | 103,794 | 84,838.900 | 18,955.100 |
| 173 | 116 | 104,672 | 84,838.900 | 19,833.100 |
| 174 | 116 | 105,368 | 84,838.900 | 20,529.100 |
| 175 | 116 | 106,007 | 84,838.900 | 21,168.100 |
| 176 | 116 | 106,672 | 84,838.900 | 21,833.100 |
| 177 | 117 | 107,400 | 85,289.360 | 22,110.640 |
| 178 | 118 | 108,236 | 85,739.210 | 22,496.790 |
| 179 | 119 | 109,157 | 86,188.450 | 22,968.550 |
| 180 | 120 | 109,991 | 86,637.090 | 23,353.920 |
| 181 | 121 | 110,699 | 87,085.110 | 23,613.890 |
| 182 | 123 | 111,293 | 87,979.350 | 23,313.650 |
| 183 | 123 | 111,982 | 87,979.350 | 24,002.650 |
| 184 | 123 | 112,877 | 87,979.350 | 24,897.650 |
| 185 | 124 | 113,680 | 88,425.550 | 25,254.450 |
| 186 | 124 | 114,404 | 88,425.550 | 25,978.450 |
| 187 | 124 | 115,044 | 88,425.550 | 26,618.450 |
| 188 | 124 | 115,832 | 88,425.550 | 27,406.450 |
| 189 | 124 | 116,640 | 88,425.550 | 28,214.450 |
| 190 | 124 | 117,335 | 88,425.550 | 28,909.450 |
| 191 | 125 | 118,014 | 88,871.160 | 29,142.840 |
| 192 | 126 | 118,972 | 89,316.150 | 29,655.850 |
| 193 | 130 | 119,627 | 91,090.110 | 28,536.890 |
| 194 | 131 | 120,232 | 91,532.090 | 28,699.910 |
| 195 | 131 | 120,860 | 91,532.090 | 29,327.910 |
| 196 | 131 | 121,492 | 91,532.090 | 29,959.910 |
| 197 | 131 | 122,269 | 91,532.090 | 30,736.910 |
| 198 | 131 | 122,838 | 91,532.090 | 31,305.910 |
| 199 | 133 | 123,467 | 92,414.250 | 31,052.750 |
| 200 | 133 | 124,221 | 92,414.250 | 31,806.750 |
| 201 | 133 | 124,738 | 92,414.250 | 32,323.750 |
| 202 | 136 | 125,349 | 93,732.980 | 31,616.020 |
| 203 | 139 | 126,156 | 95,046.310 | 31,109.690 |
| 204 | 139 | 126,901 | 95,046.310 | 31,854.690 |
| 205 | 140 | 127,588 | 95,482.890 | 32,105.110 |
| 206 | 141 | 128,340 | 95,918.870 | 32,421.130 |
| 207 | 143 | 129,247 | 96,789.050 | 32,457.950 |
| 208 | 146 | 129,996 | 98,089.840 | 31,906.160 |
| 209 | 147 | 130,616 | 98,522.240 | 32,093.760 |
| 210 | 149 | 131,342 | 99,385.280 | 31,956.720 |
| 211 | 149 | 131,996 | 99,385.280 | 32,610.720 |
| 212 | 149 | 132,731 | 99,385.280 | 33,345.720 |
| 213 | 149 | 133,722 | 99,385.280 | 34,336.720 |
| 214 | 151 | 134,736 | 100,245.900 | 34,490.070 |
| 215 | 151 | 135,807 | 100,245.900 | 35,561.070 |
| 216 | 152 | 136,590 | 100,675.400 | 35,914.630 |
| 217 | 152 | 137,434 | 100,675.400 | 36,758.630 |
| 218 | 153 | 138,272 | 101,104.200 | 37,167.780 |
| 219 | 154 | 139,087 | 101,532.500 | 37,554.520 |
| 220 | 157 | 139,868 | 102,813.700 | 37,054.300 |
| 221 | 161 | 140,811 | 104,513.700 | 36,297.250 |
| 222 | 161 | 141,482 | 104,513.700 | 36,968.250 |
| 223 | 164 | 142,219 | 105,782.600 | 36,436.390 |
| 224 | 164 | 143,256 | 105,782.600 | 37,473.390 |
| 225 | 165 | 144,038 | 106,204.400 | 37,833.610 |
| 226 | 180 | 144,963 | 112,461.100 | 32,501.940 |
| 227 | 208 | 145,900 | 123,793.000 | 22,106.990 |
| 228 | 209 | 146,597 | 124,189.500 | 22,407.530 |
| 229 | 210 | 147,567 | 124,585.400 | 22,981.650 |
| 230 | 215 | 148,465 | 126,556.300 | 21,908.670 |
| 231 | 220 | 149,213 | 128,513.200 | 20,699.750 |
| 232 | 222 | 150,242 | 129,292.100 | 20,949.900 |
| 233 | 223 | 151,044 | 129,680.700 | 21,363.320 |
| 234 | 223 | 151,954 | 129,680.700 | 22,273.320 |
| 235 | 226 | 152,727 | 130,843.100 | 21,883.910 |
| 236 | 228 | 153,449 | 131,615.200 | 21,833.760 |
| 237 | 228 | 154,084 | 131,615.200 | 22,468.760 |
| 238 | 230 | 154,900 | 132,385.200 | 22,514.830 |
| 239 | 239 | 155,972 | 135,822.400 | 20,149.570 |
| 240 | 242 | 156,905 | 136,958.200 | 19,946.750 |
| 241 | 244 | 157,767 | 137,712.700 | 20,054.280 |
| 242 | 245 | 158,474 | 138,089.100 | 20,384.870 |
| 243 | 253 | 159,337 | 141,080.700 | 18,256.270 |
| 244 | 259 | 160,570 | 143,301.600 | 17,268.440 |
| 245 | 261 | 161,277 | 144,037.500 | 17,239.500 |
| 246 | 263 | 162,127 | 144,771.300 | 17,355.720 |
| 247 | 263 | 162,837 | 144,771.300 | 18,065.720 |
| 248 | 266 | 163,589 | 145,867.900 | 17,721.090 |
| 249 | 269 | 164,385 | 146,959.700 | 17,425.310 |
| 250 | 275 | 165,089 | 149,128.800 | 15,960.210 |
| 251 | 277 | 165,881 | 149,847.500 | 16,033.450 |
| 252 | 278 | 166,846 | 150,206.100 | 16,639.870 |
| 253 | 284 | 167,636 | 152,346.400 | 15,289.560 |
| 254 | 285 | 168,508 | 152,701.300 | 15,806.690 |
| 255 | 289 | 169,409 | 154,115.500 | 15,293.520 |
| 256 | 292 | 170,240 | 155,170.600 | 15,069.430 |
| 257 | 293 | 170,955 | 155,521.200 | 15,433.790 |
| 258 | 294 | 171,745 | 155,871.300 | 15,873.670 |
| 259 | 299 | 172,498 | 157,614.100 | 14,883.930 |
| 260 | 303 | 173,328 | 158,998.900 | 14,329.150 |
| 261 | 317 | 174,007 | 163,780.200 | 10,226.780 |
| 262 | 321 | 174,952 | 165,127.800 | 9,824.223 |
| 263 | 338 | 175,722 | 170,763.900 | 4,958.133 |
| 264 | 341 | 176,484 | 171,743.300 | 4,740.710 |
| 265 | 346 | 177,256 | 173,365.600 | 3,890.382 |
| 266 | 350 | 177,965 | 174,654.500 | 3,310.523 |
| 267 | 356 | 178,662 | 176,572.800 | 2,089.169 |
| 268 | 360 | 179,467 | 177,841.800 | 1,625.180 |
| 269 | 368 | 180,351 | 180,356.100 | -5.146 |
| 270 | 373 | 181,156 | 181,911.700 | -755.671 |
| 271 | 382 | 181,971 | 184,681.000 | -2,709.959 |
| 272 | 386 | 183,059 | 185,899.200 | -2,840.175 |
| 273 | 390 | 183,623 | 187,109.700 | -3,486.694 |
| 274 | 393 | 184,571 | 188,012.500 | -3,441.550 |
| 275 | 398 | 185,518 | 189,507.800 | -3,989.761 |
| 276 | 399 | 186,258 | 189,805.400 | -3,547.375 |
| 277 | 403 | 187,141 | 190,991.100 | -3,850.086 |
| 278 | 406 | 187,879 | 191,875.400 | -3,996.401 |
| 279 | 408 | 188,557 | 192,462.600 | -3,905.586 |
| 280 | 413 | 189,297 | 193,922.300 | -4,625.319 |
| 281 | 415 | 190,021 | 194,502.900 | -4,481.930 |
| 282 | 415 | 190,716 | 194,502.900 | -3,786.930 |
| 283 | 418 | 191,711 | 195,370.300 | -3,659.344 |
| 284 | 420 | 192,531 | 195,946.300 | -3,415.289 |
| 285 | 425 | 193,257 | 197,378.000 | -4,121.024 |
| 286 | 427 | 194,106 | 197,947.500 | -3,841.476 |
| 287 | 429 | 195,012 | 198,515.100 | -3,503.082 |
| 288 | 433 | 196,021 | 199,644.800 | -3,623.768 |
| 289 | 433 | 196,940 | 199,644.800 | -2,704.768 |
| 290 | 434 | 197,679 | 199,926.000 | -2,247.041 |
| 291 | 436 | 198,479 | 200,487.200 | -2,008.214 |
| 292 | 440 | 199,272 | 201,604.100 | -2,332.073 |
| 293 | 440 | 199,944 | 201,604.100 | -1,660.073 |
| 294 | 444 | 200,748 | 202,713.600 | -1,965.644 |
| 295 | 446 | 201,470 | 203,265.700 | -1,795.707 |
| 296 | 452 | 202,260 | 204,911.000 | -2,651.045 |
| 297 | 455 | 202,902 | 205,727.600 | -2,825.635 |
| 298 | 459 | 203,643 | 206,810.100 | -3,167.146 |
| 299 | 464 | 204,594 | 208,153.200 | -3,559.243 |
| 300 | 467 | 205,532 | 208,953.800 | -3,421.768 |
| 301 | 474 | 206,369 | 210,806.200 | -4,437.187 |
| 302 | 475 | 207,149 | 211,069.100 | -3,920.057 |
| 303 | 484 | 207,857 | 213,415.200 | -5,558.193 |
| 304 | 489 | 208,976 | 214,703.400 | -5,727.367 |
| 305 | 492 | 209,686 | 215,471.100 | -5,785.080 |
| 306 | 495 | 210,527 | 216,234.900 | -5,707.916 |
| 307 | 499 | 211,348 | 217,247.400 | -5,899.353 |
| 308 | 500 | 212,036 | 217,499.400 | -5,463.393 |
| 309 | 503 | 212,866 | 218,253.000 | -5,386.950 |
| 310 | 503 | 213,676 | 218,253.000 | -4,576.950 |
| 311 | 504 | 214,647 | 218,503.300 | -3,856.284 |
| 312 | 506 | 215,525 | 219,002.700 | -3,477.677 |
| 313 | 509 | 216,389 | 219,748.600 | -3,359.586 |
| 314 | 510 | 217,146 | 219,996.400 | -2,850.376 |
| 315 | 512 | 218,208 | 220,490.700 | -2,282.690 |
| 316 | 514 | 219,085 | 220,983.300 | -1,898.318 |
| 317 | 514 | 219,869 | 220,983.300 | -1,114.318 |
| 318 | 516 | 220,826 | 221,474.300 | -648.265 |
| 319 | 519 | 221,777 | 222,207.500 | -430.540 |
| 320 | 523 | 222,840 | 223,179.400 | -339.389 |
| 321 | 526 | 223,444 | 223,903.900 | -459.905 |
| 322 | 530 | 223,904 | 224,864.100 | -960.122 |
| 323 | 531 | 224,452 | 225,103.100 | -651.143 |
| 324 | 536 | 224,972 | 226,292.100 | -1,320.071 |
| 325 | 541 | 225,551 | 227,470.800 | -1,919.750 |
| 326 | 543 | 226,084 | 227,939.400 | -1,855.365 |
| 327 | 547 | 226,584 | 228,871.700 | -2,287.718 |
| 328 | 547 | 227,117 | 228,871.700 | -1,754.718 |
| 329 | 553 | 227,628 | 230,258.100 | -2,630.113 |
| 330 | 553 | 228,260 | 230,258.100 | -1,998.113 |
| 331 | 556 | 228,761 | 230,945.900 | -2,184.878 |
| 332 | 558 | 229,297 | 231,402.400 | -2,105.385 |
| 333 | 562 | 229,730 | 232,310.600 | -2,580.608 |
| 334 | 563 | 230,223 | 232,536.700 | -2,313.668 |
| 335 | 566 | 230,885 | 233,212.500 | -2,327.467 |
| 336 | 567 | 231,450 | 233,436.900 | -1,986.942 |
| 337 | 571 | 232,010 | 234,330.900 | -2,320.891 |
| 338 | 582 | 232,600 | 236,756.900 | -4,156.885 |
| 339 | 585 | 233,170 | 237,410.300 | -4,240.342 |
| 340 | 588 | 233,643 | 238,060.300 | -4,417.317 |
| 341 | 590 | 234,261 | 238,491.700 | -4,230.706 |
| 342 | 596 | 234,798 | 239,776.700 | -4,978.664 |
| 343 | 601 | 235,319 | 240,837.000 | -5,517.964 |
| 344 | 606 | 235,866 | 241,887.800 | -6,021.785 |
| 345 | 610 | 236,252 | 242,721.700 | -6,469.657 |
| 346 | 616 | 236,705 | 243,961.200 | -7,256.225 |
| 347 | 631 | 237,214 | 247,001.800 | -9,787.769 |
| 348 | 640 | 237,743 | 248,786.600 | -11,043.580 |
| 349 | 643 | 238,296 | 249,375.000 | -11,079.000 |
| 350 | 646 | 238,868 | 249,960.200 | -11,092.190 |
| 351 | 648 | 239,404 | 250,348.500 | -10,944.530 |
| 352 | 649 | 240,115 | 250,542.200 | -10,427.170 |
| 353 | 653 | 240,656 | 251,313.100 | -10,657.140 |
| 354 | 656 | 241,112 | 251,887.600 | -10,775.650 |
| 355 | 658 | 241,611 | 252,268.900 | -10,657.880 |
| 356 | 658 | 242,180 | 252,268.900 | -10,088.880 |
| 357 | 671 | 242,627 | 254,712.800 | -12,085.760 |
| 358 | 680 | 243,093 | 256,370.400 | -13,277.360 |
| 359 | 691 | 243,755 | 258,358.800 | -14,603.760 |
| 360 | 695 | 244,267 | 259,071.700 | -14,804.680 |
| 361 | 697 | 244,813 | 259,426.100 | -14,613.140 |
| 362 | 705 | 245,362 | 260,830.600 | -15,468.640 |
| 363 | 707 | 245,948 | 261,178.500 | -15,230.450 |
| 364 | 710 | 246,530 | 261,697.700 | -15,167.710 |
| 365 | 714 | 247,067 | 262,385.500 | -15,318.470 |
| 366 | 721 | 247,633 | 263,576.500 | -15,943.540 |
| 367 | 723 | 248,120 | 263,913.900 | -15,793.930 |
| 368 | 724 | 248,702 | 264,082.200 | -15,380.150 |
| 369 | 724 | 249,205 | 264,082.200 | -14,877.150 |
| 370 | 730 | 249,735 | 265,084.700 | -15,349.730 |
| 371 | 733 | 250,321 | 265,581.700 | -15,260.710 |
| 372 | 733 | 250,866 | 265,581.700 | -14,715.710 |
| 373 | 740 | 251,467 | 266,730.300 | -15,263.290 |
| 374 | 747 | 252,068 | 267,863.500 | -15,795.510 |
| 375 | 752 | 252,577 | 268,663.700 | -16,086.650 |
| 376 | 755 | 253,212 | 269,140.000 | -15,928.040 |
| 377 | 762 | 253,834 | 270,240.900 | -16,406.910 |
| 378 | 766 | 254,514 | 270,863.300 | -16,349.310 |
| 379 | 768 | 255,112 | 271,172.700 | -16,060.700 |
| 380 | 769 | 255,964 | 271,326.900 | -15,362.950 |
| 381 | 771 | 256,538 | 271,634.500 | -15,096.540 |
| 382 | 779 | 257,172 | 272,853.000 | -15,681.000 |
| 383 | 784 | 257,758 | 273,604.900 | -15,846.940 |
| 384 | 791 | 258,389 | 274,645.400 | -16,256.390 |
| 385 | 791 | 259,074 | 274,645.400 | -15,571.390 |
| 386 | 796 | 259,674 | 275,379.900 | -15,705.890 |
| 387 | 801 | 260,286 | 276,107.200 | -15,821.210 |
| 388 | 804 | 260,868 | 276,540.200 | -15,672.190 |
| 389 | 807 | 261,504 | 276,970.600 | -15,466.620 |
| 390 | 809 | 262,237 | 277,256.200 | -15,019.170 |
| 391 | 811 | 262,810 | 277,540.600 | -14,730.590 |
| 392 | 815 | 263,520 | 278,106.100 | -14,586.090 |
| 393 | 821 | 264,094 | 278,946.000 | -14,852.010 |
| 394 | 823 | 264,688 | 279,223.800 | -14,535.780 |
| 395 | 823 | 265,725 | 279,223.800 | -13,498.780 |
| 396 | 833 | 266,402 | 280,596.300 | -14,194.260 |
| 397 | 836 | 267,018 | 281,002.700 | -13,984.740 |
| 398 | 840 | 267,597 | 281,541.000 | -13,943.950 |
| 399 | 847 | 268,294 | 282,472.600 | -14,178.610 |
| 400 | 849 | 268,823 | 282,736.400 | -13,913.430 |
| 401 | 852 | 269,395 | 283,130.200 | -13,735.200 |
| 402 | 860 | 269,927 | 284,168.800 | -14,241.830 |
| 403 | 864 | 270,515 | 284,682.000 | -14,166.980 |
| 404 | 873 | 271,137 | 285,821.700 | -14,684.710 |
| 405 | 878 | 271,702 | 286,446.100 | -14,744.110 |
| 406 | 885 | 272,213 | 287,309.900 | -15,096.880 |
| 407 | 892 | 272,898 | 288,161.700 | -15,263.660 |
| 408 | 895 | 273,595 | 288,523.100 | -14,928.080 |
| 409 | 896 | 274,177 | 288,643.100 | -14,466.080 |
| 410 | 904 | 274,857 | 289,594.500 | -14,737.450 |
| 411 | 905 | 275,473 | 289,712.300 | -14,239.310 |
| 412 | 910 | 276,109 | 290,298.100 | -14,189.070 |
| 413 | 918 | 276,694 | 291,223.200 | -14,529.220 |
| 414 | 919 | 277,332 | 291,337.800 | -14,005.830 |
| 415 | 924 | 277,968 | 291,907.400 | -13,939.450 |
| 416 | 926 | 278,756 | 292,133.700 | -13,377.710 |
| 417 | 929 | 279,408 | 292,471.400 | -13,063.420 |
| 418 | 930 | 280,037 | 292,583.500 | -12,546.540 |
| 419 | 932 | 280,851 | 292,807.100 | -11,956.110 |
| 420 | 935 | 281,506 | 293,140.800 | -11,634.810 |
| 421 | 938 | 282,332 | 293,472.500 | -11,140.510 |
| 422 | 942 | 283,171 | 293,911.700 | -10,740.720 |
| 423 | 943 | 283,892 | 294,021.000 | -10,128.970 |
| 424 | 952 | 284,668 | 294,994.600 | -10,326.560 |
| 425 | 959 | 285,413 | 295,739.900 | -10,326.860 |
| 426 | 963 | 286,121 | 296,161.100 | -10,040.130 |
| 427 | 969 | 286,842 | 296,786.800 | -9,944.811 |
| 428 | 974 | 287,666 | 297,302.600 | -9,636.577 |
| 429 | 974 | 288,349 | 297,302.600 | -8,953.577 |
| 430 | 978 | 289,044 | 297,711.500 | -8,667.544 |
| 431 | 982 | 289,889 | 298,117.300 | -8,228.303 |
| 432 | 983 | 290,685 | 298,218.200 | -7,533.245 |
| 433 | 985 | 291,456 | 298,419.500 | -6,963.536 |
| 434 | 991 | 292,097 | 299,018.700 | -6,921.689 |
| 435 | 993 | 292,850 | 299,216.800 | -6,366.847 |
| 436 | 996 | 293,626 | 299,512.600 | -5,886.635 |
| 437 | 999 | 294,383 | 299,806.700 | -5,423.695 |
| 438 | 1,004 | 295,136 | 300,293.000 | -5,156.991 |
| 439 | 1,009 | 295,948 | 300,774.600 | -4,826.583 |
| 440 | 1,017 | 296,930 | 301,535.500 | -4,605.492 |
| 441 | 1,022 | 297,702 | 302,005.100 | -4,303.130 |
| 442 | 1,030 | 298,611 | 302,747.200 | -4,136.233 |
| 443 | 1,039 | 299,542 | 303,568.700 | -4,026.659 |
| 444 | 1,049 | 300,522 | 304,465.100 | -3,943.069 |
| 445 | 1,057 | 301,513 | 305,170.200 | -3,657.161 |
| 446 | 1,065 | 302,344 | 305,864.800 | -3,520.818 |
| 447 | 1,070 | 303,276 | 306,293.800 | -3,017.787 |
| 448 | 1,077 | 304,221 | 306,887.800 | -2,666.768 |
| 449 | 1,082 | 305,028 | 307,307.400 | -2,279.424 |
| 450 | 1,086 | 305,974 | 307,640.400 | -1,666.423 |
| 451 | 1,092 | 306,655 | 308,135.500 | -1,480.453 |
| 452 | 1,093 | 307,511 | 308,217.400 | -706.443 |
| 453 | 1,097 | 308,344 | 308,543.900 | -199.946 |
| 454 | 1,106 | 309,128 | 309,270.200 | -142.195 |
| 455 | 1,117 | 309,974 | 310,142.500 | -168.487 |
| 456 | 1,119 | 310,848 | 310,299.300 | 548.683 |
| 457 | 1,126 | 311,628 | 310,844.000 | 783.964 |
| 458 | 1,143 | 312,578 | 312,140.700 | 437.342 |
| 459 | 1,151 | 313,631 | 312,738.500 | 892.538 |
| 460 | 1,162 | 314,642 | 313,548.100 | 1,093.892 |
| 461 | 1,163 | 315,907 | 313,621.000 | 2,285.977 |
| 462 | 1,172 | 316,788 | 314,272.200 | 2,515.761 |
| 463 | 1,179 | 317,671 | 314,772.600 | 2,898.356 |
| 464 | 1,193 | 318,869 | 315,758.100 | 3,110.852 |
| 465 | 1,195 | 319,736 | 315,897.300 | 3,838.673 |
| 466 | 1,197 | 320,619 | 316,036.100 | 4,582.885 |
| 467 | 1,202 | 321,617 | 316,381.400 | 5,235.599 |
| 468 | 1,210 | 322,771 | 316,929.000 | 5,842.022 |
| 469 | 1,212 | 323,631 | 317,065.000 | 6,566.042 |
| 470 | 1,221 | 324,620 | 317,672.500 | 6,947.514 |
| 471 | 1,225 | 325,676 | 317,940.300 | 7,735.740 |
| 472 | 1,236 | 326,518 | 318,669.800 | 7,848.154 |
| 473 | 1,239 | 327,545 | 318,867.200 | 8,677.845 |
| 474 | 1,244 | 328,527 | 319,194.500 | 9,332.526 |
| 475 | 1,251 | 329,527 | 319,649.600 | 9,877.391 |
| 476 | 1,255 | 330,561 | 319,908.100 | 10,652.890 |
| 477 | 1,259 | 331,504 | 320,165.500 | 11,338.490 |
| 478 | 1,269 | 332,426 | 320,804.300 | 11,621.650 |
| 479 | 1,275 | 333,406 | 321,184.600 | 12,221.380 |
| 480 | 1,285 | 334,379 | 321,813.700 | 12,565.320 |
| 481 | 1,288 | 335,387 | 322,001.300 | 13,385.700 |
| 482 | 1,299 | 336,217 | 322,685.200 | 13,531.770 |
| 483 | 1,310 | 337,022 | 323,363.300 | 13,658.690 |
| 484 | 1,317 | 337,851 | 323,792.100 | 14,058.940 |
| 485 | 1,327 | 338,765 | 324,401.200 | 14,363.790 |
| 486 | 1,337 | 339,517 | 325,006.800 | 14,510.150 |
| 487 | 1,354 | 340,323 | 326,029.700 | 14,293.270 |
| 488 | 1,371 | 341,678 | 327,046.200 | 14,631.820 |
| 489 | 1,392 | 342,482 | 328,296.400 | 14,185.630 |
| 490 | 1,393 | 343,328 | 328,355.800 | 14,972.180 |
| 491 | 1,393 | 344,164 | 328,355.800 | 15,808.180 |
| 492 | 1,393 | 345,029 | 328,355.800 | 16,673.180 |
| 493 | 1,393 | 346,043 | 328,355.800 | 17,687.180 |
| 494 | 1,400 | 346,913 | 328,771.900 | 18,141.110 |
| 495 | 1,409 | 347,623 | 329,306.800 | 18,316.160 |
| 496 | 1,416 | 348,707 | 329,723.100 | 18,983.900 |
| 497 | 1,432 | 349,457 | 330,676.100 | 18,780.870 |
| 498 | 1,463 | 350,282 | 332,535.000 | 17,746.950 |
| 499 | 1,476 | 351,079 | 333,322.200 | 17,756.810 |
| 500 | 1,502 | 351,984 | 334,915.500 | 17,068.470 |
| 501 | 1,511 | 352,564 | 335,474.200 | 17,089.820 |
| 502 | 1,524 | 353,249 | 336,288.700 | 16,960.320 |
| 503 | 1,537 | 354,111 | 337,113.000 | 16,997.980 |
| 504 | 1,552 | 354,981 | 338,077.800 | 16,903.210 |
| 505 | 1,570 | 355,788 | 339,256.900 | 16,531.080 |
| 506 | 1,583 | 356,640 | 340,124.500 | 16,515.510 |
| 507 | 1,605 | 357,494 | 341,626.400 | 15,867.560 |
| 508 | 1,617 | 358,204 | 342,465.100 | 15,738.910 |
| 509 | 1,636 | 359,354 | 343,823.400 | 15,530.570 |
| 510 | 1,660 | 360,219 | 345,597.100 | 14,621.890 |
| 511 | 1,666 | 361,190 | 346,051.300 | 15,138.680 |
| 512 | 1,685 | 362,025 | 347,520.000 | 14,505.030 |
| 513 | 1,685 | 362,743 | 347,520.000 | 15,223.030 |
| 514 | 1,717 | 363,383 | 350,105.000 | 13,277.990 |
| 515 | 1,740 | 364,225 | 352,056.800 | 12,168.150 |
| 516 | 1,756 | 365,107 | 353,464.400 | 11,642.570 |
| 517 | 1,778 | 366,116 | 355,470.600 | 10,645.410 |
| 518 | 1,791 | 366,818 | 356,696.400 | 10,121.630 |
| 519 | 1,815 | 367,744 | 359,042.200 | 8,701.798 |
| 520 | 1,834 | 368,633 | 360,979.200 | 7,653.846 |
| 521 | 1,873 | 369,559 | 365,190.700 | 4,368.305 |
| 522 | 1,890 | 370,901 | 367,131.300 | 3,769.671 |
| 523 | 1,896 | 372,115 | 367,832.100 | 4,282.868 |
| 524 | 1,924 | 373,179 | 371,215.900 | 1,963.124 |
| 525 | 1,944 | 373,968 | 373,751.200 | 216.783 |
| 526 | 1,960 | 374,857 | 375,853.400 | -996.431 |
| 527 | 1,995 | 375,765 | 380,691.200 | -4,926.173 |
| 528 | 2,023 | 376,835 | 384,808.700 | -7,973.713 |
| 529 | 2,041 | 377,775 | 387,577.100 | -9,802.102 |
| 530 | 2,053 | 378,603 | 389,477.100 | -10,874.130 |
| 531 | 2,085 | 379,545 | 394,763.600 | -15,218.640 |
| 532 | 2,108 | 380,502 | 398,767.500 | -18,265.480 |
| 533 | 2,136 | 381,388 | 403,881.600 | -22,493.550 |
| 534 | 2,160 | 382,361 | 408,482.000 | -26,121.030 |
| 535 | 2,184 | 383,259 | 413,289.700 | -30,030.730 |
| 536 | 2,209 | 384,208 | 418,525.200 | -34,317.160 |

----------------------------------------

Analysis of Variance Table - Cubic Model

=============================================================================================

| | Df | Sum Sq | Mean Sq | F value | Pr(> F) |
|---|---|---|---|---|---|
| Linear effect | 1 | 6,040,593,804,231.000000 | 6,040,593,804,231.000000 | 21,067.270000 | 0 |
| Quadratic effect | 1 | 620,523,772,463.000000 | 620,523,772,463.000000 | 2,164.148000 | 0 |
| Cubic effect | 1 | 97,715,724,176.000000 | 97,715,724,176.000000 | 340.794800 | 0 |
| Residuals | 532 | 152,539,767,778.000000 | 286,728,887.000000 | | |

-----------------------------------------------------------------------------------------

```
Cubic Model
================================================
    Estimate     Std. Error   t value   Pr(> | t| )
--------------------------------------------------
b0 28,329.280000 1,430.630000 19.801960      0
b1  524.603900    9.151839   57.322240      0
b2   -0.332026    0.012391   -26.795290     0
b3   0.000079     0.000004   18.460630      0
--------------------------------------------------

R-square - Cubic Model
========
0.977929
--------

Equation - Cubic Model
==========================================
28329.2791+524.603943x-0.332026x2+0.000079x3
------------------------------------------

#Selecting Quadratic given R^2
regr_easy_graf(CumMPIGuides, CumMPI, model="Q", plot_eq=0, plot_R2=0, title_y="Cumulative FDA-Issued Medical Product Guidelines", title_x="Cumulative FDA-Registered Medical Products")

#Structual analysis
library(WaveletComp);citation("WaveletComp")
  Roesch A, Schmidbauer H (2018). _WaveletComp: Computational Wavelet Analysis_. R package
  version 1.1, <https://CRAN.R-project.org/package=WaveletComp>.

one=two=three=four=five=six=seven=eight=0;

Month<-0;Month<-seq(as.Date("1976/5/1"), by = "month", length.out=536)
df<-0;df<-data.frame(date=Month, x=Ucurve[,1])
one<-analyze.wavelet(df,c("x"), dt=1/12, upperPeriod=64, make.pval=TRUE, n.sim=250)
X11();wt.image(one, main="FDA-Registered Medical Devices Wavelet Power", legend.params=list(lab="", lab.line=2.0, label.digits=3), periodlab="Period (Years)", show.date=TRUE, date.format="%F %T", timelab="Year", lwd=3)

Month<-0;Month<-seq(as.Date("1976/5/1"), by = "month", length.out=536)
df<-0;df<-data.frame(date=Month, x=Ucurve[,2])
two<-analyze.wavelet(df,c("x"), dt=1/12, upperPeriod=64, make.pval=TRUE, n.sim=250)
X11();wt.image(two, main="FDA-Registered Medicines Wavelet Power", legend.params=list(lab="", lab.line=2.0, label.digits=3), periodlab="Period (Years)", show.date=TRUE, date.format="%F %T", timelab="Year", lwd=3)

Month<-0;Month<-seq(as.Date("1976/5/1"), by = "month", length.out=536)
df<-0;df<-data.frame(date=Month, x=Ucurve[,3])
three<-analyze.wavelet(df,c("x"), dt=1/12, upperPeriod=64, make.pval=TRUE, n.sim=250)
```

```
X11();wt.image(three, main="FDA-Issued Medical Device Guidelines Wavelet Power",
legend.params=list(lab="", lab.line=2.0, label.digits=3), periodlab="Period (Years)", show.date=TRUE,
date.format="%F %T", timelab="Year", lwd=3)

Month<-0;Month<-seq(as.Date("1976/5/1"), by = "month", length.out=536)
df<-0;df<-data.frame(date=Month, x=Ucurve[,4])
four<-analyze.wavelet(df,c("x"), dt=1/12, upperPeriod=64, make.pval=TRUE, n.sim=250)
X11();wt.image(four, main="FDA-Issued Medicines Guidelines Wavelet Power",
legend.params=list(lab="", lab.line=2.0, label.digits=3), periodlab="Period (Years)", show.date=TRUE,
date.format="%F %T", timelab="Year", lwd=3)

Month<-0;Month<-seq(as.Date("1976/5/1"), by = "month", length.out=536)
df<-0;df<-data.frame(date=Month, x=Ucurve[,5])
five<-analyze.wavelet(df,c("x"), dt=1/12, upperPeriod=64, make.pval=TRUE, n.sim=250)
X11();wt.image(five, main="FDA-Registered Medical Devices and Medicines Wavelet Power",
legend.params=list(lab="", lab.line=2.0, label.digits=3), periodlab="Period (Years)", show.date=TRUE,
date.format="%F %T", timelab="Year", lwd=3)

Month<-0;Month<-seq(as.Date("1976/5/1"), by = "month", length.out=536)
df<-0;df<-data.frame(date=Month, x=Ucurve[,6])
six<-analyze.wavelet(df,c("x"), dt=1/12, upperPeriod=64, make.pval=TRUE, n.sim=250)
X11();wt.image(six, main="FDA-Issued Medical Devices and Medicines Guidelines Wavelet Power",
legend.params=list(lab="", lab.line=2.0, label.digits=3), periodlab="Period (Years)", show.date=TRUE,
date.format="%F %T", timelab="Year", lwd=3)

Month<-0;Month<-seq(as.Date("1976/5/1"), by = "month", length.out=536)
df<-0;df<-data.frame(date=Month, x=Ucurve[,7])
seven<-analyze.wavelet(df,c("x"), dt=1/12, upperPeriod=64, make.pval=TRUE, n.sim=250)
X11();wt.image(seven, main="Cumulative FDA-Registered Medical Devices and Medicines Wavelet
Power", legend.params=list(lab="", lab.line=2.0, label.digits=3), periodlab="Period (Years)",
show.date=TRUE, date.format="%F %T", timelab="Year", lwd=3)

Month<-0;Month<-seq(as.Date("1976/5/1"), by = "month", length.out=536)
df<-0;df<-data.frame(date=Month, x=Ucurve[,8])
eight<-analyze.wavelet(df,c("x"), dt=1/12, upperPeriod=64, make.pval=TRUE, n.sim=250)
X11();wt.image(eight, main="Cumulative FDA-Issued Medical Devices and Medicines Guidelines Wavelet
Power", legend.params=list(lab="", lab.line=2.0, label.digits=3), periodlab="Period (Years)",
show.date=TRUE, date.format="%F %T", timelab="Year", lwd=3)

#Wavelet Coherency
Month<-seq(as.Date("1976/5/1"), by = "month", length.out=536)
my.data<-0;my.data<-data.frame(date=Month, x=Ucurve[,8], y=Ucurve[,7])

str(my.data)
'data.frame':   536 obs. of  3 variables:
 $ date: Date, format:  ...
 $ x   : int  255 442 721 1016 1380 1673 2068 2449 2769 3090 ...
 $ y   : int  0 0 0 2 2 4 5 5 6 6 ...
```

```
out<-analyze.coherency(my.data,c("x", "y"), dt=1/12, upperPeriod=64, make.pval=TRUE, n.sim=250)
wc.image(out, main="Wavelet Power", legend.params=list(lab="Wavelet Power (Quantiles)",
lab.line=2.0, label.digits=5), periodlab="Period (Years)", show.date=TRUE, date.format="%F %T",
timelab="Year", lwd=3)
wc.avg(out, siglvl = 0.05, sigcol = "red", periodlab = "Period (Years)")
#Indication of lag: wc.sel.phases(out, only.coi = FALSE)


#END
```